\documentclass[11pt]{article}

\usepackage{graphics}
\usepackage{graphicx}
\graphicspath{{./}{figures/}}
\usepackage{amsmath}
\usepackage{cite}
\usepackage{axodraw}
\usepackage{rotating}
\usepackage{url}
\usepackage{xspace}
\usepackage{color}
\usepackage{wasysym}
\usepackage{multirow}
\usepackage{comment}        
\makeatletter
\if@twoside \oddsidemargin -17pt \evensidemargin 00pt
\else \oddsidemargin 00pt \evensidemargin 00pt
\fi
\expandafter\ifx\csname mathrm\endcsname\relax\def\mathrm#1{{\rm #1}}\fi

\renewcommand{\theequation}{\thesection.\arabic{equation}}
\newcounter{saveeqn}

\@addtoreset{equation}{section}

\makeatother

\def\nl{\nonumber\\}

\def\beq{\begin{equation}}
\def\eeq{\end{equation}}
\def\ba{\begin{eqnarray}}
\def\ea{\end{eqnarray}}
\def\beqar{\begin{eqnarray}}
\def\eeqar{\end{eqnarray}}
\def\barr#1{\begin{array}{#1}}
\def\earr{\end{array}}
\def\bfi{\begin{figure}}
\def\efi{\end{figure}}
\def\btab{\begin{table}}
\def\etab{\end{table}}
\def\bce{\begin{center}}
\def\ece{\end{center}}
\def\nn{\nonumber}

\def\text{\textstyle}
\def\arraystretch{1.0}

\def\pTo#1{p_{\mathrm{T},\, #1}}
\def\k#1{k_{#1}}
\def\kt#1{\tilde{k}_{#1}}
\newcommand{\HT}{H_\mathrm{T}}
\newcommand{\phill}{\phi_{\Plm\Plp}}

\def\al{\alpha}

\def\ga{\gamma}
\def\de{\delta}

\def\si{\sigma}
\def\Ga{\Gamma}
\def\De{\Delta}

\def\refeq#1{\mbox{(\ref{#1})}}
\def\reffi#1{\mbox{Fig.~\ref{#1}}}

\def\refta#1{\mbox{Table~\ref{#1}}}

\def\refse#1{\mbox{Section~\ref{#1}}}
\def\refses#1{\mbox{Sections~\ref{#1}}}
\def\refapp#1{\mbox{App.~\ref{#1}}}

\def\citere#1{\mbox{Ref.~\cite{#1}}}
\def\citeres#1{\mbox{Refs.~\cite{#1}}}

\newcommand{\GeV}{\unskip\,\mathrm{GeV}}

\newcommand{\TeV}{\unskip\,\mathrm{TeV}}
\newcommand{\fba}{\unskip\,\mathrm{fb}}

\newcommand{\pba}{\unskip\,\mathrm{pb}}

\def\mathswitch#1{\relax\ifmmode#1\else$#1$\fi}
\def\mathswitchr#1{\relax\ifmmode{\mathrm{#1}}\else$\mathrm{#1}$\fi}
\def\mathswitchit#1{\relax\ifmmode{#1}\else$#1$\fi}
\newcommand{\muF}{\mu_{\rm F}}
\newcommand{\zg}{z_\gamma}

\newcommand{\PW}{\mathswitchr W}
\newcommand{\PZ}{\mathswitchr Z}
\newcommand{\Pg}{\mathswitchr g}
\newcommand{\PH}{\mathswitchr H}
\newcommand{\Pe}{\mathswitchr e}
\newcommand{\Pd}{\mathswitchr d}
\newcommand{\Pj}{\mathswitchr j}
\newcommand{\Pl}{\mathswitch \ell}
\newcommand{\Plp}{\mathswitch \ell^+}
\newcommand{\Plm}{\mathswitch \ell^-}
\newcommand{\Pu}{\mathswitchr u}
\newcommand{\Ps}{\mathswitchr s}
\newcommand{\Pb}{\mathswitchr b}
\newcommand{\Pc}{\mathswitchr c}

\newcommand{\Pp}{\mathswitchr p}
\newcommand{\Pq}{\mathswitchit q}

\newcommand{\MW}{\mathswitch {M_{\PW}}}

\newcommand{\MZ}{\mathswitch {M_{\PZ}}}
\newcommand{\MH}{\mathswitch {M_{\PH}}}

\newcommand{\Mt}{\mathswitch {m_{\PQt}}}

\newcommand{\GZ}{\mathswitch {\Gamma_{\PZ}}}

\newcommand{\zcut}{z_{\gamma}^{\rm cut}}
\newcommand{\dd}{{\rm d}}
\newcommand{\eps}{\epsilon}

\newcommand{\GF}{\mathswitch {G_\mu}}

\newcommand{\alphas}{\alpha_{\mathrm{s}}}
\newcommand{\gs}{g_{\mathrm{s}}}

\newcommand{\ub}{\bar{u}}
\newcommand{\db}{\bar{d}}

\def\ie{i.e.\ }

\newcommand{\ord}{{\cal O}}
\renewcommand{\O}{{\cal O}}

\newcommand{\ri}{{\mathrm{i}}}

\newcommand{\rT}{{\mathrm{T}}}

\newcommand{\rd}{{\mathrm{d}}}

\newcommand{\M}{{\cal {M}}}

\newcommand{\Pcal}{{\cal {P}}}

\newcommand{\cut}{{\mathrm{cut}}}

\newcommand{\EW}{{\mathrm{EW}}}

\newcommand{\OS}{{\mathrm{OS}}}

\newcommand{\LO}{{\mathrm{LO}}}

\newcommand{\NLO}{{\mathrm{NLO}}}

\mathchardef\mhyphen="2D 

\def\draftdate{\relax}
\def\mda{\relax}
\def\mua{\relax}
\def\mla{\relax}
\def\draft{
\def\thtystars{******************************}
\def\sixtystars{\thtystars\thtystars}
\typeout{}
\typeout{\sixtystars**}
\typeout{* Draft mode!
         For final version remove \protect\draft\space in source file *}
\typeout{\sixtystars**}
\typeout{}
\def\draftdate{\today}
\def\mua{\marginpar[\boldmath\hfil$\uparrow$]%
                   {\boldmath$\uparrow$\hfil}%
                    \typeout{marginpar: $\uparrow$}\ignorespaces}
\def\mda{\marginpar[\boldmath\hfil$\downarrow$]%
                   {\boldmath$\downarrow$\hfil}%
                    \typeout{marginpar: $\downarrow$}\ignorespaces}
\def\mla{\marginpar[\boldmath\hfil$\rightarrow$]%
                   {\boldmath$\leftarrow $\hfil}%
                    \typeout{marginpar: $\leftrightarrow$}\ignorespaces}
\def\Mua{\marginpar[\boldmath\hfil$\Uparrow$]%
                   {\boldmath$\Uparrow$\hfil}%
                    \typeout{marginpar: $\Uparrow$}\ignorespaces}
\def\Mda{\marginpar[\boldmath\hfil$\Downarrow$]%
                   {\boldmath$\Downarrow$\hfil}%
                    \typeout{marginpar: $\Downarrow$}\ignorespaces}
\def\Mla{\marginpar[\boldmath\hfil$\Rightarrow$]%
                   {\boldmath$\Leftarrow $\hfil}%
                    \typeout{marginpar: $\Leftrightarrow$}\ignorespaces}
\overfullrule 5pt
\oddsidemargin -15mm
\marginparwidth 29mm
}

\makeatletter

\def\eqnarray{\stepcounter{equation}\let\@currentlabel=\theequation
\global\@eqnswtrue
\global\@eqcnt\z@\tabskip\@centering\let\\=\@eqncr
$$\halign to \displaywidth\bgroup\hskip\@centering
  $\displaystyle\tabskip\z@{##}$\@eqnsel&\global\@eqcnt\@ne
  \hskip 2\arraycolsep \hfil${##}$\hfil
  &\global\@eqcnt\tw@ \hskip 2\arraycolsep $\displaystyle\tabskip\z@{##}$\hfil
   \tabskip\@centering&\llap{##}\tabskip\z@\cr}
\def\appendix{\par
 \setcounter{section}{0} \setcounter{subsection}{0}
 \def\thesection{\Alph{section}}}

\newcommand{\lsim}
{\;\raisebox{-.3em}{$\stackrel{\displaystyle <}{\sim}$}\;}
\newcommand{\gsim}
{\;\raisebox{-.3em}{$\stackrel{\displaystyle >}{\sim}$}\;}

\def\dsl{\mathpalette\make@slash}
\def\make@slash#1#2{\setbox\z@\hbox{$#1#2$}%
  \hbox to 0pt{\hss$#1/$\hss\kern-\wd0}\box0}

\newcommand{\Pubar}{{\bar{\Pu}}}
\newcommand{\Pqbar}{{\bar{\Pq}}}

\newcommand{\RL}{\mathrm{L}}

\makeatother

\providecommand{\PQt}{\mathswitchr {t}}

\providecommand{\process}{\Pp \Pp \to \Pj\Pj\Plm\Plp }

\newcommand{\recola}{{\sc Recola}}
\newcommand{\collier}{{\sc Collier}}

\newcommand{\zgacut}{z_{\gamma, {\rm cut}}}

\begin{document}

\thispagestyle{empty}
\def\thefootnote{\fnsymbol{footnote}}
\setcounter{footnote}{1}
\null
\draftdate\hfill 
\\
\vskip 1cm
\begin{center}
  {\Large \boldmath{\bf Electroweak corrections to lepton pair
      production in association with two hard jets at the LHC}
\par} \vskip 2.5em
{\large
{\sc A.~Denner$^{1}$, L.~Hofer$^{2}$, A.~Scharf$^{1}$, S.~Uccirati$^{3}$}\\[3ex]
{\normalsize \it
$^1$Universit\"at W\"urzburg, 
Institut f\"ur Theoretische Physik und Astrophysik, \\
D-97074 W\"urzburg, Germany}\\[1ex]
{\normalsize \it
$^2$Institut de F\'isica d'Altes Energies (IFAE) Edifici Cn, \\
Universitat Aut\`onoma de Barcelona (UAB), \\
E-08193 Bellaterra (Barcelona), Spain}\\[1ex]
{\normalsize \it
$^3$Universit\`a di Torino, Dipartimento di Fisica, Italy\\
INFN, Sezione di Torino, Italy}\\[1ex]
}
\par \vskip 1em
\end{center}\par
\vfill \vskip .0cm \vfill {\bf Abstract:} 
\par 
We compute the next-to-leading order corrections of
$\O(\alphas^2\alpha^3)$ to the hadronic production of two oppositely
charged leptons and two hard jets, $\process$, using {\recola} and
{\collier}.  We include electroweak and QCD corrections at the given
order and all off-shell effects.  We provide detailed predictions for
the LHC operating at $13\TeV$ and obtain per-cent-level corrections
for the total cross section.  For differential distributions we find
significant non-uniform distortions in high-energy tails {at the level
  of several ten per cent} due to electroweak Sudakov logarithms and
deformations at the level of a few per cent for angular variables.
\par
\vskip 1cm
\noindent
November 2014
\par
\null
\setcounter{page}{0}
\clearpage
\def\thefootnote{\arabic{footnote}}
\setcounter{footnote}{0}

\section{Introduction}

After the discovery of the Higgs boson, the search for physics beyond
the Standard Model (SM) is the primary goal of the Large Hadron
Collider (LHC).  The lack of evident phenomena of new physics at the
TeV scale {calls for} a precise study of the SM in order to reveal
possible small deviations between theory and experiment.  This can be
achieved through sophisticated experimental analyses, capable of
highlighting a small signal on a huge background, whose knowledge is
essential to interpret the data.  Sometimes data are not sufficient
for an accurate estimation of the background and theoretical
descriptions become important.  Moreover, data-driven determinations
of the background often rely on extrapolations to the signal region
based on theoretical distributions.  Minimising theoretical
uncertainties becomes therefore necessary not only for the signal
processes, but also for all those processes that can contribute to the
background.

In order to achieve the needed precision, leading-order (LO)
predictions in perturbation theory are not sufficient for most cases.
At hadron colliders QCD corrections can be of the order of several ten
per cent and have been carefully studied for many processes.
Electroweak (EW) corrections are often small for inclusive
observables; nevertheless they can have an important impact and should
thus be studied carefully.  In particular, they are typically strongly
enhanced in high-energy tails of distributions, where for the first
time the LHC will collect enough data to see the effects of sizable
logarithms of EW origin.  Moreover, in particular cases like Higgs
production in vector-boson fusion, EW corrections can be of the same
order of magnitude as QCD corrections~\cite{Ciccolini:2007ec}.  For
these reasons EW corrections represent the next frontier of
next-to-leading-order (NLO) calculations for LHC physics, as pointed
out by the most recent Les Houches
Wish-list~\cite{Butterworth:2014efa}.

In the past years, many groups have concentrated their efforts to make
NLO calculations feasible, and a lot of codes have been
designed~\cite{
  Hahn:1998yk,Arnold:2008rz,Berger:2008sj,Becker:2010ng,Badger:2010nx,
  Hirschi:2011pa,Bevilacqua:2011xh,Cullen:2011ac,Cascioli:2011va} with
a high level of automatisation and impressive performances, however
with their range of applicability mostly restricted to the QCD sector
of the SM.  For EW corrections, the situation is more involved and a
complete automatisation has not been achieved yet, while different
groups are working in this direction
\cite{Frixione:2014qaa,Cullen:2014yla}.  Recently we have developed
the code \recola~\cite{Actis:2012qn,Uccirati:2014fda} which {performs
  efficient calculations of tree-level and one-loop amplitudes in the
  full SM. {\recola} uses} an alternative approach to Feynman
diagrams, based on recursion relations for off-shell currents.  The
algorithm, originally proposed by Andreas van Hameren
in~\citere{vanHameren:2009vq} for gluonic amplitudes, is based on the
decomposition of one-loop amplitudes as linear combinations of tensor
integrals, whose coefficients are calculated recursively.  The tensor
integrals are computed by linking {\recola} to the \collier\ 
library~\cite{collier,Denner:2014gla}, which provides one-loop scalar
and tensor integrals for arbitrary scattering processes.

A class of SM background processes, particularly important for
searches of new physics, is the production of a weak boson accompanied
by jets ($\Pp\Pp \to \PW/\PZ +{}$jets).  If for example the
$\PZ$~boson decays into neutrinos, the process $\PZ+2\,$jets has the
same signature (missing energy plus 2 jets) as the production of a
pair of squark and anti-squark, each 
subsequently decaying into a jet
and an invisible neutralino.  Such events are mainly searched for in
high-energy regions, where EW corrections are usually sizable.  
The experimental estimation of the irreducible SM background is
obtained by data-driven extrapolations from measured control samples,
where the gauge boson decays into charged leptons.  The process
$\Pp\Pp \to \PZ + 2\,$jets $\to \Plm\Plp + 2\,$jets is a typical
ingredient in these studies.  Moreover, the production of a Z~boson
with two jets provides an important background to Higgs-boson
production in vector-boson fusion~\cite{Khoze:2002fa,Oleari:2003tc}.
The signature of the Higgs signal consists typically of
two jets in forward and backward rapidity regions and a Higgs boson
decaying in the central region of the detector.  Analysing the process
$\Pp\Pp \to \PZ + 2\,$jets $\to \Plm\Plp + 2$ jets in this kinematic
region offers the possibility to carefully study the systematics of
the $\PH + 2\,$jets final state.  Analyses of LHC data with an
integrated luminosity of $5\fba^{-1}$ at $7\TeV$ and $20\fba^{-1}$ at
$8\TeV$ have appeared in~\citeres{Aad:2013ysa,CMS:2014gba,CMS:2014lga}
for Z~production in association with jets. Moreover, the pure EW
contribution to Z~production with two jets has been measured by the
ATLAS and CMS collaboration at $7\TeV$ and $8\TeV$
\cite{Chatrchyan:2013jya,Khachatryan:2014dea,Aad:2014dta}.

The LO amplitude of the process $\Pp\Pp \to \Plm\Plp + 2\,$jets gets
contributions from pure EW tree-level diagrams on top of the dominant
diagrams involving gluons (QCD tree level).  The QCD corrections to
the LO QCD contributions have been investigated
in~\citeres{Campbell:2002tg,Campbell:2003hd}, while QCD corrections to
EW LO contributions with vector-boson-fusion topology have been
computed in~\citere{Oleari:2003tc}.  The QCD corrections to the total
cross section turn out to be of the order of $10\%$.  The NLO QCD
calculations have been matched to parton showers both for the QCD
mediated processes \cite{Re:2012zi,Campbell:2013vha} and the
vector-boson-fusion-mediated processes
\cite{Jager:2012xk,Schissler:2013nga}.
In~\citere{Actis:2012qn} we performed a first study of EW corrections
to the process $\Pp\Pp \to \PZ + 2\,$jets with an on-shell Z~boson.
Restricting our attention to the dominant partonic processes involving
{external} gluons, $\Pq\,\Pg \to \Pq\,\Pg\,\PZ$, {$\Pqbar\,\Pg \to
  \Pqbar\,\Pg\,\PZ$,} $\Pg\,\Pg \to \Pq\,\Pqbar\,\PZ$, $\Pq\,\Pqbar
\to \Pg\,\Pg\,\PZ$, we found small EW effects on the total cross
section at the level of $-1\%$, while transverse-momentum
distributions received enhanced corrections at high $p_\rT$ (up to
$-25\%$ for $p_{\rm T}\simeq 1\TeV$), owing to EW Sudakov logarithms.
The large effects of EW logarithms have been also studied in
next-to-leading logarithmic approximation {for
  $\Pp\Pp\to\Pj\Pj\nu\bar{\nu}$} in~\citere{Chiesa:2013yma}. For the
production of a vector boson with one jet the complete EW corrections
are available \cite{Denner:2009gj,Denner:2011vu,Denner:2012ts}.

In this paper we perform a complete study of EW corrections of ${\cal
  O}(\alpha_s^2\alpha^3)$ for the process $\Pp\Pp$ $\to \Plm\Plp +
2\,$jets.  In \refse{sec: features} the framework of our calculation
is presented (\refse{sec: setup},\refse{sec: LO pole approximation}),
as well as the results of the leading-order computation
(Sections~\ref{sec: LO standard cuts},~\ref{sec: LO VBF cuts}).  The
EW NLO corrections are analysed in Section~\ref{sec: NLO}: the
calculation of the virtual and real corrections is sketched in
\refses{sec: virtual} and \ref{sec: real} respectively; in
\refses{sec: NLO standard cuts} and \ref{sec: NLO VBF cuts} the
results at NLO are presented for standard acceptance cuts and
vector-boson-fusion cuts.  Finally, \refse{sec: conclusions} contains
our conclusions, and in \refapp{app: photon fragmentation} our
implementation of photon fragmentation is described.

\section{Production of  $\process$ in LO at the LHC}
\label{sec: features}
In this section we define the general setup of our computation and describe
basic features of lepton pair production in association with two hard
jets at the LHC.
\subsection{General setup}
\label{sec: setup}
The hadronic production of two oppositely charged leptons and two hard
jets $\process$ proceeds via the partonic
subprocesses 
\ba
\Pq_i\,\Pg&\to&\Pq_i\,\Pg\,\Plm\,\Plp,\label{eq:born_qg}\\
\Pq_i\,\Pq_j&\to&\Pq_i\,\Pq_j\,\Plm\,\Plp,
\label{eq:born_qq}{\qquad
  q_i,q_j=\Pu,\Pc,\Pd,\Ps,\Pb} \ea and their crossing-related
counterparts.  Since we neglect flavour mixing as well as the masses
of light quarks ($\Pu,\Pc,\Pd,\Ps,\Pb$), the LO amplitudes do not
depend on the quark generation, and the contributions of the various
generations to the cross section differ only by their parton
luminosities.  All partonic processes can be constructed from the six
basic channels $\Pu\Pg\to \Pu\Pg\,\Plm\Plp$, $\Pd\Pg\to
\Pd\Pg\,\Plm\Plp$, $\Pu\Ps\to \Pu\Ps\,\Plm\Plp$, $\Pu\Pc\to
\Pu\Pc\,\Plm\Plp$, $\Pd\Ps\to \Pd\Ps\,\Plm\Plp$, and $\Pu\Ps\to
\Pd\Pc\,\Plm\Plp$ via crossing symmetry and combination.  While the
mixed quark--gluon (gluonic) channels \refeq{eq:born_qg} contribute to
the cross section exclusively at order
$\mathcal{O}(\alpha^2\alphas^2)$, the four-quark channels
\refeq{eq:born_qq} develop LO diagrams of strong as well as of EW
nature leading to contributions of order
$\mathcal{O}(\alpha^2\alphas^2)$, $\mathcal{O}(\alpha^3\alphas)$, and
$\mathcal{O}(\alpha^4)$ to the cross section (see \reffi{fig:LOdiags}
first two lines for sample diagrams).  Owing to the colour structure,
nonzero contributions of $\mathcal{O}(\alpha^3\alphas)$ only appear in
interferences between diagrams with different fermion number flow, and
therefore only in partonic channels with identical or
weak-isospin-partner quarks (see \reffi{fig:tree_interferences} for an
example).  Additional contributions arise from photon-induced
production mechanisms \ba \Pg\,\ga &\to&
\Pq_i\,\bar{\Pq}_i\,\Pl^-\,\Pl^+,\nl {\Pq}_i\,\ga &\to&
\Pq_i\,\Pg\,\Pl^-\,\Pl^+ ,\nl {\bar{\Pq}_i\,\ga} &{\to}&
{\bar{\Pq}_i\,\Pg\,\Pl^-\,\Pl^+ ,}\nl \ga\,\ga &\to&
\Pq_i\,\bar{\Pq}_i\,\Pl^-\,\Pl^+
\label{eq:gamma-LO}
\ea
(see \reffi{fig:LOdiags} last line for sample diagrams).
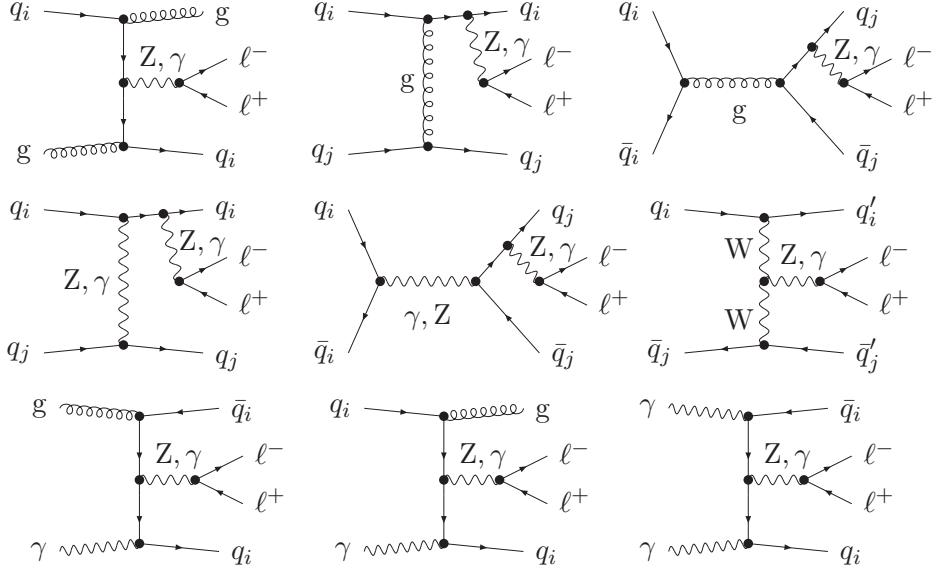
\begin{figure}
\centerline{
{\unitlength .6pt 
\begin{picture}(125,100)(0,0)       
\SetScale{.6}
\ArrowLine( 15,95)( 65,90)
\Gluon( 15, 5)( 65,10){3}{7}
\ArrowLine( 65,10)(115, 5)
\Gluon(115,95)( 65,90){3}{7}
\ArrowLine( 65,90)( 65,40)
\ArrowLine( 65,40)( 65,10)
\Photon(100,50)( 65,50){3}{4}
\ArrowLine( 100,50)( 130,65)
\ArrowLine( 130,35)( 100,50)
\Vertex(65,90){3}
\Vertex(65,10){3}
\Vertex(65,50){3}
\Vertex(100,50){3}
\Text(  7,95)[r]{$\Pq_i$}
\Text(  7,5)[r]{$\Pg$}
\put(123,90){{$\Pg$}}
\put(123,0){{$\Pq_i$}}
\put(75,61){{$\PZ,\ga$}}
\put(138,60){{$\Plm$}}
\put(138,30){{$\Plp$}}
\SetScale{1}
\end{picture}}
\hspace*{3em}
{\unitlength .6pt 
\begin{picture}(125,100)(0,0)       
\SetScale{.6}
\ArrowLine( 15,95)( 65,90)
\ArrowLine( 15, 5)( 65,10)
\ArrowLine( 65,10)(115, 5)
\ArrowLine( 65,90)(90,92.5)
\ArrowLine( 90,92.5)(115,95)
\Gluon( 65,10)( 65,90){3}{8}
\Photon(100,50)( 90,92.5){3}{4}
\ArrowLine( 100,50)( 130,65)
\ArrowLine( 130,35)( 100,50)
\Vertex(65,90){3}
\Vertex(65,10){3}
\Vertex(90,92.5){3}
\Vertex(100,50){3}
\Text(  7,95)[r]{$\Pq_i$}
\Text(  7,5)[r]{${\Pq}_j$}
\Text(123,95)[l]{{$q_i$}}
\Text(123,0)[l]{{${\Pq}_j$}}
\put(138,60){{$\Plm$}}
\put(138,30){{$\Plp$}}
\put(100,70){{$\PZ,\ga$}}
\Text(57,50)[r]{{$\Pg$}}
\SetScale{1}
\end{picture}}
\hspace*{3em}
{\unitlength .6pt 
\begin{picture}(145,100)(0,0)       
\SetScale{.6}
\ArrowLine( 15,95)( 35,50)
\ArrowLine( 35,50)( 15, 5)
\ArrowLine(95,50)(115,72.5)
\ArrowLine(115,72.5)(135,95)
\ArrowLine(135, 5)(95,50)
\Gluon( 35,50)(95,50){3}{7}
\Photon(115,72.5)(135,50){3}{4}
\ArrowLine( 135,50)( 165,65)
\ArrowLine( 165,35)( 135,50)
\Vertex( 35,50){3}
\Vertex(95,50){3}
\Vertex( 115,72.5){3}
\Vertex(135,50){3}
\Text(  7,95)[r]{$\Pq_i$}
\Text(  7,5)[r]{$\bar{\Pq}_i$}
\put(143,90){{$\Pq_j$}}
\put(143,0){{$\bar{\Pq}_j$}}
\put(173,60){{$\Plm$}}
\put(173,30){{$\Plp$}}
\put(129,65){{$\PZ,\ga$}}
\put( 65,28){{$\Pg$}}
\SetScale{1}
\end{picture}}
}
\vspace{3ex}
\centerline{
{\unitlength .6pt 
\begin{picture}(125,100)(0,0)       
\SetScale{.6}
\ArrowLine( 15, 5)( 65,10)
\ArrowLine( 65,10)(115, 5)
\ArrowLine( 15,95)( 65,90)
\ArrowLine( 65,90)(90,92.5)
\ArrowLine( 90,92.5)(115,95)
\Photon( 65,10)( 65,90){3}{8}
\Photon(100,50)( 90,92.5){3}{4}
\ArrowLine( 100,50)( 130,65)
\ArrowLine( 130,35)( 100,50)
\Vertex(65,90){3}
\Vertex(65,10){3}
\Vertex(90,92.5){3}
\Vertex(100,50){3}
\Text(  7,95)[r]{$\Pq_i$}
\Text(  7,5)[r]{$\Pq_j$}
\Text(123,95)[l]{{$q_i$}}
\Text(123,0)[l]{{$\Pq_j$}}
\put(138,60){{$\Plm$}}
\put(138,30){{$\Plp$}}
\put(100,70){{$\PZ,\ga$}}
\Text(57,50)[r]{{$\PZ,\ga$}}
\SetScale{1}
\end{picture}}
\hspace*{3em}
{\unitlength .6pt 
\begin{picture}(145,100)(0,0)       
\SetScale{.6}
\ArrowLine( 15,95)( 35,50)
\ArrowLine( 35,50)( 15, 5)
\ArrowLine(95,50)(115,72.5)
\ArrowLine(115,72.5)(135,95)
\ArrowLine(135, 5)(95,50)
\Photon( 35,50)(95,50){3}{7}
\Photon(115,72.5)(135,50){3}{4}
\ArrowLine( 135,50)( 165,65)
\ArrowLine( 165,35)( 135,50)
\Vertex( 35,50){3}
\Vertex(95,50){3}
\Vertex( 115,72.5){3}
\Vertex(135,50){3}
\Text(  7,95)[r]{$\Pq_i$}
\Text(  7,5)[r]{$\bar{\Pq}_i$}
\put(143,90){{$\Pq_j$}}
\put(143,0){{$\bar{\Pq}_j$}}
\put(173,60){{$\Plm$}}
\put(173,30){{$\Plp$}}
\put(129,66){{$\PZ,\ga$}}
\Text( 65,28)[c]{{$\ga,\PZ$}}
\SetScale{1}
\end{picture}}
\hspace{3em}
{\unitlength .6pt 
\begin{picture}(125,100)(0,0)       
\SetScale{.6}
\ArrowLine( 15,95)( 65,90)
\ArrowLine( 65,10)( 15, 5)
\ArrowLine(115, 5)( 65,10)
\ArrowLine( 65,90)(115,95)
\Photon( 65,10)( 65,90){3}{7}
\Photon(100,50)( 65,50){3}{4}
\ArrowLine( 100,50)( 130,65)
\ArrowLine( 130,35)( 100,50)
\Vertex(65,90){3}
\Vertex(65,10){3}
\Vertex(65,50){3}
\Vertex(100,50){3}
\Text(  7,95)[r]{$\Pq_i$}
\Text(  7,5)[r]{$\bar{\Pq}_j$}
\put(123,90){{$q'_i$}}
\put(123,0){{$\bar{\Pq}'_j$}}
\put(138,60){{$\Plm$}}
\put(138,30){{$\Plp$}}
\put( 75,61){{$\PZ,\ga$}}
\put( 40,65){{$\PW$}}
\put( 40,20){{$\PW$}}
\SetScale{1}
\end{picture}}
}
\vspace{3ex}
\centerline{
{\unitlength .6pt 
\begin{picture}(125,100)(0,0)       
\SetScale{.6}
\ArrowLine(115,95)( 65,90)
\Photon( 15, 5)( 65,10){3}{7}
\ArrowLine( 65,10)(115, 5)
\Gluon(15,95)( 65,90){3}{7}
\ArrowLine( 65,90)( 65,40)
\ArrowLine( 65,40)( 65,10)
\Photon(100,50)( 65,50){3}{4}
\ArrowLine( 100,50)( 130,65)
\ArrowLine( 130,35)( 100,50)
\Vertex(65,90){3}
\Vertex(65,10){3}
\Vertex(65,50){3}
\Vertex(100,50){3}
\Text(  7,95)[r]{$\Pg$}
\Text(  7,5)[r]{$\ga$}
\put(123,90){{$\bar{\Pq}_i$}}
\put(123,0){{$\Pq_i$}}
\put(75,61){{$\PZ,\ga$}}
\put(138,60){{$\Plm$}}
\put(138,30){{$\Plp$}}
\SetScale{1}
\end{picture}}
\hspace*{3em}
{\unitlength .6pt 
\begin{picture}(125,100)(0,0)       
\SetScale{.6}
\ArrowLine( 15,95)( 65,90)
\Photon( 15, 5)( 65,10){3}{7}
\ArrowLine( 65,10)(115, 5)
\Gluon(115,95)( 65,90){3}{7}
\ArrowLine( 65,90)( 65,40)
\ArrowLine( 65,40)( 65,10)
\Photon(100,50)( 65,50){3}{4}
\ArrowLine( 100,50)( 130,65)
\ArrowLine( 130,35)( 100,50)
\Vertex(65,90){3}
\Vertex(65,10){3}
\Vertex(65,50){3}
\Vertex(100,50){3}
\Text(  7,95)[r]{$\Pq_i$}
\Text(  7,5)[r]{$\ga$}
\put(123,90){{$\Pg$}}
\put(123,0){{$\Pq_i$}}
\put(75,61){{$\PZ,\ga$}}
\put(138,60){{$\Plm$}}
\put(138,30){{$\Plp$}}
\SetScale{1}
\end{picture}}
\hspace*{3em}
{\unitlength .6pt 
\begin{picture}(125,100)(0,0)       
\SetScale{.6}
\Photon( 15,95)( 65,90){3}{7}
\Photon( 15, 5)( 65,10){3}{7}
\ArrowLine( 65,10)(115, 5)
\ArrowLine(115,95)( 65,90)
\ArrowLine( 65,90)( 65,40)
\ArrowLine( 65,40)( 65,10)
\Photon(100,50)( 65,50){3}{4}
\ArrowLine( 100,50)( 130,65)
\ArrowLine( 130,35)( 100,50)
\Vertex(65,90){3}
\Vertex(65,10){3}
\Vertex(65,50){3}
\Vertex(100,50){3}
\Text(  7,95)[r]{$\ga$}
\Text(  7,5)[r]{$\ga$}
\put(123,90){{$\bar{\Pq}_i$}}
\put(123,0){{$\Pq_i$}}
\put(75,61){{$\PZ,\ga$}}
\put(138,60){{$\Plm$}}
\put(138,30){{$\Plp$}}
\SetScale{1}
\end{picture}}
}
\caption{Sample tree diagrams for the QCD
  contributions to $\Pq_i\,\Pg\to\Pq_i\,\Pg\,\Plm\,\Plp$,
  $\Pq_i\,{\Pq}_j\to\Pq_i\,\Pq_j\,\Plm\,\Plp$, and
  $\Pq_i\,\bar{\Pq}_i\to\Pq_j\,\bar{\Pq}_j\,\Plm\,\Plp$ (first line)
  the EW contributions to $\Pq_i\,{\Pq}_j\to\Pq_i\,\Pq_j\,\Plm\,\Plp$,
  $\Pq_i\,\bar{\Pq}_i\to\Pq_j\,\bar{\Pq}_j\,\Plm\,\Plp$ and
  $\Pq_i\,\bar{\Pq}_j\to\Pq'_i\,\bar{\Pq}_j'\,\Plm\,\Plp$ (second
  line) and the contributions to
  $\Pg\,\ga\to\bar{\Pq_i}\,\Pq_i\,\Plm\,\Plp$,
  $\Pq_i\,\ga\to\Pq_i\,\Pg\,\Plm\,\Plp$ and
  $\ga\,\ga\to\Pq_i\,\bar{\Pq_i}\,\Plm\,\Plp$.  }
\label{fig:LOdiags}
\end{figure}
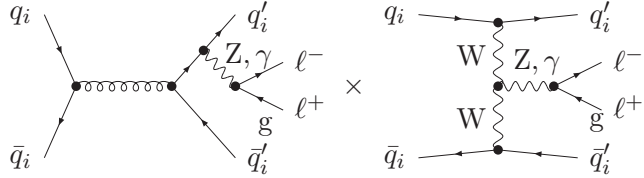
\begin{figure}
\centerline{
{\unitlength .6pt 
\begin{picture}(175,100)(0,0)
\SetScale{.6}
\ArrowLine( 15,95)( 35,50)
\ArrowLine( 35,50)( 15, 5)
\ArrowLine(95,50)(115,72.5)
\ArrowLine(115,72.5)(135,95)
\ArrowLine(135, 5)(95,50)
\Gluon( 35,50)(95,50){3}{7}
\Photon(115,72.5)(135,50){3}{4}
\ArrowLine( 135,50)( 165,65)
\ArrowLine( 165,35)( 135,50)
\Vertex( 35,50){3}
\Vertex(95,50){3}
\Vertex( 115,72.5){3}
\Vertex(135,50){3}
\Text(  7,95)[r]{$\Pq_i$}
\Text(  7,5)[r]{$\bar{\Pq}_i$}
\put(143,90){{$\Pq'_i$}}
\put(143,0){{$\bar{\Pq}'_i$}}
\put(173,60){{$\Plm$}}
\put(173,30){{$\Plp$}}
\put(129,65){{$\PZ,\ga$}}
\Text( 150,25)[l]{{$\Pg$}}
\Text(210,50)[]{$\times$}
\SetScale{1}
\end{picture}}
\qquad\quad
{\unitlength .6pt 
\begin{picture}(125,100)(0,0)
\SetScale{.6}
\ArrowLine( 15,95)( 65,90)
\ArrowLine( 65,10)( 15, 5)
\ArrowLine( 65,90)(115,95)
\ArrowLine(115, 5)( 65,10)
\put(123,25){{$\Pg$}}
\Photon( 65,10)( 65,90){3}{7}
\Photon(100,50)( 65,50){3}{4}
\ArrowLine( 100,50)( 130,65)
\ArrowLine( 130,35)( 100,50)
\Vertex(65,90){3}
\Vertex(65,10){3}
\Vertex(65,50){3}
\Vertex(100,50){3}
\Text(  7,95)[r]{$\Pq_i$}
\Text(  7,5)[r]{$\bar{\Pq}_i$}
\put(123,90){{$\Pq'_i$}}
\put(123,0){{$\bar{\Pq}'_i$}}
\put(75,61){{$\PZ,\ga$}}
\Text( 58,70)[r]{{$\rm W$}}
\Text( 58,30)[r]{{$\rm W$}}
\put(138,60){{$\Plm$}}
\put(138,30){{$\Plp$}}
\SetScale{1}
\end{picture}}
}
\caption{Sample tree diagrams for interferences of QCD and EW diagrams.} 
\label{fig:tree_interferences}
\end{figure}

For the calculation of the differential cross section at LO and NLO we
follow closely the implementation of $\PZ +2\,{\rm jet}$ production in
\citere{Actis:2012qn}.  Potentially resonant $\PZ$-boson
propagators are described attributing a complex mass
\begin{equation}
\mu_{\PZ}^2 = \MZ^2 -\ri \MZ \GZ.
\end{equation}
To this end we consistently use the complex-mass scheme
\cite{Denner:1999gp,Denner:2005fg,Denner:2006ic} where $\mu^2_\PW$ and
$\mu^2_\PZ$ are defined as the poles of the
$\PW$- and $\PZ$-boson propagators in the complex plane. %
The pole values $M_V$ and $\Gamma_V$ ($V=\PW,\PZ$) for the mass and
width of the $\PW$ and $\PZ$~boson are related to the on-shell quantities
$M_V^{\mathrm{OS}}$ and $\Gamma_V^{\mathrm{OS}}$ obtained from the LEP
and Tevatron experiments by \cite{Bardin:1988xt}
\beq\label{eq:m_ga_pole}
M_V = M_V^{\OS}/ \sqrt{1+(\Ga_V^{\OS}/M_V^{\OS})^2}\,, \qquad \Ga_V =
\Ga_V^{\OS}/ \sqrt{1+(\Ga_V^{\OS}/M_V^{\OS})^2}\,.
\eeq
We define the electromagnetic coupling constant $\alpha$
within the $G_\mu$ scheme, {i.e.\ }we fix the value of
$\alpha$ via its tree-level relation with the Fermi constant $G_\mu$:
\beq\label{eq:Gmu}
 \alpha_{\GF} = \frac{\sqrt{2}\GF\MW^2}{\pi}\left(1-\frac{\MW^2}{\MZ^2}\right).
\eeq
Compared to the Thomson-limit definition of $\alpha$, 
\refeq{eq:Gmu} incorporates effects of the renormalisation-group
running from the scale $Q^2=0$ to the scale $Q^2=\MW^2$.  Moreover,
using \refeq{eq:Gmu}
the renormalisation of $\al$ becomes independent of light quark masses
and the hadronic vacuum polarisation.

For the analysis we 
use the following
numerical input parameters \cite{Beringer:1900zz,Lancaster:2011wr}:
\begin{equation}
       \begin{array}[b]{rclrcl}
          \GF & = & 1.1663787 \times 10^{-5} \GeV^{-2}, \quad\\[.5ex]
          \MW^{\OS} & = & 80.385\GeV,\quad &
          \Gamma_\PW^{\OS} & = & 2.085\GeV, \\[.5ex]
          \MZ^{\OS} & = & 91.1876\GeV, &
          \Gamma_\PZ^{\OS} & = & 2.4952\GeV, \\[.5ex]
          \MH & = & 125\GeV, & \Mt&=&173.2\GeV.
       \end{array}
       \label{eq:SMinput}
\end{equation}
The superscript $\OS$ denotes on-shell values, and the corresponding
pole masses and widths entering our calculation are obtained by
\refeq{eq:m_ga_pole}. 

We are mainly interested in the relative size of the
$\mathcal{O}(\alpha_s^2\alpha^3)$ corrections compared to the LO
prediction, which is dominated by $\mathcal{O}(\alpha_s^2\alpha^2)$
contributions. In the corresponding ratio $\alpha_s$ enters
exclusively through subdominant contributions to the LO quark channels
and any $\alpha_s$ dependence is thus strongly suppressed.  Since we
do not include the leading QCD corrections, we resort to LO parton
distribution functions (PDFs) using, if not stated otherwise, the
LHAPDF implementation of the central MSTW2008LO PDF set
\cite{Martin:2009iq}. From there we infer the value of the strong
coupling constant to
\beq
\alpha_{\rm s}^{\LO}(\MZ)=0.139395\ldots\;.
\eeq
We choose the QCD factorisation scale $\mu_{\mathrm{F}}$ and the
renormalisation scale $\mu_{\mathrm{R}}$ as
\beq
  \mu_{\mathrm{F}}=\mu_{\mathrm{R}}=\MZ\,.
\eeq
The considered relative corrections depend only weakly on these scales.

For the jet-reconstruction we use the anti-$k_\mathrm{T}$ clustering
algorithm \cite{Cacciari:2008gp} with separation parameter $R=0.4$.
The spatial distance between partons $i$ and $j$, $\Delta R_{ij} =
\sqrt{(y_i-y_j)^2 + (\phi_i - \phi_j)^2}$, is defined in terms of
rapidity $y = \frac{1}{2} \ln[(E + p_\RL)/(E - p_\RL)]$, where $p_\RL$
is the momentum component along the beam axis, and azimuthal angle
$\phi$ of the partons. Only partons with $|y_i|<5$ are clustered.  We
include also photons and leptons in the jet clustering according to
the anti-$k_\mathrm{T}$ description with $R=0.4$. While quarks and/or
gluons are combined into jets, leptons and photons are recombined into
leptons, and quarks/gluons and photons are combined into jets.

We apply two different sets of phase-space cuts to define the
production cross section.  The first set is inspired by an ATLAS
analysis \cite{Aad:2013ysa} for the production of jets in association
with a $\PZ$ boson at $\sqrt{s} = 7\TeV$. The second set of cuts
constitutes a typical vector-boson-fusion (VBF) setup, i.e. similar
cuts are usually applied in order to enhance in $\PH\Pj\Pj$ signatures
the production channel via vector-boson-fusion production.

Denoting the momenta of the incoming partons by $\k{1},\k{2}$ and the
final-state momenta by $\k{i} = \{\k{\Pj_1}, \k{\Pj_2}, \k{\Plm},
\k{\Plp}$\}, the LO cross section $\si_{\LO}$ is obtained as
\beq
\sigma_{\LO} = \frac{1}{2\k{1}\k{2}}\: \int {\dd \Pcal(x_i)} 
\:\int {\dd}\Phi(\k{i})\: \Theta_\cut(\k{i}) \: \left|{\M}_{\LO}(\k{1},\k{2},\k{i})\right|^2,
\eeq
where ${\dd \Pcal(x_i)}$ incorporates the convolution with the parton
distributions functions, ${\dd}\Phi(\k{i})$ represents the phase space
measure, and $\Theta_\cut(\k{i})$ defines the acceptance function.

\subsection{The pole approximation}
\label{sec: LO pole approximation}

We also consider the LO total cross section in the pole approximation
$\Pp\Pp \to \Pj \Pj \PZ \to \Pj \Pj \Plp \Plm$.  { From the full LO
  amplitude ${\M}_{\LO}$ we define the subamplitude
  ${\M}^{\PZ}_{\LO}$} where only diagrams corresponding to the
production of a $\PZ$ boson and its subsequent leptonic decay are
taken into account and all non-resonant diagrams (including those with
a virtual photon decaying into a lepton pair) are neglected. While the
phase space is populated with
lepton momenta $\k{\Plm},\k{\Plp}$ of generic invariant mass
$M_{\Plm\Plp}^2=(\k{\Plm}+\k{\Plp})^2$, the matrix element
$\cal{M}^\PZ_{\LO}$ is calculated with on-shell-projected momenta, \ie
momenta where the invariant mass of the lepton pair equals exactly the
Z-boson mass. Only the resonant $\PZ$-boson propagator is replaced by
its off-shell variant by means of a correcting prefactor. The LO cross
section $\si^\PZ_{\LO}$ in the pole approximation thus reads
\beq
\sigma^{\PZ}_{\LO} = \frac{1}{2 \k{1}\k{2}}\: \int {\dd \Pcal(x_i)} \:
\int {\dd}\Phi(\k{i})\: \Theta_\cut(\k{i}) \: 
\left|\frac{\MZ\GZ}{\k{\PZ}^2 - \MZ^2 - \ri\MZ\GZ}\right|^2\: 
\left|{\M}^{\PZ}_{\LO}(\k{1},\k{2},\kt{i})\right|^2,
\eeq
where 
\beq
\k{\PZ} = \k{\Plm} + \k{\Plp}.
\eeq


The on-shell projected final-state momenta $\kt{i}$ are constructed
from the off-shell momenta $\k{i}$ through 
\ba
\kt{\Pj_1} &=& \beta \k{\Pj_1}, \qquad \kt{\Pj_2} = \beta \k{\Pj_2}, \nn \\
\kt{\PZ} &=& \k{\PZ} + (1-\beta)(\k{\Pj_1} + \k{\Pj_2}) ,\nn\\
\kt{\Plm} &=& \frac{\MZ^2}{2(\kt{\PZ}\k{\Plm})} \k{\Plm}, \qquad
\kt{\Plp} = \kt{\PZ} - \kt{\Plm}, 
\ea 
where $\beta$ is chosen in such
a way that $\kt{\PZ}^2=\MZ^2$. Of the two solutions of the quadratic
equation
\beqar
0&=&\kt{\PZ}^2 - \MZ^2 \nn\\
&=& 
2(\k{\Pj_1}\k{\Pj_2})\beta^2 -[4(\k{\Pj_1}\k{\Pj_2}) +
2(\k{\Pj_1}\k{\PZ}) + 2(\k{\Pj_2}\k{\PZ})]\beta \nn\\ 
&&{}+ 2(\k{\Pj_1}\k{\Pj_2}) + 2(\k{\Pj_1}\k{\PZ}) + 2(\k{\Pj_2}\k{\PZ}) + \k{\PZ}^2 - \MZ^2
\eeqar
$\beta$ is the one closer to 1.

\subsection{Leading-order analysis with standard acceptance cuts}
\label{sec: LO standard cuts}

In this section we investigate the production of $\process$, where
$\Pl=\Pe$ or $\mu$ (not summed), at LO for a set of standard
acceptance cuts (called {\em basic cuts} in the following).
We require two hard jets with
\beq
   \label{eq:jetcuts}
   p_{\mathrm{T},\Pj}>30\GeV, \qquad |y_\Pj|<4.5
\eeq
for the transverse momenta $p_{\mathrm{T}}$ and rapidities $y$
and two hard leptons with
\beqar
   \label{eq:leptoncuts}
   p_{\mathrm{T},\Pl}>20\GeV, \qquad |y_\Pl|<2.5.
\eeqar
We then apply to the two jets {(at NLO to the two or three jets)}
and the two charged leptons passing the cuts \refeq{eq:jetcuts} and
\refeq{eq:leptoncuts} the rapidity--azimuthal angle separation cuts
\beqar
   \label{eq:DRcuts}
\De R_{\Plm\Plp}>0.2, \qquad \De R_{\Pl\Pj}>0.5,
\eeqar
and finally a cut on the invariant mass of the lepton pair
\beqar\label{eq:Mcuts}
66\GeV<M_{\Plm\Plp}<116\GeV.
\eeqar

The total cross section for the $13\TeV$ LHC and the set of cuts
listed above can be found in \refta{tab:LO-LHC13-xsection}, where it
is split into various contributions.  Neglecting photon-induced
contributions we find
\beq
 \sigma^{13\TeV}_{\rm tot} = 
51.209(8)\pba.
\label{eq:LO-x-section-LHC13}
\eeq

\begin{table}
\begin{center}
\renewcommand{\arraystretch}{1.2}
   \begin{tabular}{|c|c|c|c|c|c|c|}
     \hline
process class  & $\sigma $   & $\sigma^{\PZ} $& $ \sigma/\sigma_{\rm tot}$& $\sigma_{\alphas^2\alpha^2}/\sigma$ & $\sigma_{\alphas\alpha^3}/\sigma$ & $\sigma_{\alpha^4}/\sigma$ \\
             & $[\rm{ fb}]$  &$[\rm{ fb}]$   & $[\%]$ & $[\%]$ & $[\%]$ & $[\%]$ \\
\hline 
$u\Pg\to u\Pg \Pl^-\Pl^+$, $d\Pg\to d\Pg \Pl^-\Pl^+$ & \multirow{2}{*}{34584(8)}& \multirow{2}{*}{34105(10)} & \multirow{2}{*} {67.5}&\multirow{2}{*}{100} & \multirow{2}{*}{---}& \multirow{2}{*}{---} \\
$\ub \Pg\to \ub \Pg \Pl^-\Pl^+$, $\db \Pg\to \db \Pg \Pl^-\Pl^+$ &&&&& &\\
\hline 
$u\ub \to \Pg\Pg \Pl^-\Pl^+$, $d\db \to \Pg\Pg \Pl^-\Pl^+$  & 2713(1)& 2671(1)      &    5.3        & {100}  & {---}& {---} \\
\hline 
$\Pg\Pg \to u\ub \Pl^-\Pl^+$, $ \Pg\Pg \to d\db \Pl^-\Pl^+$ & 3612(1)& 3574(1)      &    7.1        & {100}  & {---}& {---} \\
\hline 
$uu\to uu \Pl^-\Pl^+$, $dd\to dd \Pl^-\Pl^+$ & \multirow{2}{*}{1315.1(3)}  & \multirow{2}{*}{1291.4(4)} & \multirow{2}{*}{2.6}    &\multirow{2}{*}{97.4} & \multirow{2}{*}{+2.0}& \multirow{2}{*}{0.5} \\
$\ub\ub\to \ub\ub \Pl^-\Pl^+$, $\db\db\to \db\db \Pl^-\Pl^+$ &&&&&&\\
\hline 
$u\ub\to u'\ub' \Pl^-\Pl^+$,  $d\db\to d'\db' \Pl^-\Pl^+$ & \multirow{3}{*}{2463.7(5)}& \multirow{3}{*}{2420.5(7)}&\multirow{3}{*}{4.8}&\multirow{3}{*}{98.3} & \multirow{3}{*}{$-1.3$}& \multirow{3}{*}{2.9}\\
$u\ub'\to u\ub' \Pl^-\Pl^+$,  $d\db'\to d\db' \Pl^-\Pl^+$ &&&&&&\\
$u\ub\to u\ub \Pl^-\Pl^+$,  $d\db\to d\db \Pl^-\Pl^+$ &&&&&&\\
\hline 
$u\ub\to d\db \Pl^-\Pl^+$, $d\db\to u\ub \Pl^-\Pl^+$ & \multirow{2}{*}{438.82(7)} & \multirow{2}{*}{432.6(1)} & \multirow{2}{*}{0.9} &\multirow{2}{*}{76.6} & \multirow{2}{*}{$-9.0$}& \multirow{2}{*}{32.3}\\
$u\ub'\to d\db' \Pl^-\Pl^+$,  $d\db'\to u\ub' \Pl^-\Pl^+$ &&&&&&\\
\hline 
$ud\to u'd' \Pl^-\Pl^+$, $\ub\db\to \ub'\db' \Pl^-\Pl^+$& \multirow{4}{*}{3856.8(7)}& \multirow{4}{*}{3800(1)} & \multirow{4}{*}{7.5} &\multirow{4}{*}{92.9} & \multirow{4}{*}{+2.8}& \multirow{4}{*}{4.3}\\
$ud\to ud \Pl^-\Pl^+$, $\ub\db\to \ub\db \Pl^-\Pl^+$&&&&&& \\
$uu'\to uu' \Pl^-\Pl^+$, $\ub\ub'\to \ub\ub' \Pl^-\Pl^+$&&&&&& \\
$dd'\to dd' \Pl^-\Pl^+$, $\db\db'\to \db\db' \Pl^-\Pl^+$&&&&&& \\
     \hline 
$u\db\to u'\db' \Pl^-\Pl^+$, $\ub d\to \ub'd' \Pl^-\Pl^+$ & \multirow{2}{*}{2224.9(4)}& \multirow{2}{*}{2197.1(6)} & \multirow{2}{*}{4.3} &\multirow{2}{*}{95.9} & \multirow{2}{*}{$-1.1$}& \multirow{2}{*}{5.2}\\
$u\db\to u\db \Pl^-\Pl^+$, $\ub d\to \ub d \Pl^-\Pl^+$&&&&&&\\
     \hline
     \hline gluonic       & 40910(8)& 40349(11) &  79.9 & 100  & ---  & --- \\
     \hline four-quark    & 10299(1)& 10141(1)  &  20.1 & 94.7 & +0.4  & 4.8 \\
     \hline bottom quarks &  4376(3)&   ---     &  8.54 & ---  & --- & ---    \\
     \hline sum           & 51209(8)& 50490(11) &  100  & 98.9 & $<0.1$ & 1.0   \\
     \hline
   \end{tabular}
\end{center}
  \caption{Composition of the LO cross section for 
   $\process$ at the LHC operating at
    $13\TeV$ for basic cuts. In the first column the contributing
    partonic processes are listed, where $u,u'\ne u$ denote the up-type
    quarks $\Pu,\Pc$ and $d,d'\ne d$ the down-type quarks
    $\Pd,\Ps,\Pb$. The second column 
   provides the corresponding cross section where the numbers in
   parenthesis give the integration error on the last digit. The cross
   section for the pole 
   approximation is given in the third column. 
   The fourth column contains the relative contribution to the total
   cross section in per cent. In the fifth, sixth and seventh column 
   we provide the relative contribution to a partonic channel from
   strong and EW contributions and their interference.}
  \label{tab:LO-LHC13-xsection}
\end{table}

We note that about 80\% of $\Pj\Pj\Plm\Plp$ events will be produced in
parton interactions involving external gluons. This is true for the
full LO calculation shown in the second column as well as for the pole
approximation shown in the third column of
\refta{tab:LO-LHC13-xsection}. On average the pole approximation
underestimates the size of the cross section for the various processes
by 1.5\%, {in agreement with the expected accuracy of order $\GZ/\MZ$
  given the cut on $M_{\Plm\Plp}$.}  In addition,
\refta{tab:LO-LHC13-xsection} shows the composition of the total cross
section for different partonic
processes 
{in terms of} the various orders in the strong and the electromagnetic
coupling constant. By far the most dominant contribution results from
quark--gluon initiated processes, due to the high quark--gluon
luminosities in proton--proton collisions. The dominant production
mechanism for four-quark processes is given by strong interactions
between initial-state partons and jets [$\O(\alphas^2\alpha^2)$]. The
relative contribution at $\O(\alpha^4)$ varies between a few per mille
and a few per cent, except for the process class $u\ub\to d\db
\Pl^-\Pl^+$. Since $t$-channel gluon exchange does not contribute in
this process class, the absolute contribution is small, but the
relative contribution of the EW diagrams (involving $t$-channel
\PW-boson exchange) is enhanced.  Interferences between strong and EW
amplitudes [$\O(\alphas\alpha^3)$] are comparable to $\O(\alpha^4)$
contributions in absolute size. Since they are not positive definite
they lower the cross section slightly for certain partonic processes.
As a consequence there are cancellations between different partonic
processes, and the relative contribution of $\O(\alphas\alpha^3)$ to
the total cross section is less than one per mille in our set-up.
Partonic processes with external bottom quarks contribute $8.5\%$ to
the total cross section.

Before we turn to differential distributions {we elaborate on} the
impact of photon-induced reactions [see \refeq{eq:gamma-LO}]. For that
purpose we have redone our LO analysis for the LHC operating at $13
\TeV$ employing the NNPDF 2.3 \cite{Ball:2013hta} parton distribution
functions, using the same input parameters and phase-space cuts as
before.  With this setup we find the relative contribution of photonic
processes to the total cross section to be less than 0.5\permil.
Assuming that this represents the order of magnitude for
photon-induced processes, we neglect these contributions in the following.

\begin{figure}
  \includegraphics[width=8cm]{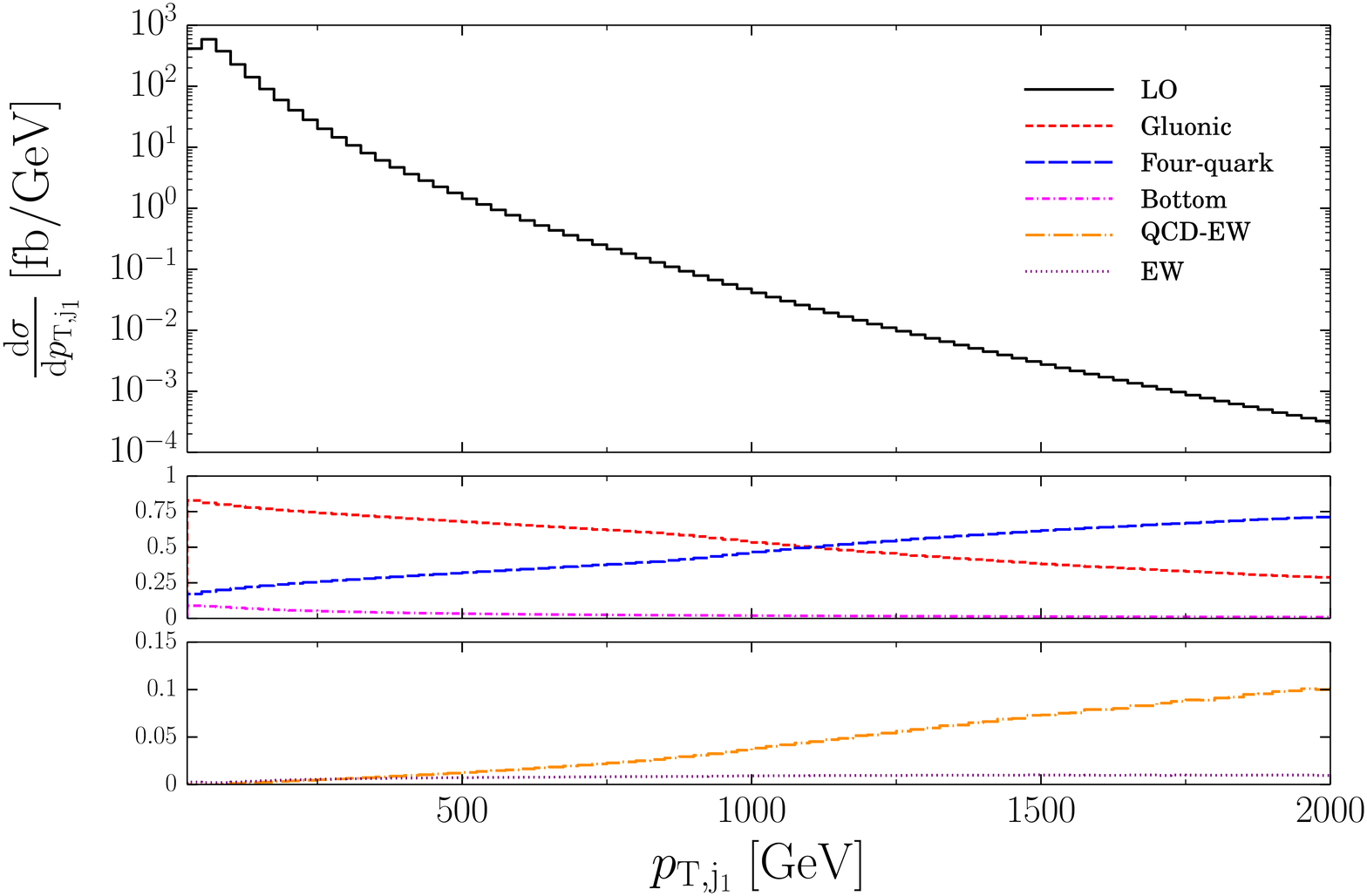}
  \includegraphics[width=8cm]{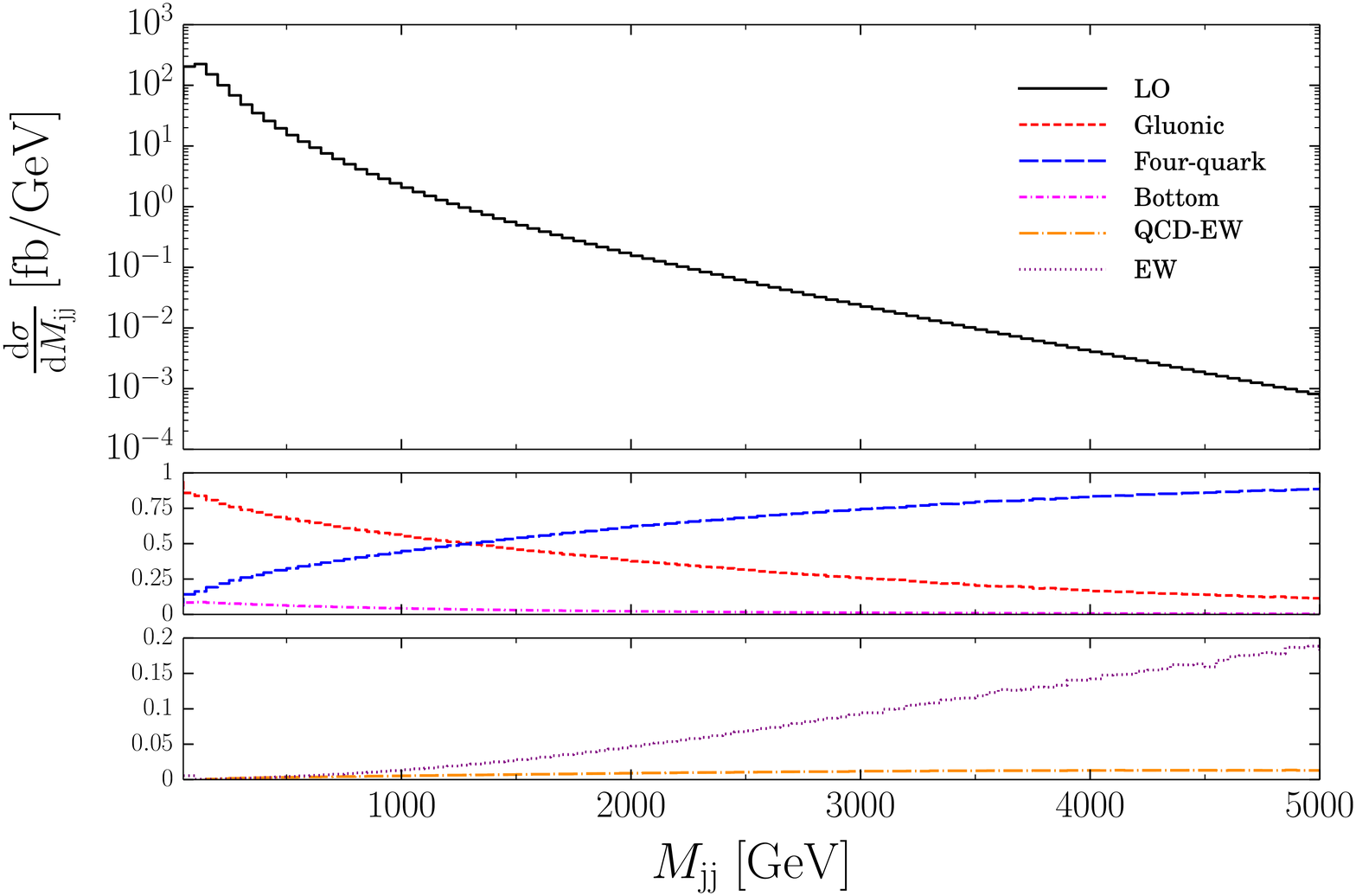}
  \includegraphics[width=8cm]{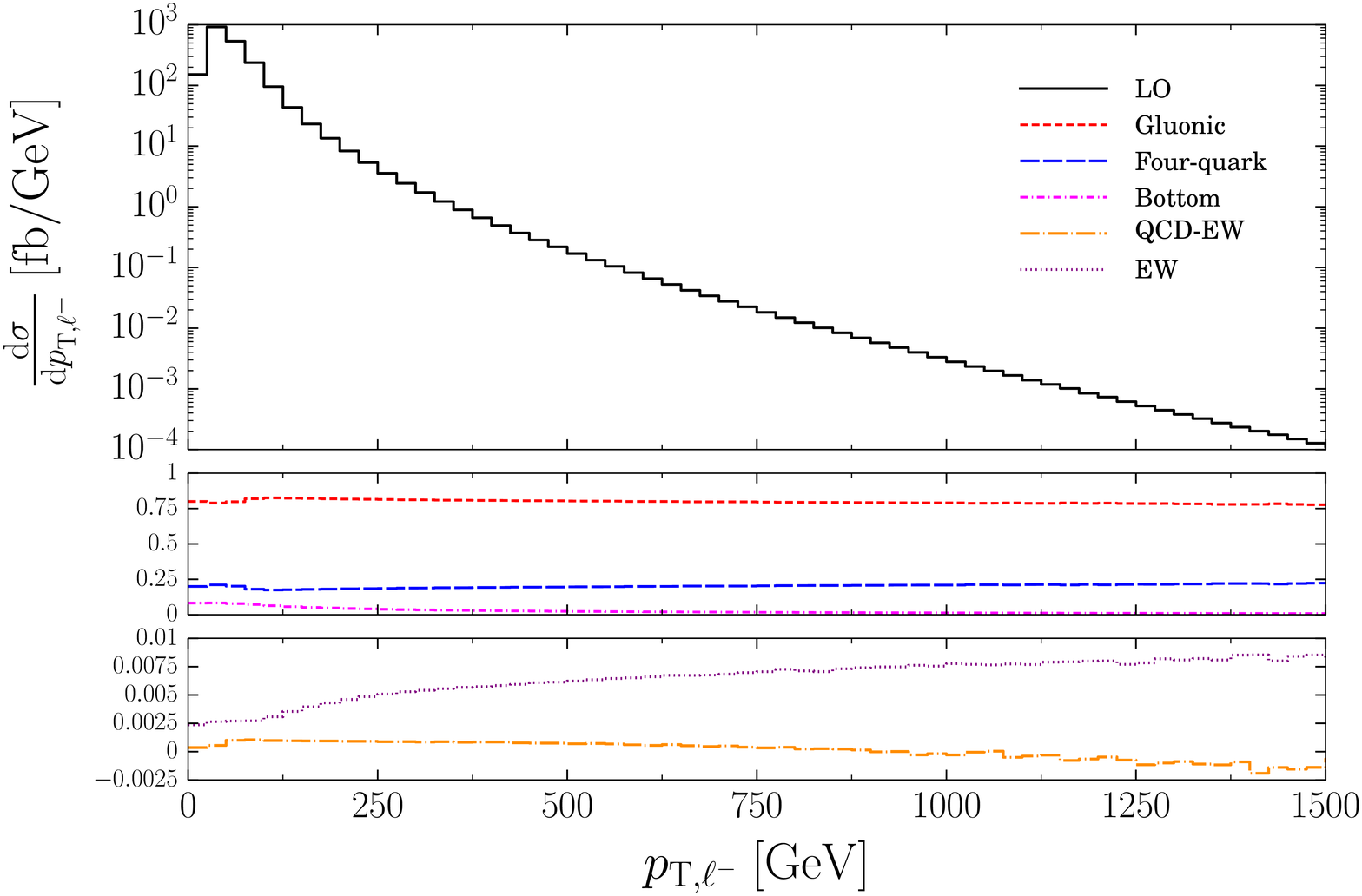}
  \includegraphics[width=8cm]{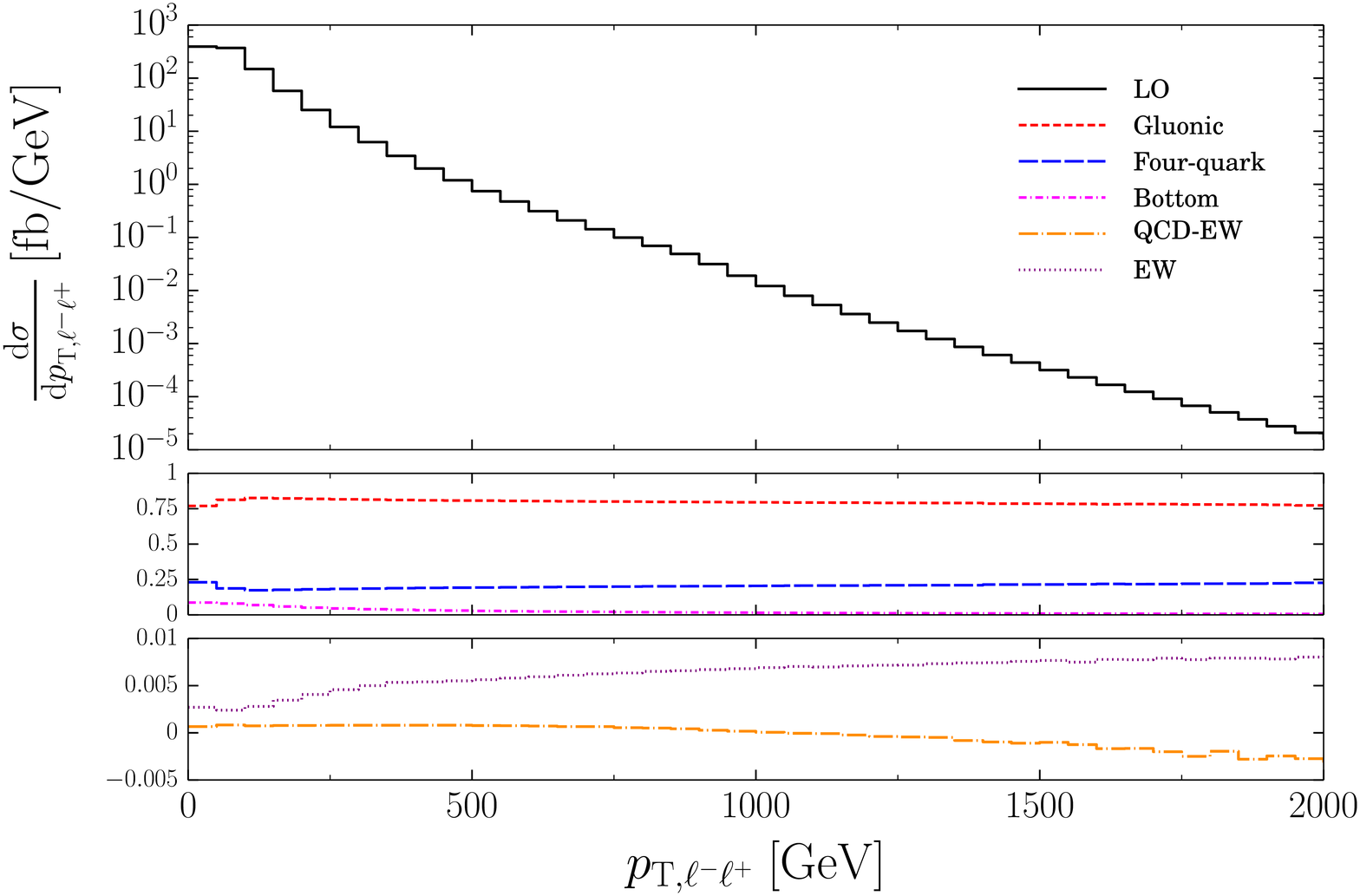}
  \includegraphics[width=8cm]{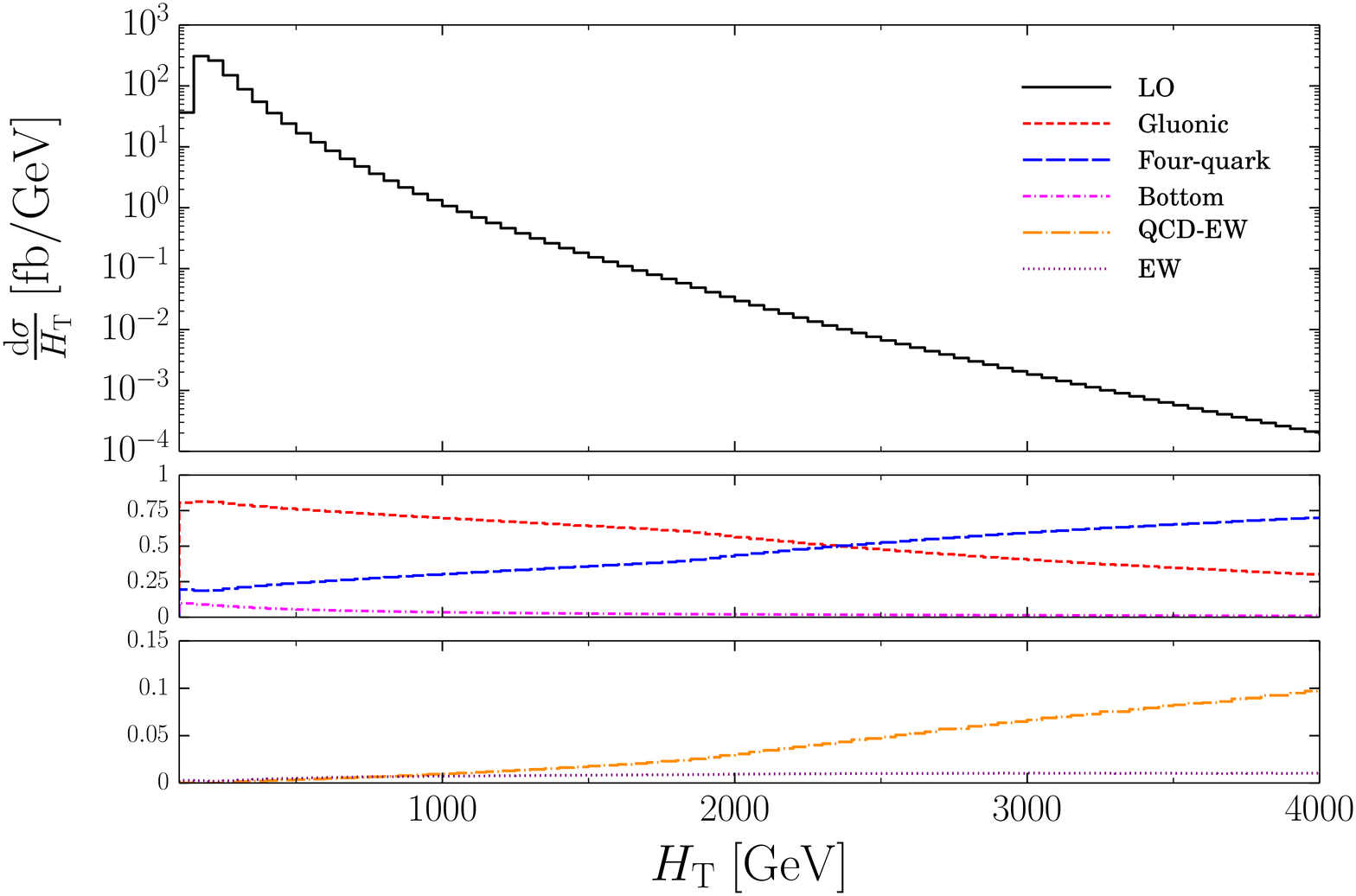}
  \includegraphics[width=8cm]{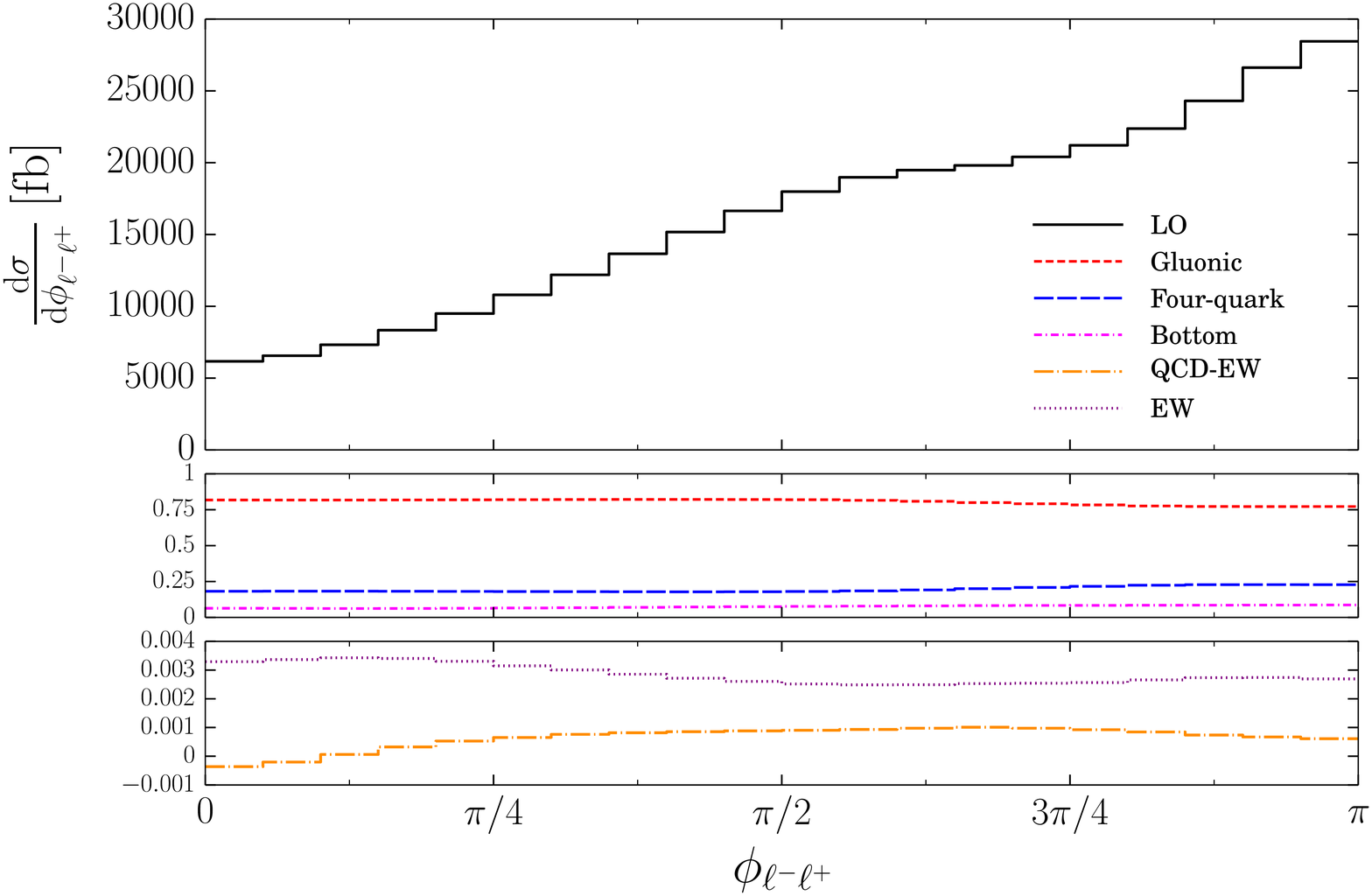}
  \caption{Distributions of the transverse momentum of the harder jet
    ${\rm j_1}$, the di-jet invariant mass $M_{\Pj\Pj}$, the
    transverse momenta of the negatively charged lepton $\Plm$ and the
    lepton pair $\Plm\Plp$, the scalar sum of all transverse momenta,
    and the relative azimuthal angle between the leptons at the
    $13\TeV$ LHC for basic cuts.  The upper panels show the corresponding
    distributions, the middle ones the composition of the cross
    section and the lower ones the contributions of the EW diagrams.
    Further details are described in the text.}
  \label{fig:LO_LHC13_Std}
\end{figure}

In \reffi{fig:LO_LHC13_Std} we present LO differential distributions
for the transverse momenta of the hardest jet {$\pTo{\Pj_1}$} (jet
with highest transverse momentum), the negatively charged lepton
{$\pTo{\Plm}$}, and the lepton pair {$\pTo{\Plm\Plp}$}, as well as the
di-jet invariant mass {$M_{\Pj\Pj}$}, the scalar
sum of all transverse momenta 
\beq\label{eq:def-HT}
\HT = \pTo{j_1} + \pTo{j_2} + \pTo{\Plm}+ \pTo{\Plp},
\eeq
and the relative
azimuthal angle $\phill$ between the two leptons. The upper panels
depict the absolute distributions at leading order. In the middle
panels we illustrate the composition of these distributions in terms
of gluonic (red, short-dashed) and four-quark (blue, long-dashed) processes
including all light flavours ($\Pu,\,\Pd,\,\Ps,\,\Pc,\,\Pb$) and
show the relevance of processes involving at least one
external bottom quark (magenta, dashed-dotted). In the lower panels we present
the relative contributions of the squared EW diagrams (purple, dotted)
and the QCD--EW interference (orange, long-dashed dotted).
The distribution of the transverse momentum of the hardest jet (upper
left plot of \reffi{fig:LO_LHC13_Std}) drops by about six orders of
magnitude in the depicted range $\pTo{\Pj_1}\le 2\TeV$.  The
composition of the distribution changes significantly with increasing
$\pTo{\Pj_1}$.  For low transverse momentum, gluonic processes
dominate while for higher $\pTo{\Pj_1}$ values the four-quark
processes become more important and dominate for $\pTo{\Pj_1} \gsim
1100 \GeV$. We find a similar composition for other jet observables
which are energy dependent like the transverse momentum distribution
of the second hardest jet {(not shown)} or the invariant jet mass of
the two hardest jets (upper right plot of \reffi{fig:LO_LHC13_Std}).
This behaviour is reminiscent of the relative gluon and quark-induced
contributions in di-jet production \cite{Moretti:2006ea,Scharf:2009sp}
and is related to the different characteristics of quark and gluon
parton distribution functions. We have checked that the events with
high $\pTo{\Pj_1}$ are dominated by events with two hard jets and
relatively soft leptons.

We consider the differential cross section as a function of the
transverse momentum of the negatively charged lepton in the middle
left plot of \reffi{fig:LO_LHC13_Std}.  The distribution shows a drop
over seven orders of magnitude for $20 \GeV < \pTo{\Plm} < 1500 \GeV$.
Since the leptons result mainly from Z-boson decays their
transverse-momentum distributions show a stronger drop compared to
those of the jets, which are produced directly in the collision.
Events with high $\pTo{\Plm}$ are typically accompanied by
a hard jet balancing the transverse momentum and a soft jet.  It is
striking that the relative composition of this distribution in terms
of gluonic and four-quark contributions is largely independent of
$\pTo{\Plm}$. For any $\pTo{\Plm}$ value, about three quarters of the
$\process$ events are produced in interactions involving external
gluons.
For the distributions in the transverse momentum of the positively
charged lepton (not shown) and the lepton pair (middle right plot of
\reffi{fig:LO_LHC13_Std}) we find similar results.

The lower left plot of \reffi{fig:LO_LHC13_Std} depicts the $\HT$
distribution, which is sensitive to both, lepton and jet transverse
momenta.  With increasing $\HT$ the relative contributions behave
similarly as for the $\pTo{j_1}$ distribution.  Four-quark and gluonic
contributions break even around $\HT = 2300 \GeV$.  Finally, we
consider the azimuthal angle between the leptons.  We find that the
two leptons prefer to be anticollinear in the transverse plane. This
comes about as the lepton pairs result mainly from the decays of
relatively soft Z~bosons. We observe a rather constant ratio between
gluonic and four-quark processes as a function of $\phi_{\Plm\Plp}$.

We note that partonic contributions involving external bottom quarks
have only a minor impact in all studied differential distributions.
They are below ten per cent in all cases and even much smaller in the
high-energy tails of the distributions.

While the EW contributions stay at the level of $1\%$ in the
$\pTo{\Pj_1}$ distribution, they increase with $M_{\Pj\Pj}$, reaching
almost $20\%$ at $M_{\Pj\Pj}=5\TeV$.  The QCD--EW interference grows
to $10\%$ for $\pTo{\Pj_1}=2\TeV$.  While the EW contributions are
generally small for other distributions, a sizable QCD--EW
interference is also visible in the $H_\rT$ distribution reaching
$10\%$ at $4\TeV$ (lower left plot of \reffi{fig:LO_LHC13_Std}).  For
distributions in lepton variables and for rapidity distributions of
jets, both EW contributions and QCD--EW interferences are below one
per cent. EW contributions are larger for large rapidity or
rapidity--azimuthal-angle separation of the jets.

\subsection{Leading-order analysis with VBF cuts}
\label{sec: LO VBF cuts}
We now examine the LO cross section for the process $\process$ using
the same input parameters as in the previous section and employing the
cuts \refeq{eq:jetcuts}, \refeq{eq:leptoncuts}, \refeq{eq:DRcuts} as
before, but instead of \refeq{eq:Mcuts} the following VBF cuts
\ba
&&M_{\Pj\Pj} > 600\GeV,\quad |y_{\Pj_1} - y_{\Pj_2}| > 4, \quad y_{\Pj_1}\cdot y_{\Pj_2} < 0, \nn\\
&& \min{(y_{\Pj_1},y_{\Pj_2})} < y_\Pl <   \max{(y_{\Pj_1},y_{\Pj_2})}.
\label{eq:VBFcuts}
\ea
Here $\Pj_1$ and $\Pj_2$ are the leading and subleading jets, \ie the
jets with the highest transverse momenta.

The results for the cross section with VBF cuts are given in
\refta{tab:LO-LHC13-VBF-xsection}.  Compared with the result for
standard acceptance cuts in \refeq{eq:LO-x-section-LHC13} the total
cross section decreased by a factor of 50. This reduction of the
signal is a result of the strong VBF constraints to the phase space.
While gluonic processes still dominate the cross section their
relative contribution decreased to 59\%. The pole approximation does
not work as well as in the previous setup and underestimates the full
result at LO by about 11\%. This is due to the fact that we did not
apply the cut \refeq{eq:Mcuts} thus allowing for a larger contribution
from photon exchanges.  When imposing the cut \refeq{eq:Mcuts} in the
VBF set-up, the pole approximation is again accurate at the level of
$1{-}2$ per cent. The EW contributions of $\O(\alpha^4)$ become more
significant (17\% in the four-quark processes and 7\% in the total
sum), while interferences between EW and QCD amplitudes are below
$1\%$ for all four-quark processes for this setup apart from the
suppressed channels of type $u\ub\to d\db \Plm\Plp$.
\begin{table}%
\begin{center}
\renewcommand{\arraystretch}{1.2}
   \begin{tabular}{|c|c|c|c|c|c|c|}
     \hline
process class  & $\sigma $  & $\sigma^{\PZ}$ & $ \sigma/\sigma_{\rm tot}$ & $\sigma_{\alphas^2\alpha^2}/\sigma$ & $\sigma_{\alphas\alpha^3}/\sigma$ & $\sigma_{\alpha^4}/\sigma$ \\
             & $[\rm{ fb}]$  &  $[\rm{ fb}]$  & $[\%]$ & $[\%]$ & $[\%]$ & $[\%]$ \\
\hline 
$u\Pg\to u\Pg \Pl^-\Pl^+$, $d\Pg\to d\Pg \Pl^-\Pl^+$& \multirow{2}{*}{540.9(3)} & \multirow{2}{*}{482.4(3)} & \multirow{2}{*} {52.0}  &\multirow{2}{*}{100}  & \multirow{2}{*}{---}& \multirow{2}{*}{---} \\
$\ub \Pg\to \ub \Pg \Pl^-\Pl^+$, $\db \Pg\to \db \Pg \Pl^-\Pl^+$ &&&&&&\\
\hline 
$u\ub \to \Pg\Pg \Pl^-\Pl^+$, $d\db \to \Pg\Pg \Pl^-\Pl^+$  & 22.35(1)  & 19.80(1) & 2.2   & {100}  & {---}& {---} \\
\hline 
$\Pg\Pg \to u\ub \Pl^-\Pl^+$, $ \Pg\Pg \to d\db \Pl^-\Pl^+$ &  54.53(4) & 50.56(3) & 5.2  & {100}  & {---}& {---} \\
\hline 
$uu\to uu \Pl^-\Pl^+$, $dd\to dd \Pl^-\Pl^+$ & \multirow{2}{*}{86.22(5)} & \multirow{2}{*}{73.70(5)} & \multirow{2}{*}{8.3}    &\multirow{2}{*}{97.0} & \multirow{2}{*}{0.1}& \multirow{2}{*}{2.8} \\
$\ub\ub\to \ub\ub \Pl^-\Pl^+$, $\db\db\to \db\db \Pl^-\Pl^+$ &&&&&&\\
\hline 
$u\ub\to u'\ub' \Pl^-\Pl^+$,  $d\db\to d'\db' \Pl^-\Pl^+$ & \multirow{3}{*}{65.98(3)}& \multirow{3}{*}{57.64(3)} & \multirow{3}{*}{6.3} &\multirow{3}{*}{98.2} & \multirow{3}{*}{-0.1}& \multirow{2}{*}{2.0}    \\
$u\ub'\to u\ub' \Pl^-\Pl^+$,  $d\db'\to d\db' \Pl^-\Pl^+$ &&&&&&\\
$u\ub\to u\ub \Pl^-\Pl^+$,  $d\db\to d\db \Pl^-\Pl^+$ &&&&&&\\
\hline 
$u\ub\to d\db \Pl^-\Pl^+$, $d\db\to u\ub \Pl^-\Pl^+$ & \multirow{2}{*}{21.198(7)}& \multirow{2}{*}{20.102(7)} & \multirow{2}{*}{2.0} &\multirow{2}{*}{1.9} & \multirow{2}{*}{-4.6}& \multirow{2}{*}{102.7}\\
$u\ub'\to d\db' \Pl^-\Pl^+$,  $d\db'\to u\ub' \Pl^-\Pl^+$ &&&&&&\\
\hline 
$ud\to u'd' \Pl^-\Pl^+$, $\ub\db\to \ub'\db' \Pl^-\Pl^+$ & \multirow{4}{*}{180.61(8)}& \multirow{4}{*}{163.94(8)} & \multirow{4}{*}{17.3} &\multirow{4}{*}{74.0} & \multirow{4}{*}{1.1}& \multirow{4}{*}{24.9}\\
$ud\to ud \Pl^-\Pl^+$, $\ub\db\to \ub\db \Pl^-\Pl^+$&&&&&& \\
$uu'\to uu' \Pl^-\Pl^+$, $\ub\ub'\to \ub\ub' \Pl^-\Pl^+$&&&&&& \\
$dd'\to dd' \Pl^-\Pl^+$, $\db\db'\to \db\db' \Pl^-\Pl^+$&&&&&& \\
     \hline 
$u\db\to u'\db' \Pl^-\Pl^+$, $\ub d\to \ub'd' \Pl^-\Pl^+$ & \multirow{2}{*}{ 67.73(3)} & \multirow{2}{*}{61.01(3)} & \multirow{2}{*}{6.5} &\multirow{2}{*}{99.0} &\multirow{2}{*}{-0.1} & \multirow{2}{*}{1.1}\\
$u\db\to u\db \Pl^-\Pl^+$, $\ub d\to \ub d \Pl^-\Pl^+$&&&&&&\\
     \hline
     \hline gluonic       &  617.8(4)  & 552.8(3) &  59.4 & 100 & ---  & --- \\
     \hline four-quark    &  421.7(1)  & 376.4(1) &  40.6 & 82.9& 0.2  & 16.9 \\
     \hline bottom quarks &  51.82(2)  & ---      &  4.98 & --- & ---  & --- \\
     \hline sum           &  1039.6(4) & 929.2(3) &  100  & 93.1& 0.01 & 6.9\\
     \hline
   \end{tabular}
\end{center}
  \caption{Composition of the LO cross section for 
   $\process$ at the LHC operating at $13\TeV$ for
    VBF cuts. In the first column the contributing
    partonic processes are listed, where $u,u'\ne u$ denote the up-type
    quarks $\Pu,\Pc$ and $d,d'\ne d$ the down-type quarks
    $\Pd,\Ps,\Pb$. The second column 
   provides the corresponding cross section where the numbers in
   parentheses give the integration error on the last digit. The third
   column contains the relative contribution to the total cross section
   in per cent. In the fourth, fifth and sixth column 
   we provide the relative contribution to a partonic channel from
   strong and EW contributions and their interference.}
  \label{tab:LO-LHC13-VBF-xsection}
\end{table}%

\begin{figure}
  \includegraphics[width=8cm]{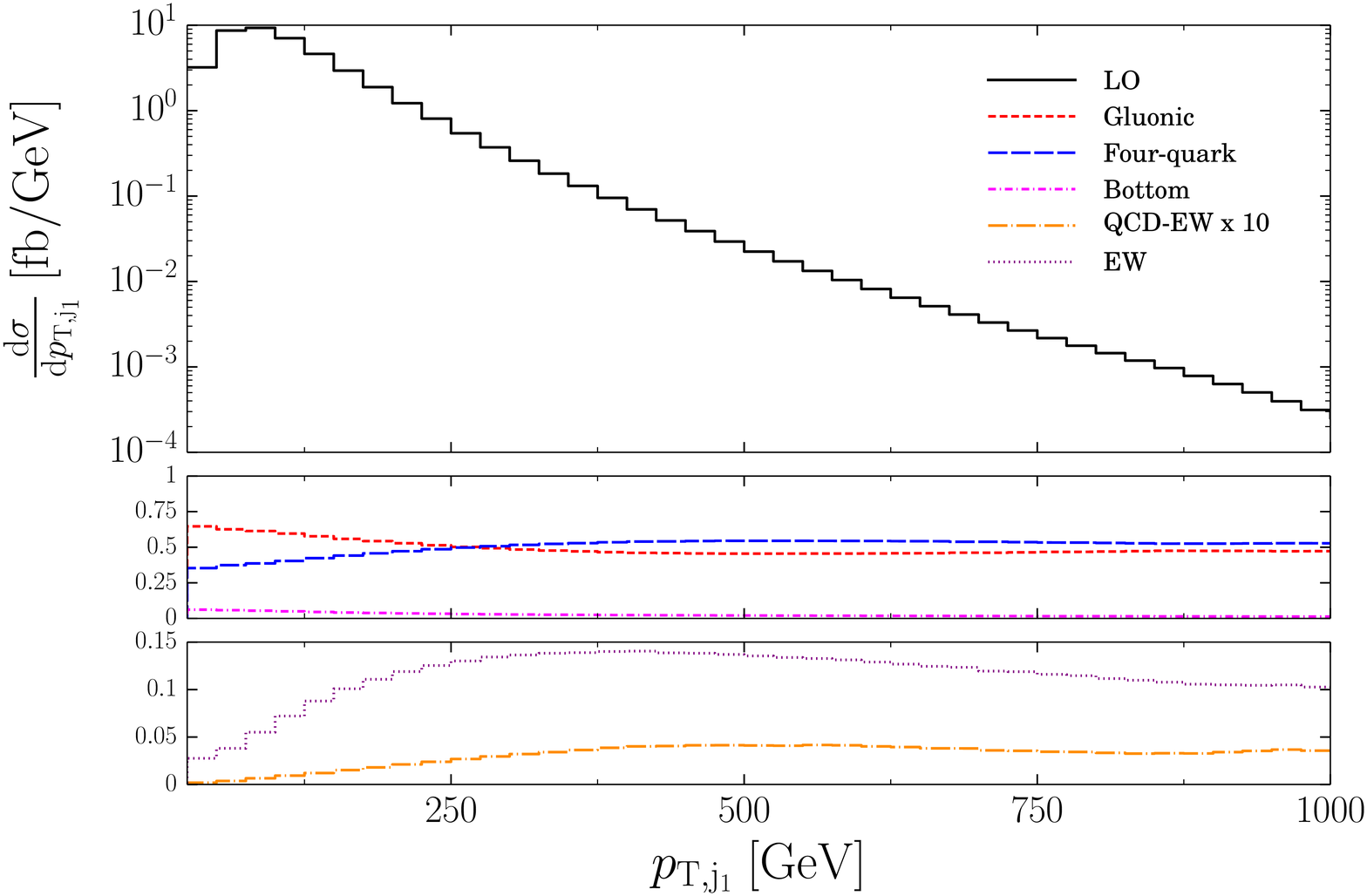}
  \includegraphics[width=8cm]{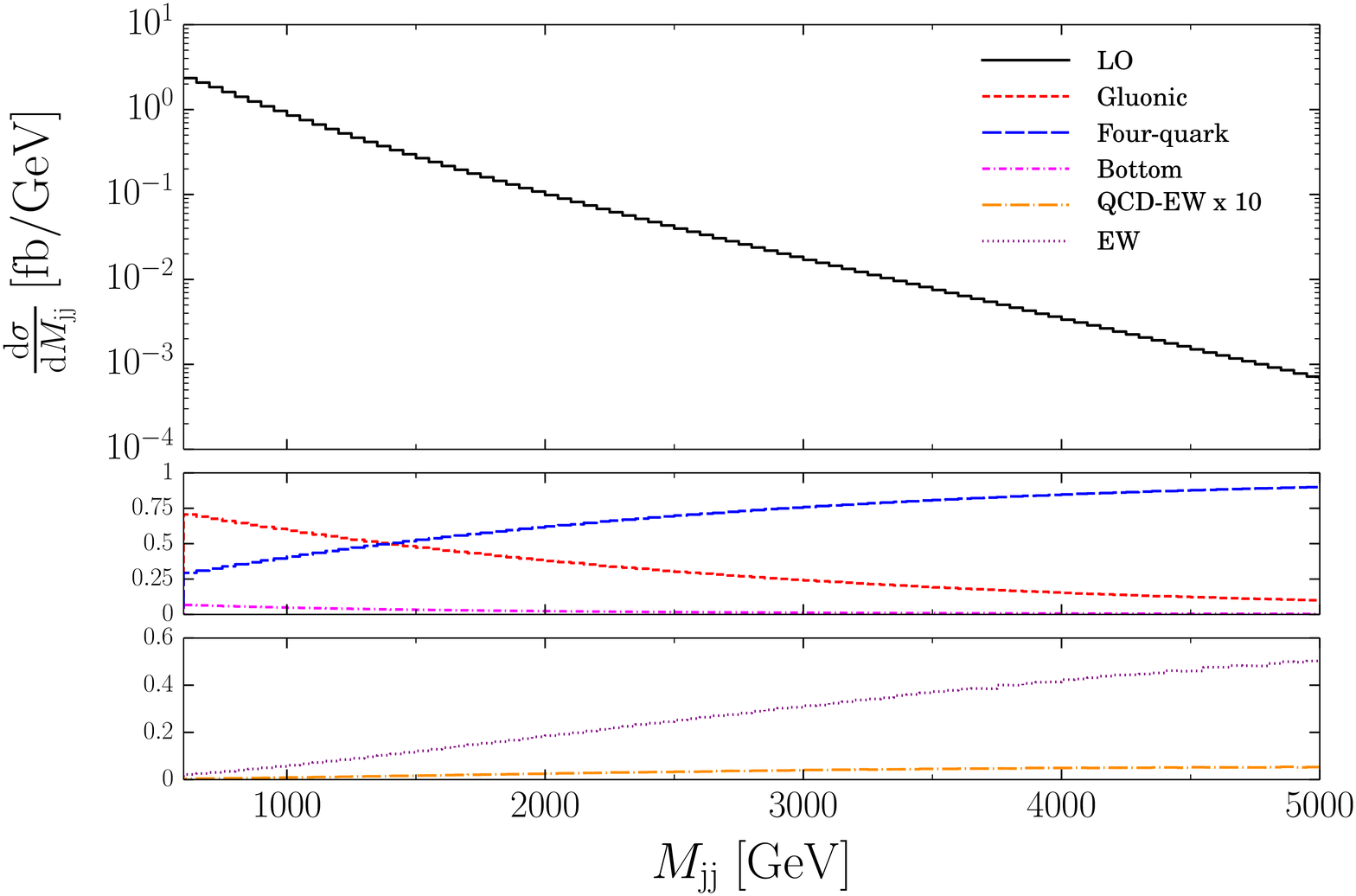}
  \includegraphics[width=8cm]{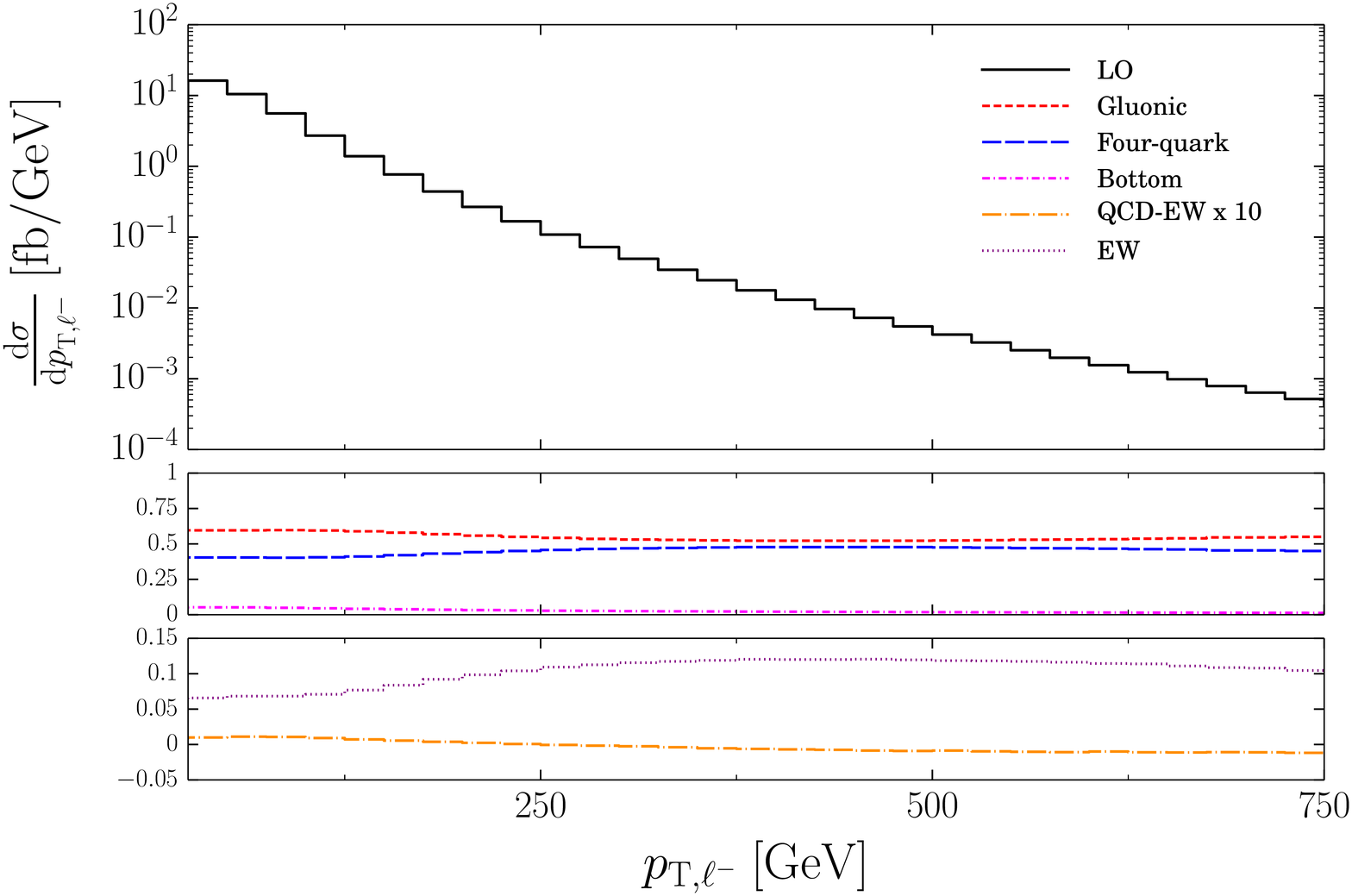}
  \includegraphics[width=8cm]{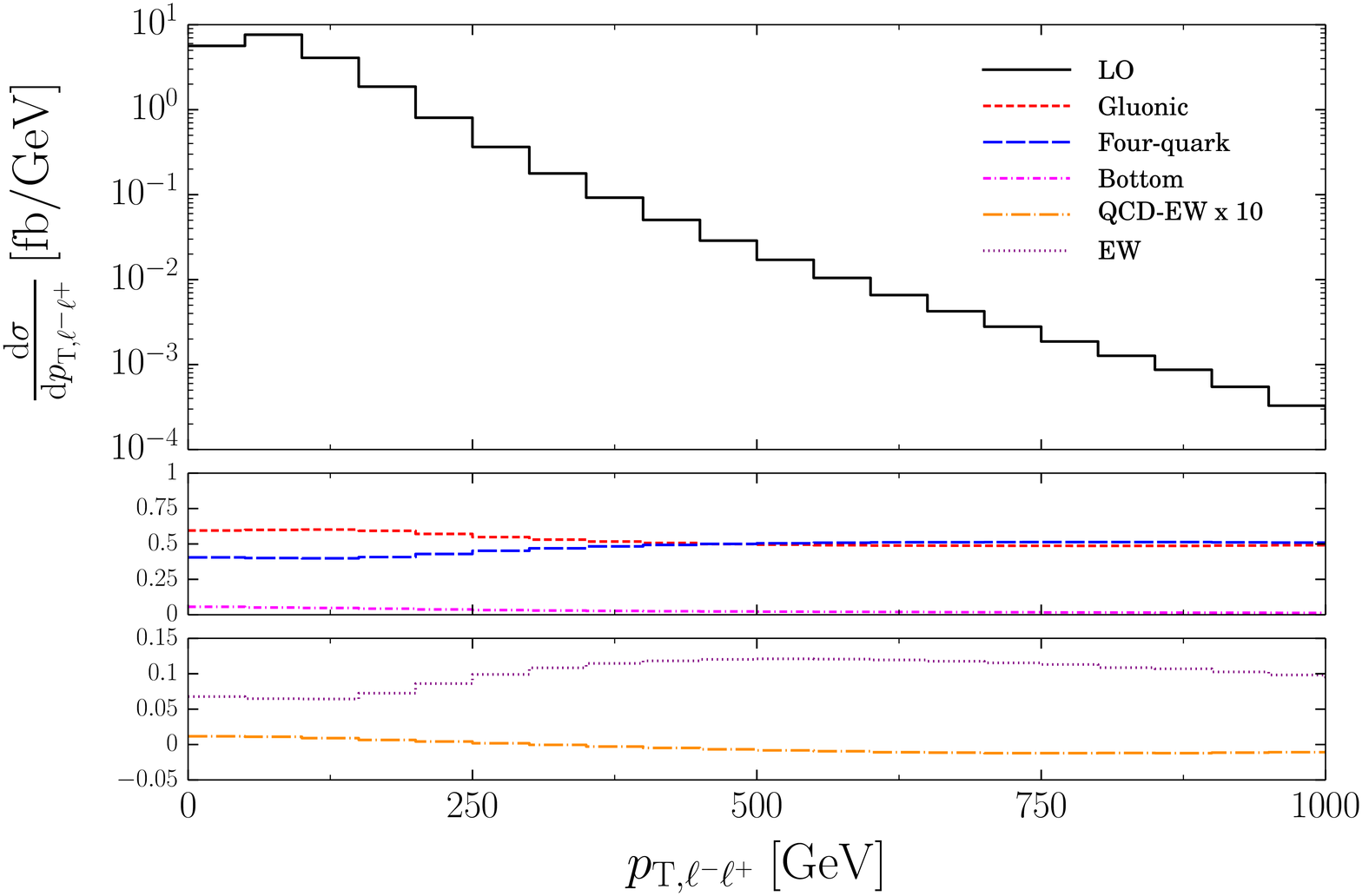}
  \includegraphics[width=8cm]{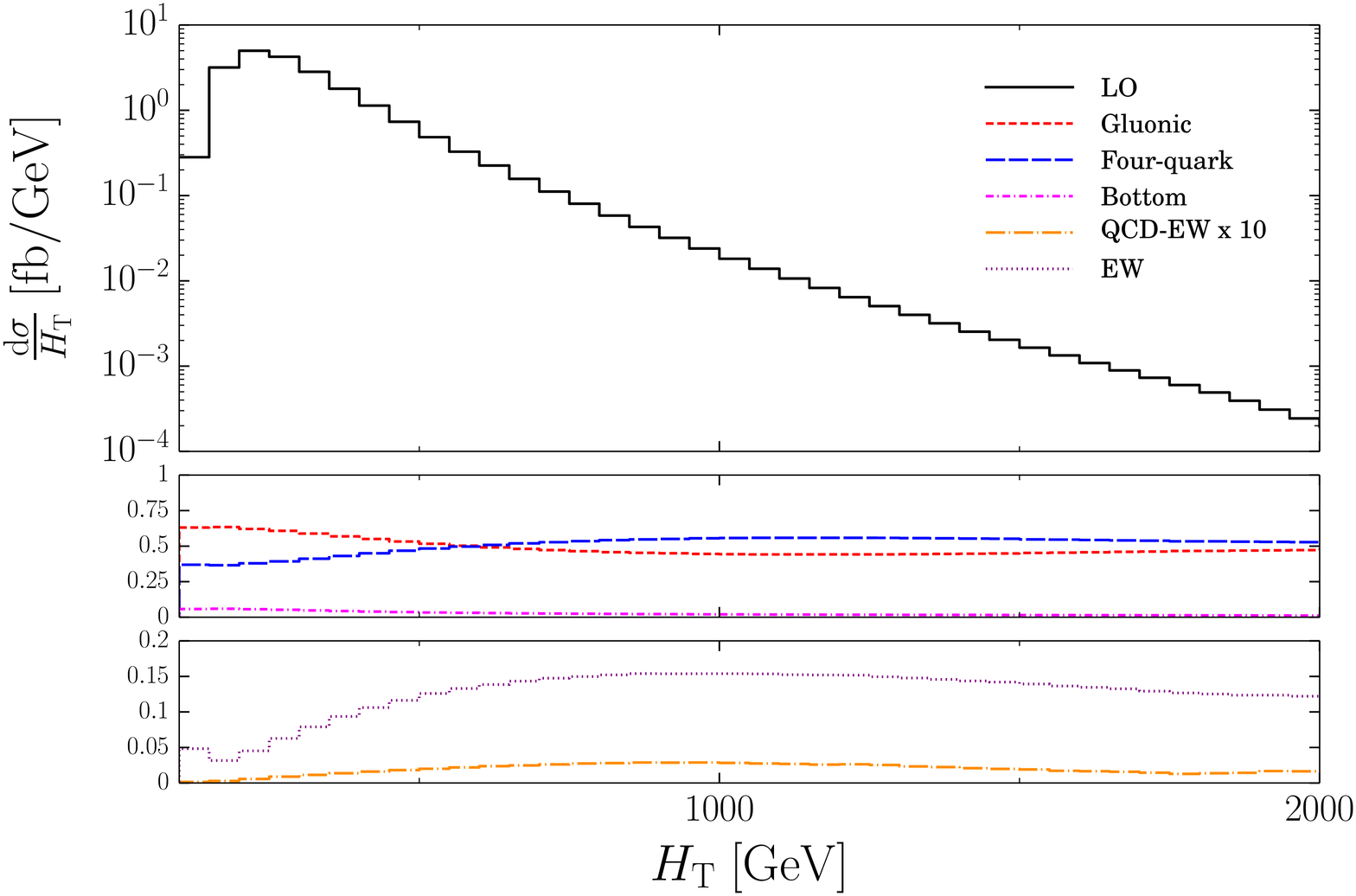}
  \includegraphics[width=8cm]{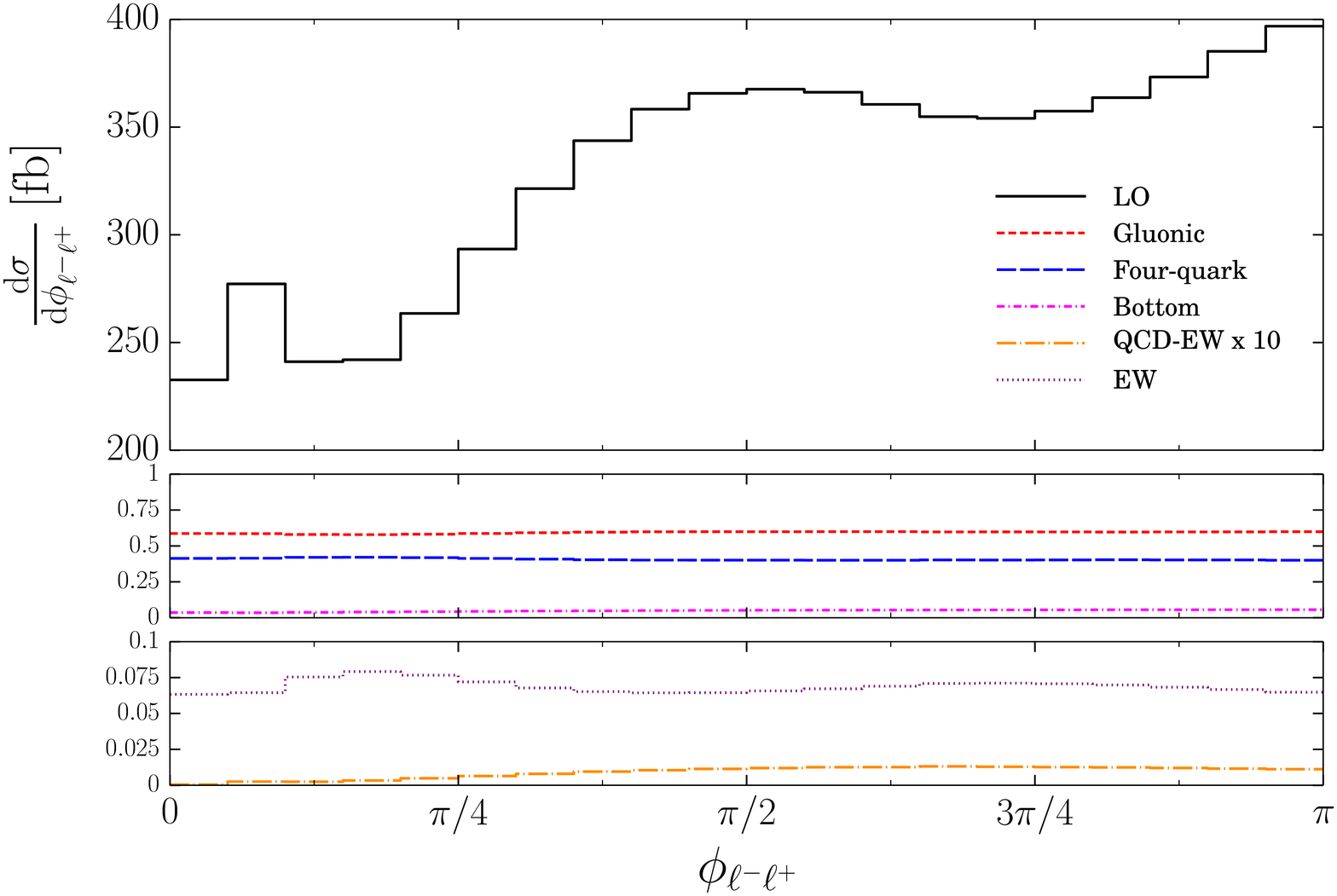}
  \caption{Distributions of the transverse momentum of the harder jet
    ${\rm j_1}$, the di-jet invariant mass $M_{\Pj\Pj}$, the
    transverse momenta of the negatively charged lepton $\Plm$ and the
    lepton pair $\Plm\Plp$, the scalar sum of all transverse momenta,
    and the relative azimuthal angle between the leptons at $13 \TeV$
    LHC for the VBF setup. The upper panels show the corresponding
    distributions, the middle ones the composition of the cross
    section and the lower ones the contributions of the EW diagrams.
 Further details are described in the text.}
  \label{fig:LO_LHC13_VBF}
\end{figure}

The corresponding differential distributions are shown in
\reffi{fig:LO_LHC13_VBF}. For the transverse momentum distributions of
the hardest jet, the negatively charged lepton and the lepton pair as
well as for the $\HT$ distribution we find no significant qualitative
changes beside the reduced normalisation.  The distribution in the
di-jet invariant mass decreases much slower as events with large
$M_{\Pj\Pj}$ tend to pass VBF cuts.  The variation in the $\phill$
distribution is smaller than for basic cuts.
However, VBF kinematics also affects the composition of the
distributions. For the $\pTo{\Pj_1}$ distribution the relative gluonic
contribution drops from 60\% at $\pTo{\Pj_1} = 30\GeV$ below 50\% at
$\pTo{\Pj_1}=270\GeV$. This is in line with the fact that VBF cuts
tend to enhance the four-quark contributions. The resulting balance
between four-quark and gluonic contributions is valid for higher
transverse momenta up to $1\TeV$. Thus for large $\pTo{\Pj_1}$ the
four-quark contributions are less relevant as for basic cuts. The
composition of the distribution in the di-jet invariant mass is very
similar as for basic cuts. For large $\pTo{\Plm}$ or large
$\pTo{\Plm\Plp}$ four-quark and gluonic contributions are about equal.
In the case of the $\HT$ distribution we find gluonic and four-quark
contributions similar as for the $\pTo{\Pj_1}$ distribution. The
composition does hardly depend on the azimuthal angle between the
leptons $\phill$.
As for basic cuts, the EW diagrams contribute sizeably ($50\%$ for
$M_{\Pj\Pj}=5\TeV$) for large $M_{\Pj\Pj}$ (as well as large $\Delta
y_{\Pj\Pj}$ or $\Delta R_{\Pj\Pj}$), while they stay at the level of
10\% for other distributions. QCD--EW interferences (multiplied by a
factor $10$ in \reffi{fig:LO_LHC13_VBF}) are at the per-mille level
for all considered distributions in the VBF set-up.

The relative contribution of bottom quarks is always below 5\% and
drops to the per-mille level in the high-energy tails of
distributions.

\section{NLO electroweak corrections to $\process$}
\label{sec: NLO}

\subsection{General remarks}
In this section we consider NLO electroweak corrections to the process
$\Pp\Pp \to \Pj\Pj\Plm\Plp$.  Since the dominant contributions in LO
perturbation theory are of $\O(\alphas^2\alpha^2)$ we focus on
electroweak corrections to these dominant contributions and study the
complete set of $\O(\alphas^2\alpha^3)$ contributions. LO gluonic
processes contribute exclusively at order $\O(\alphas^2\alpha^2)$
requiring the calculation of electroweak $\O(\alpha)$ one-loop
corrections and real photon emission
\beq
\delta\sigma^{\rm NLO, gluonic}_{\process} = \delta\sigma^{\rm virtual, gluonic}_{\process} + \delta\sigma^{\rm real, gluonic}_{\process\gamma}.
\eeq
Four-quark processes feature various powers of the
electromagnetic and the strong coupling constant at LO and thus the
computation is more involved.  A complete treatment in perturbation
theory at $\O(\alphas^2\alpha^3)$ for these processes requires the
incorporation of
\begin{itemize}
\item[a)]corrections of $\O(\alpha)$ to LO QCD contributions of
    $\O(\alphas^2\alpha^2)$, and
\item[b)]corrections of $\O(\alphas)$ to LO QCD--EW interferences of
    $\O(\alphas\alpha^3)$.
\end{itemize}
Consequently, we need to incorporate photon and gluon real emission
\beq
\delta\sigma^{\rm NLO, four\mhyphen{}quark}_{\process} = \delta\sigma^{\rm virtual, four\mhyphen{}quark}_{\process} + \delta\sigma^{\rm real, four\mhyphen{}quark}_{\process\gamma} + \delta\sigma^{\rm real, four\mhyphen{}quark}_{\process \Pg}
\eeq
at the relevant order.

Contributions involving external bottom quarks contribute less than
10\% to the total cross section and differential distributions,
and we do not consider NLO electroweak corrections to these processes.

\subsection{Virtual corrections}
\label{sec: virtual}
\begin{figure}
\centerline{
{\unitlength .6pt 
\begin{picture}(170,120)(0,0)
\SetScale{.6}
\ArrowLine( 15,95)( 45,90)
\Gluon( 45,10)( 15, 5){3}{3}
\ArrowLine( 95,90)( 125,95)
\Gluon( 125,5)(95, 10){3}{3}
\ArrowLine(45,90)(45,10)
\ArrowLine(45,10)(95,10)
\ArrowLine(95,10)(130,30)
\ArrowLine(130,70)(95,90)
\ArrowLine(165,70)(130,70)
\ArrowLine(135,30)(165,30)
\Photon(45,90)(95,90){3}{6}
\Photon(130,70)( 130,30){3}{4}
\Vertex(45,90){3}
\Vertex(45,10){3}
\Vertex(95,90){3}
\Vertex(95,10){3}
\Vertex(130,70){3}
\Vertex(130,30){3}
\put(-3,90){$q_i$}
\put(-3,0){$\rm g$}
\put(168,30){{$q_i$}}
\put(128,0){{$\rm g$}}
\put(137,45){{$\ga,\PZ,\PW$}}
\put(40,100){{$\ga,\PZ,\PW$}}
\put(168,66){{$\Plp$}}
\put(128,90){{$\Plm$}}
\SetScale{1}
\end{picture}
}
\qquad\qquad
{\unitlength .6pt
\begin{picture}(170,120)(0,0)
\SetScale{.6}
\ArrowLine( 15,60)( 45,55)
\ArrowLine(45,55)(75,55)
\ArrowLine(75,55)(105,55)
\ArrowLine( 105,55)( 135,60)
\ArrowLine( 15,5)( 45,10)
\ArrowLine(45,10)(105,10)
\ArrowLine( 105,10)( 135,05)
\Gluon( 45,10)( 45, 55){3}{3}
\Gluon( 105,10)( 105, 55){-3}{3}
\ArrowLine(105,85)(135,95)
\ArrowLine(135,75)(105,85)
\Photon(75,55)(105,85){3}{5}
\Vertex(45,55){3}
\Vertex(45,10){3}
\Vertex(105,55){3}
\Vertex(105,10){3}
\Vertex(75,55){3}
\Vertex(105,85){3}
\put(-3,55){$q_i$}
\put(-3,0){$q_j$}
\put(143,55){{$q_i$}}
\put(143,0){{$q_j$}}
\put(143,92){{$\Plm$}}
\put(143,72){{$\Plp$}}
\Text(80,80)[r]{{$\ga,\PZ$}}
\Text(35,30)[r]{{$\Pg$}}
\Text(115,30)[l]{{$\Pg$}}
\SetScale{1}
\end{picture}}
}\vspace{2ex}
\centerline{
{\unitlength .6pt 
\begin{picture}(170,120)(0,0)
\SetScale{.6}
\ArrowLine( 15,95)( 45,90)
\ArrowLine(45,90)(95,90)
\ArrowLine( 95,90)( 125,95)
\ArrowLine( 15, 5)( 45,10)
\ArrowLine(45,10)(95,10)
\ArrowLine( 95,10)( 125,5)
\ArrowLine(165,30)(130,30)
\ArrowLine(130,70)(165,70)
\ArrowLine(130,30)(130,70)
\Photon(95,10)(130,30){3}{4}
\Photon(130,70)(95,90){3}{4}
\Gluon(45,90)(45,10){-3}{6}
\Vertex(45,90){3}
\Vertex(45,10){3}
\Vertex(95,90){3}
\Vertex(95,10){3}
\Vertex(130,70){3}
\Vertex(130,30){3}
\put(-3,90){$q_i$}
\put(-3,0){$q_j$}
\put(128,0){{$q_j$}}
\put(168,66){{$\Plm$}}
\put(168,26){{$\Plp$}}
\put(128,90){{$q_i$}}
\Text(115,65)[r]{{$\ga,\PZ$}}
\Text(115,35)[r]{{$\ga,\PZ$}}
\Text(35,50)[r]{{$\Pg$}}
\SetScale{1}
\end{picture}}
\qquad\qquad
{\unitlength .6pt 
\begin{picture}(170,120)(0,0)
\SetScale{.6}
\ArrowLine( 15,95)( 45,90)
\ArrowLine(45,90)(95,90)
\ArrowLine( 95,90)( 125,95)
\ArrowLine( 15, 5)( 45,10)
\ArrowLine(45,10)(95,10)
\ArrowLine( 95,10)( 125,5)
\ArrowLine(165,30)(130,30)
\ArrowLine(130,70)(165,70)
\ArrowLine(130,30)(130,70)
\Photon(95,10)(130,30){3}{4}
\Photon(130,70)(95,90){3}{4}
\Gluon(45,90)(45,10){-3}{6}
\Vertex(45,90){3}
\Vertex(45,10){3}
\Vertex(95,90){3}
\Vertex(95,10){3}
\Vertex(130,70){3}
\Vertex(130,30){3}
\put(-3,90){$q_i$}
\put(-3,0){$q_j$}
\put(128,0){{$q'_j$}}
\put(168,66){{$\Plm$}}
\put(168,26){{$\Plp$}}
\put(128,90){{$q'_i$}}
\Text(115,65)[r]{{$\PW$}}
\Text(115,35)[r]{{$\PW$}}
\Text(35,50)[r]{{$\Pg$}}
\SetScale{1}
\end{picture}}
}
\caption{Sample diagrams for virtual corrections: hexagon of $\O(\gs^2
  e^4)$ for {$q_i\,\Pg\to q_i\,\Pg\,\Plm\,\Plp$}
  (upper left), pentagon of $\O(\gs^4
  e^2)$ for {$q_i\,q_j\to q_i\,q_j\,\Plm\,\Plp$}
  (upper right), hexagon of $\O(\gs^2
  e^4)$ for {$q_i\,q_j\to q_i\,q_j\,\Plm\,\Plp$}
  (lower left), hexagon of $\O(\gs^2
  e^4)$ for {$q_i\,q_j\to q_i^\prime \,q_j^\prime\,\Plm\,\Plp$}
  (lower right).} 
\label{fig:NLOdiags}
\end{figure}
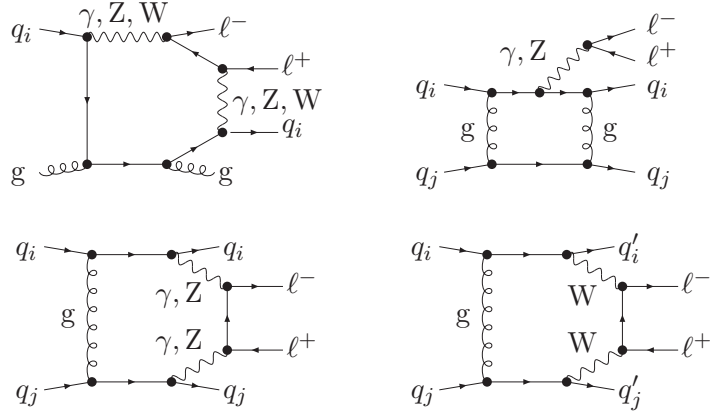

The virtual amplitudes are calculated using the 't~Hooft--Feynman
gauge.  The amplitudes for the various partonic processes can be
constructed from the same basic channels as at LO. Sample diagrams
  are given in \reffi{fig:NLOdiags}.  The virtual electroweak
corrections for a gluonic process like $\Pu \Pg \to \Pu \Pg \Plm \Plp$
involve $\O(1200)$ diagrams, including 18 hexagons and 85 pentagons,
all contributing to the cross section at $\O(\alphas^2\alpha^3)$. For
the $\Pu \Ps \to \Pu \Ps \Plm\Plp$ channel there are about 150
diagrams of order $\O(\gs^4 e^2)$ (QCD corrections to LO QCD
contributions), and some 800 diagrams of order $\O(\gs^2e^4)$,
including 32 hexagons and 50 pentagons.  At $\O(\alphas^2\alpha^3)$
the former contribute via interference with LO
EW diagrams, the latter via interference with LO QCD diagrams.
Additional diagrams of order $\O(e^6)$, the EW corrections to LO EW
diagrams, do not contribute at the considered order.  The $\Pu \Ps \to
\Pd \Pc\Plm\Plp$ channel furnishes no contributions of order $\O(\gs^4
e^2)$ but about 120 diagrams of order $\O(\gs^2e^4)$, including 24
hexagons and 4 pentagons, as well as diagrams of order $\O(e^6)$ that
are irrelevant in our approximation.

The most complicated topologies involve 6-point functions up to rank
4.  For the calculation of tensor integrals we use the library
\collier\ \cite{collier,Denner:2014gla}. It implements the recursive
numerical reduction methods of \citeres{Denner:2002ii,Denner:2005nn},
where numerical instabilities from small Gram determinants are avoided
by choosing suitable expansion algorithms depending on the actual
input momenta.  The scalar integrals are evaluated according to the
explicit results of
\citeres{Beenakker:1990jr,Denner:1991qq,Denner:2010tr}.  Both, in the
case of ultraviolet divergences as well as
in the case of infrared (IR) divergences, dimensional regularisation
is applied to extract the corresponding singularities. The EW sector
of the SM is renormalised using an on-shell prescription for the
$\PW$- and $\PZ$-boson masses in the framework of the complex-mass
scheme \cite{Denner:2005fg}.  As the coupling $\alpha_{\GF}$ is
derived from $\MW$, $\MZ$ and $\GF$ its counterterm inherits a
correction term $\Delta r$ from the weak corrections to muon decay.

\subsection{Real corrections}
\label{sec: real}
\subsubsection{Gluonic processes}
The real corrections to the gluonic subprocesses are induced by photon
Bremsstrahlung (see \reffi{fig:real_gamma_diags} left for a sample
diagram) and are all of $\O(\alphas^2\alpha^3)$.  IR divergences
resulting from emission of a soft or a collinear photon from an
external quark are regularised dimensionally. For an IR-save event
definition, the final-state singularities cancel with corresponding IR
poles from the virtual corrections. For the initial-state
singularities this cancellation is incomplete but the remnant can be
absorbed into a redefinition of the quark distribution function.
Technically we make use of the Catani--Seymour dipole formalism
\cite{Catani:1996vz}, with the extension as formulated in
\citeres{Nagy:1998bb,Nagy:2003tz,Campbell:2004ch}, which we
transferred in a straightforward way to the case of dimensionally
regularised photon emission.

In combination with photon radiation also final-state gluons, present
in the LO processes, cause IR divergences (see \citere{Denner:2009gj})
when they become soft.  While isolated soft gluons do not pass the
selection cuts because the requirement of two hard jets is not
fulfilled, in IR-safe observables quarks and thus all QCD partons have
to be recombined with photons if they are sufficiently collinear.
Thus, a soft gluon still passes the selection cuts if it is recombined
with a sufficiently hard collinear photon, giving rise to a soft-gluon
divergence, which would be cancelled by the virtual QCD corrections to
$\Pp\Pp \to \Plm\Plp\Pj+\gamma$ production.
Following \citeres{Denner:2009gj,Denner:2010ia} we eliminate this
singularity by discarding events which contain a jet consisting of a
hard photon recombined with a soft parton $a$ ($a=q_i,\bar{q}_i,\Pg$):
Photonic jets with $z_\gamma=
E_{\gamma}/(E_{\gamma}+E_{a})$ above a critical value $\zcut$ are
attributed to the process $\Pp\Pp\to \Plm\Plp \Pj+\gamma$ and
therefore excluded.  This event definition is still not IR-save
because the application of the $z_\gamma$ cut to recombined
quark--photon jets spoils the cancellation of final-state collinear
singularities with the virtual photonic corrections. The left-over
singularities can be absorbed into the quark--photon fragmentation
function \cite{Glover:1993xc,Buskulic:1995au}. 
The additional cut on $z_\gamma$ implies a modification of the
integrated dipole terms. The corresponding expressions and further
details can be found in \refapp{app: photon fragmentation}.

In the following numerical analysis, a recombined photon--parton jet
is subjected to the cut $z_\gamma=E_\gamma/(E_\gamma+E_a)< 0.7$. 

\subsubsection{Four-quark processes}
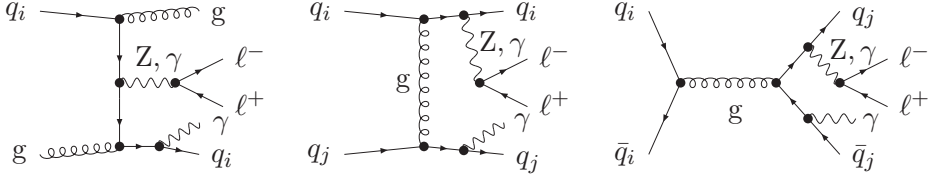
\begin{figure}
\centerline{
{\unitlength .6pt 
\begin{picture}(125,100)(0,0)
\SetScale{.6}
\ArrowLine( 15,95)( 65,90)
\Gluon( 15, 5)( 65,10){-3}{6}
\ArrowLine( 65,10)(90,10)
\ArrowLine( 90,10)(115, 5)
\Gluon(115,95)( 65,90){-3}{6}
\ArrowLine( 65,90)( 65,40)
\ArrowLine( 65,40)( 65,10)
\Photon(100,50)( 65,50){3}{4}
\ArrowLine( 100,50)( 130,65)
\ArrowLine( 130,35)( 100,50)
\Photon(90,10)(115,25){3}{4}
\Vertex(65,90){3}
\Vertex(65,10){3}
\Vertex(65,50){3}
\Vertex(100,50){3}
\Vertex(90,10){3}
\Text(  7,95)[r]{$\Pq_i$}
\Text(  7,5)[r]{$\Pg$}
\put(123,90){{$\Pg$}}
\put(123,0){{$\Pq_i$}}
\put(75,61){{$\PZ,\ga$}}
\put(138,60){{$\Plm$}}
\put(138,30){{$\Plp$}}
\Text( 124,23)[l]{{$\gamma$}}
\SetScale{1}
\end{picture}}
\hspace*{3em}
{\unitlength .6pt 
\begin{picture}(125,100)(0,0)       
\SetScale{.6}
\ArrowLine( 15,95)( 65,90)
\ArrowLine( 15, 5)( 65,10)
\ArrowLine( 65,10)(90,7.5)
\ArrowLine( 90,7.5)(115, 5)
\ArrowLine( 65,90)(90,92.5)
\ArrowLine( 90,92.5)(115,95)
\Gluon( 65,10)( 65,90){3}{8}
\Photon(100,50)( 90,92.5){3}{4}
\ArrowLine( 100,50)( 130,65)
\ArrowLine( 130,35)( 100,50)
\Photon(115,25)( 90,7.5){3}{4}
\Vertex(65,90){3}
\Vertex(65,10){3}
\Vertex(90,92.5){3}
\Vertex(90,7.5){3}
\Vertex(100,50){3}
\Text(  7,95)[r]{$\Pq_i$}
\Text(  7,5)[r]{${\Pq}_j$}
\Text(123,95)[l]{{$q_i$}}
\Text(123,0)[l]{{${\Pq}_j$}}
\put(138,60){{$\Plm$}}
\put(138,30){{$\Plp$}}
\put(100,70){{$\PZ,\ga$}}
\Text(57,50)[r]{{$\Pg$}}
\Text(122,25)[l]{{$\ga$}}
\SetScale{1}
\end{picture}}
\hspace*{3em}
{\unitlength .6pt 
\begin{picture}(175,100)(0,0)
\SetScale{.6}
\ArrowLine( 15,95)( 35,50)
\ArrowLine( 35,50)( 15, 5)
\ArrowLine(95,50)(115,72.5)
\ArrowLine(115,72.5)(135,95)
\ArrowLine(115,27.55)(95,50)
\ArrowLine(135, 5)(115,27.5)
\Gluon( 35,50)(95,50){3}{7}
\Photon(115,72.5)(135,50){3}{4}
\ArrowLine( 135,50)( 165,65)
\ArrowLine( 165,35)( 135,50)
\Photon(115,27.5)(145,27.5){3}{3}
\Vertex( 35,50){3}
\Vertex(95,50){3}
\Vertex( 115,72.5){3}
\Vertex(135,50){3}
\Vertex(115,27.5){3}
\Text(  7,95)[r]{$\Pq_i$}
\Text(  7,5)[r]{$\bar{\Pq}_i$}
\put(143,90){{$\Pq_j$}}
\put(143,0){{$\bar{\Pq}_j$}}
\put(173,60){{$\Plm$}}
\put(173,30){{$\Plp$}}
\put(129,65){{$\PZ,\ga$}}
\put( 65,28){{$\Pg$}}
\Text( 150,25)[l]{{$\gamma$}}
\SetScale{1}
\end{picture}}
\hspace*{3em}
}
\caption{Sample diagrams for real photon radiation:
  contributions to $\Pq_i\,\Pg\to\Pq_i\,\Pg\,\Plm\,\Plp\,\gamma$,  $\Pq_i\,\Pq_j\to\Pq_i\,\Pq_j\,\Plm\,\Plp\,\gamma$ and
  $\Pq_i\,\bar{\Pq}_i\to\Pq_j\,\bar{\Pq}_j\,\Plm\,\Plp\,\gamma$.  } 
\label{fig:real_gamma_diags}
\end{figure}

\begin{figure}
\centerline{
{\unitlength .6pt 
\begin{picture}(175,100)(0,0)
\SetScale{.6}
\ArrowLine( 15,95)( 35,50)
\ArrowLine( 35,50)( 15, 5)
\ArrowLine(95,50)(115,72.5)
\ArrowLine(115,72.5)(135,95)
\ArrowLine(115,27.55)(95,50)
\ArrowLine(135, 5)(115,27.5)
\Gluon( 35,50)(95,50){3}{7}
\Photon(115,72.5)(135,50){3}{4}
\ArrowLine( 135,50)( 165,65)
\ArrowLine( 165,35)( 135,50)
\Gluon(115,27.5)(145,27.5){3}{3}
\Vertex( 35,50){3}
\Vertex(95,50){3}
\Vertex( 115,72.5){3}
\Vertex(135,50){3}
\Vertex(115,27.5){3}
\Text(  7,95)[r]{$\Pq_i$}
\Text(  7,5)[r]{$\bar{\Pq}_i$}
\put(143,90){{$\Pq'_i$}}
\put(143,0){{$\bar{\Pq}'_i$}}
\put(173,60){{$\Plm$}}
\put(173,30){{$\Plp$}}
\put(129,65){{$\PZ,\ga$}}
\put( 65,32){{$\Pg$}}
\Text( 150,25)[l]{{$\Pg$}}
\Text(210,50)[]{$\times$}
\SetScale{1}
\end{picture}}
\qquad\quad
{\unitlength .6pt 
\begin{picture}(125,100)(0,0)
\SetScale{.6}
\ArrowLine( 15,95)( 65,90)
\ArrowLine( 65,10)( 15, 5)
\ArrowLine( 65,90)(115,95)
\ArrowLine(115, 5)( 90,7.5)
\ArrowLine( 90,7.5)( 65,10)
\Gluon(90,7.5)(115,25){3}{3}
\Photon( 65,10)( 65,90){3}{7}
\Photon(100,50)( 65,50){3}{4}
\ArrowLine( 100,50)( 130,65)
\ArrowLine( 130,35)( 100,50)
\Vertex(65,90){3}
\Vertex(65,10){3}
\Vertex(65,50){3}
\Vertex(100,50){3}
\Vertex(90,7.5){3}
\Text(  7,95)[r]{$\Pq_i$}
\Text(  7,5)[r]{$\bar{\Pq}_i$}
\put(123,90){{$\Pq'_i$}}
\put(123,0){{$\bar{\Pq}'_i$}}
\put(75,61){{$\PZ,\ga$}}
\Text( 58,70)[r]{{$\rm W$}}
\Text( 58,30)[r]{{$\rm W$}}
\put(138,60){{$\Plm$}}
\put(138,30){{$\Plp$}}
\put(123,23){{$\Pg$}}
\SetScale{1}
\end{picture}}
}\vspace{4ex}
\centerline{
{\unitlength .6pt 
\begin{picture}(175,100)(0,0)
\SetScale{.6}
\ArrowLine( 15,95)( 35,50)
\ArrowLine( 35,50)( 25,27.5)
\ArrowLine( 25,27.5)( 15, 5)
\ArrowLine(95,50)(115,72.5)
\ArrowLine(115,72.5)(135,95)
\ArrowLine(135,5)(95,50)
\Gluon( 35,50)(95,50){3}{7}
\Photon(115,72.5)(135,50){3}{4}
\ArrowLine( 135,50)( 165,65)
\ArrowLine( 165,35)( 135,50)
\Gluon(25,27.5)(75,15){3}{6}
\Vertex( 35,50){3}
\Vertex(95,50){3}
\Vertex( 115,72.5){3}
\Vertex(135,50){3}
\Vertex(25,27.5){3}
\Text(  7,95)[r]{$\Pq_i$}
\Text(  7,5)[r]{$\bar{\Pq}_i$}
\put(143,90){{$\Pq'_i$}}
\put(143,0){{$\bar{\Pq}'_i$}}
\put(173,60){{$\Plm$}}
\put(173,30){{$\Plp$}}
\put(129,65){{$\PZ,\ga$}}
\put( 65,64){{$\Pg$}}
\Text( 80,15)[l]{{$\Pg$}}
\Text(210,50)[]{$\times$}
\SetScale{1}
\end{picture}}
\qquad\quad
{\unitlength .6pt 
\begin{picture}(125,100)(0,0)
\SetScale{.6}
\ArrowLine( 15,95)( 35,50)
\ArrowLine( 35,50)( 15, 5)
\ArrowLine(95,50)(115,72.5)
\ArrowLine(115,72.5)(135,95)
\ArrowLine(115,27.55)(95,50)
\ArrowLine(135, 5)(115,27.5)
\Photon( 35,50)(95,50){3}{7}
\Photon(115,72.5)(135,50){3}{4}
\ArrowLine( 135,50)( 165,65)
\ArrowLine( 165,35)( 135,50)
\Gluon(115,27.5)(145,27.5){3}{3}
\Vertex( 35,50){3}
\Vertex(95,50){3}
\Vertex( 115,72.5){3}
\Vertex(135,50){3}
\Vertex(115,27.5){3}
\Text(  7,95)[r]{$\Pq_i$}
\Text(  7,5)[r]{$\bar{\Pq}_i$}
\put(143,90){{$\Pq'_i$}}
\put(143,0){{$\bar{\Pq}'_i$}}
\put(173,60){{$\Plm$}}
\put(173,30){{$\Plp$}}
\put(129,65){{$\PZ,\ga$}}
\put( 65,28){{$\PZ$}}
\Text( 150,25)[l]{{$\Pg$}}
\SetScale{1}
\end{picture}}
}
\caption{Sample diagrams for interferences of QCD and EW diagrams with
  gluon radiation.} 
\label{fig:real_gluon_diags}
\end{figure}
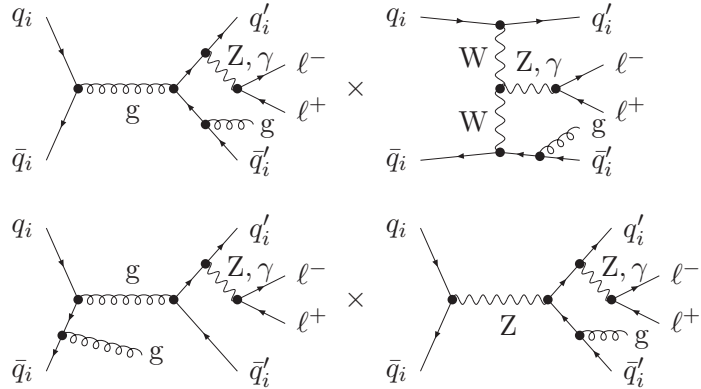

The real corrections to the four-quark processes feature soft and
collinear emission of a photon or a gluon. At $\O(\alphas^2\alpha^3)$
we have contributions with photon emission from the LO QCD diagrams (see
\reffi{fig:real_gamma_diags} for examples) and interferences from
diagrams with real gluon emission from QCD and EW diagrams (see
\reffi{fig:real_gluon_diags} for examples). 
All channels that result from crossing of two quarks or a quark and
the gluon have to be taken into account. Crossing the gluon into the
initial state leads to partonic subprocesses which do not have a LO
counterpart (such as $\Pg \Pu \to \Pu \Pu \Pubar \Plm\Plp$) and are IR
finite.

In the case of photon emission we employ the same recombination
procedure as in the previous subsection and thus ensure the equal
treatment of gluons and quarks in the framework of photon
fragmentation.  For gluon emission amplitudes we apply the standard
Catani--Seymour formalism with phase-space restricted subtraction
terms \cite{Nagy:1998bb,Nagy:2003tz,Campbell:2004ch}. 

\subsubsection{Checks}

The calculation has been performed with
\recola~\cite{Actis:2012qn,Uccirati:2014fda}, which provides all
relevant amplitudes for tree level, one loop {and} real radiation, as
well as the colour-correlated squared matrix elements required for the
subtraction terms. For the evaluation of the loop integrals \recola\ 
uses the tensor-integral library \collier\ 
\cite{collier,Denner:2014gla}. The phase-space integration is carried
out with an in-house multi-channel Monte-Carlo generator \cite{Motz}.

We performed several consistency checks of the calculation.  The
tensor integrals are evaluated by {\sc Collier}, which includes a
second independent implementation of all its building blocks. We have
checked that the cross section and the distributions do not change
within integration errors when switching between the different
implementations in {\sc Collier}.  Making use of our implementation of
phase-space restricted subtraction terms we performed the calculation
of all cross sections and distributions both for $\alpha=1$ and
$\alpha=0.01$ and found agreement within integration errors. The
distributions shown in the following have been obtained with
$\alpha=0.01$, which yields somewhat smoother results.

We have performed various cross-checks of the calculation with
conventional methods. To this end we have used {\sc FeynArts}~3.2
\cite{Hahn:2000kx,Hahn:2001rv}, {\sc FormCalc}~3.1 \cite{Hahn:1998yk}
and {\sc Pole} \cite{Accomando:2005ra} for the generation,
simplification and calculation of the Feynman amplitudes.  
The phase-space integration is performed with the multi-channel
generator {\sc Lusifer} \cite{Dittmaier:2002ap}.  For the gluonic
channels a complete second calculation was performed using this second
setup in the case of an on-shell $\PZ$ boson \cite{Actis:2012qn}. For
the off-shell calculation the individual contributions of real photon
emission and the corresponding dipoles and integrated dipoles
(including the contribution from photon fragmentation) have been
checked against the second implementation taking the channel
$\Pu\Pg\to\Pu\Pg\Plm\Plp$ as reference. In the case of the four-quark
channels, a full integration of the sample channel
$\Pu\Pd\to\Pu\Pd\Plm\Plp$ has been performed within the second setup
yielding perfect agreement with the results (including differential
distributions) obtained by \recola.

\subsection{Results for standard acceptance cuts}
\label{sec: NLO standard cuts}

In this section we present NLO results for the total cross section and
various differential distributions using the numerical input
parameters \refeq{eq:SMinput} and acceptance cuts \refeq{eq:jetcuts},
\refeq{eq:leptoncuts}, \refeq{eq:DRcuts}, and \refeq{eq:Mcuts}, as
introduced in \refse{sec: setup}.  In \refta{tab:NLO_LHC13_basic} we
present the impact of the NLO corrections on the various subprocesses.
\begin{table}
\begin{center}
\renewcommand{\arraystretch}{1.2}
   \begin{tabular}{|c|c|c|c|}
     \hline
&&&\\[-.4cm]
Process class  & $\sigma^\LO~ [\rm{ fb}] $     & $\sigma_\EW^\NLO~ [\rm{ fb}] $  & $ \frac{\sigma_\EW^\NLO}{\sigma^\LO}-1$ $[\%]$ \\[+.2cm]
\hline 
$u\Pg\to u\Pg \Pl^-\Pl^+$, $d\Pg\to d\Pg \Pl^-\Pl^+$,& \multirow{2}{*}{34584(8)}  &\multirow{2}{*}{33729(8)}  & \multirow{2}{*}{-2.41}\\
$\ub \Pg\to \ub \Pg \Pl^-\Pl^+$, $\db \Pg\to \db \Pg \Pl^-\Pl^+$ &&&\\
\hline 
$u\ub \to \Pg\Pg \Pl^-\Pl^+$, $d\db \to \Pg\Pg \Pl^-\Pl^+$ & 2713(1)    & 2646(1) & -2.47 \\
\hline 
$\Pg\Pg \to u\ub \Pl^-\Pl^+$, $ \Pg\Pg \to d\db \Pl^-\Pl^+$ & 3612(1)   & 3532(1) & -2.21 \\
\hline 
$uu\to uu \Pl^-\Pl^+$, $dd\to dd \Pl^-\Pl^+$, & \multirow{2}{*}{1315.1(3)}  &\multirow{2}{*}{1276.1(4)}  & \multirow{2}{*}{-2.97}\\
$\ub\ub\to \ub\ub \Pl^-\Pl^+$, $\db\db\to \db\db \Pl^-\Pl^+$ &&&\\
\hline 
$u\ub\to u'\ub' \Pl^-\Pl^+$,  $d\db\to d'\db' \Pl^-\Pl^+$, & 
\multirow{2}{*}{2463.7(5)}  &\multirow{2}{*}{2388.1(7)}  & \multirow{2}{*}{-3.07}\\
$u\ub'\to u\ub' \Pl^-\Pl^+$,  $d\db'\to d\db' \Pl^-\Pl^+$ &&&\\
\hline 
$u\ub\to d\db \Pl^-\Pl^+$, $d\db\to u\ub \Pl^-\Pl^+$, & \multirow{2}{*}{438.82(7)} & \multirow{2}{*}{425.2(2) }  & \multirow{2}{*}{-3.10 }\\
$u\ub'\to d\db' \Pl^-\Pl^+$,  $d\db'\to u\ub' \Pl^-\Pl^+$ &&&\\
\hline 
$ud\to u'd' \Pl^-\Pl^+$, $\ub\db\to \ub'\db' \Pl^-\Pl^+$,& \multirow{4}{*}{3856.8(7)}  &\multirow{4}{*}{3766.5(8)}  & \multirow{4}{*}{-2.34}\\
$ud\to ud \Pl^-\Pl^+$, $\ub\db\to \ub\db \Pl^-\Pl^+$&&& \\
$uu'\to uu' \Pl^-\Pl^+$, $\ub\ub'\to \ub\ub' \Pl^-\Pl^+$&&& \\
$dd'\to dd' \Pl^-\Pl^+$, $\db\db'\to \db\db' \Pl^-\Pl^+$&&& \\
     \hline 
$u\db\to u'\db' \Pl^-\Pl^+$, $\ub d\to \ub'd' \Pl^-\Pl^+$, & \multirow{2}{*}{2224.9(4)} &\multirow{2}{*}{2174.5(5)}  & \multirow{2}{*}{-2.27}\\
$u\db\to u\db \Pl^-\Pl^+$, $\ub d\to \ub d \Pl^-\Pl^+$&&&\\
     \hline
     \hline gluonic    & 40910(8) & 39907(8)   &  -2.45 \\
     \hline four-quark & 10299(1) & 10029(1)   &  -2.62 \\
     \hline sum        & 51209(8) & 49936(8)   &  -2.48 \\
     \hline
   \end{tabular}
\end{center}
  \caption{NLO corrections to the total cross section for 
   $\process$ at the $13\TeV$ LHC for basic cuts. In the first column the
    partonic processes are listed at LO. The second column
    provides the corresponding cross section including the NLO corrections. The third column contains the relative contribution to the total cross section
   in per cent.}
  \label{tab:NLO_LHC13_basic}
\end{table}
The corrections to the partonic total cross sections are negative for
all process classes. Both gluonic and four-quark processes receive
rather small relative corrections between $-3.1\%$ and $-2.2\%$. They
sum up to a correction of $-2.5\%$ for the total hadronic
cross section. For the LHC operating at $8\TeV$ we found the total
cross section corrected by $-2.3\%$ \cite{Denner:2013fca}.

\begin{figure}
  \includegraphics[width=8cm]{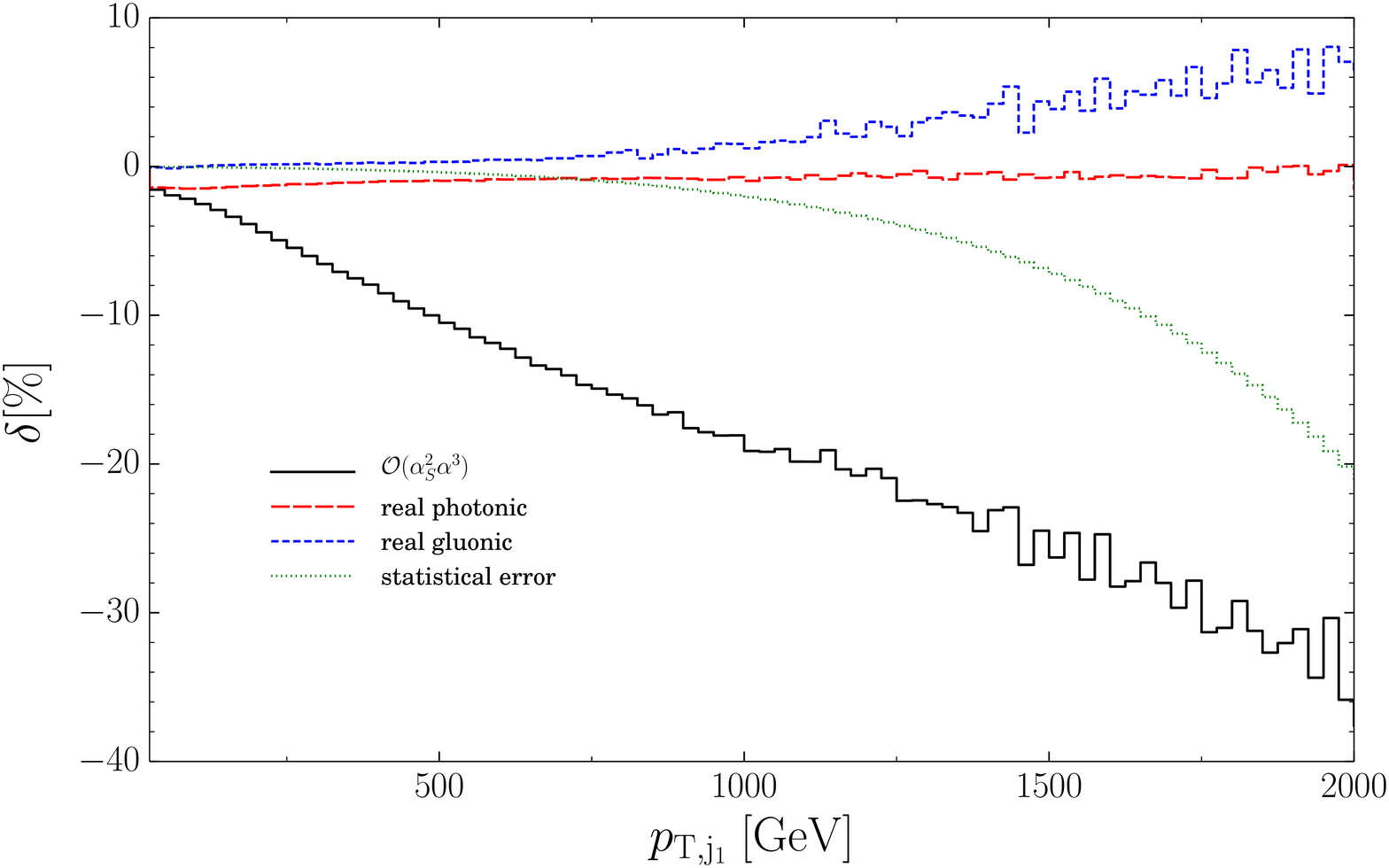}
  \includegraphics[width=8cm]{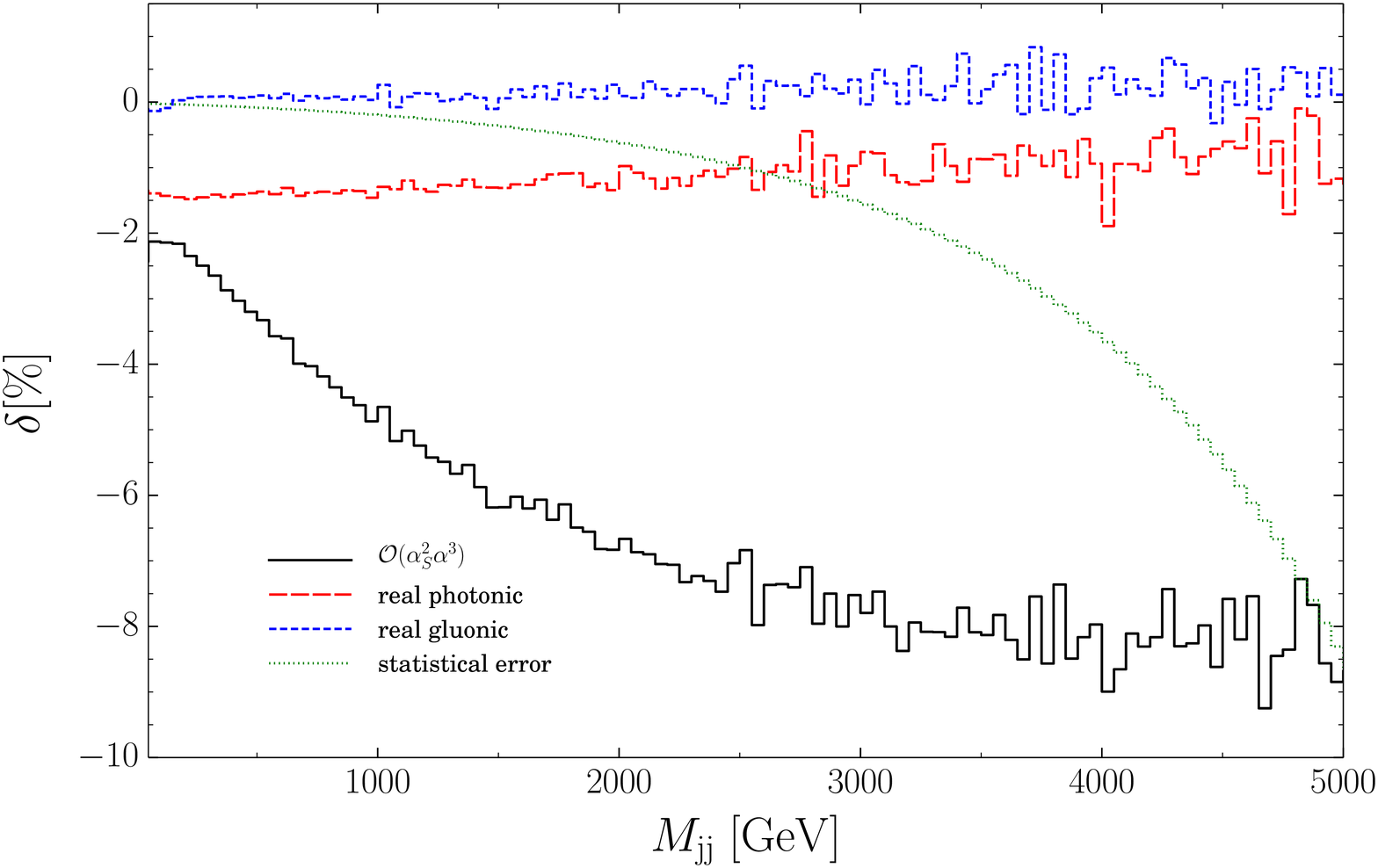}
  \includegraphics[width=8cm]{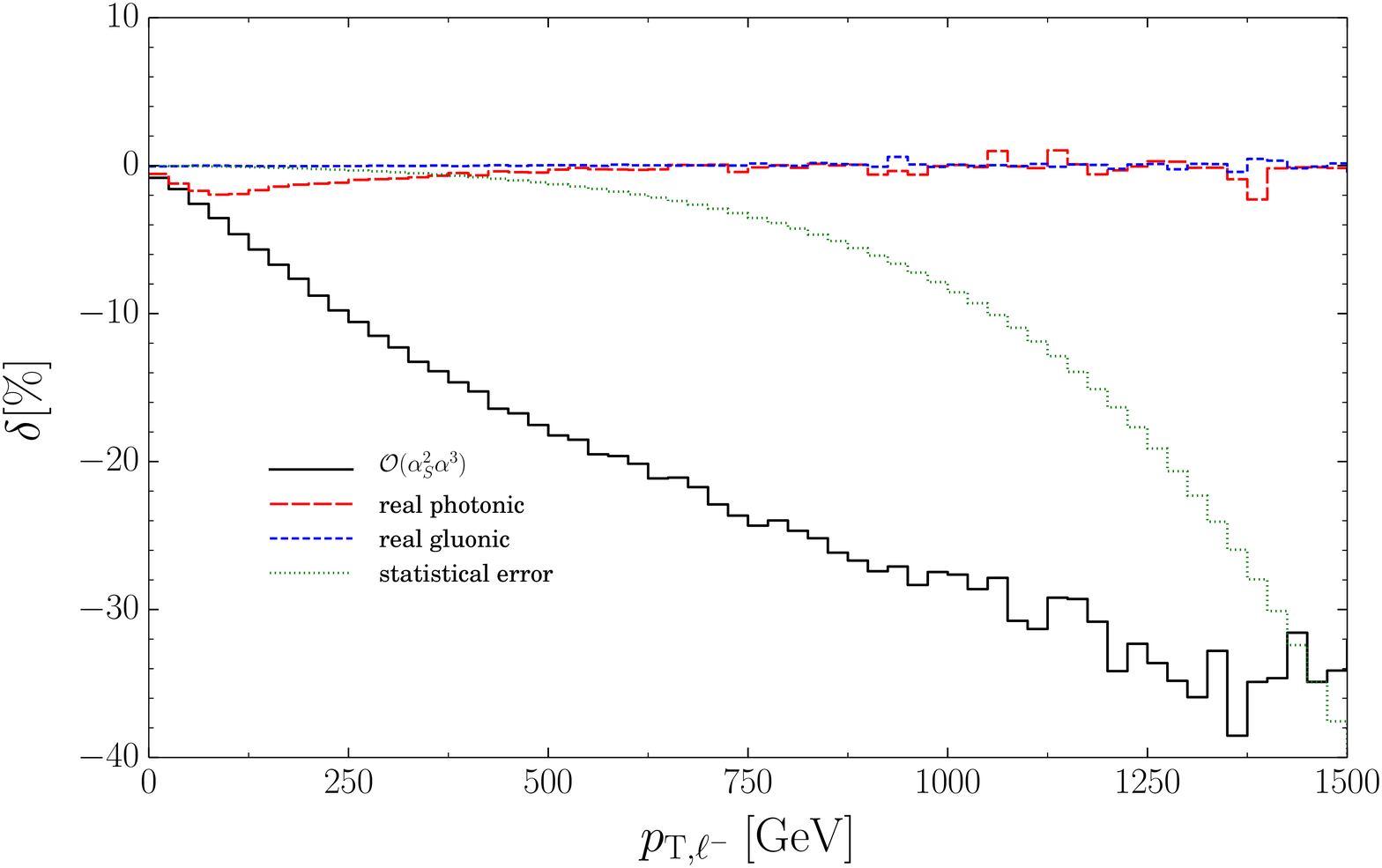}
  \includegraphics[width=8cm]{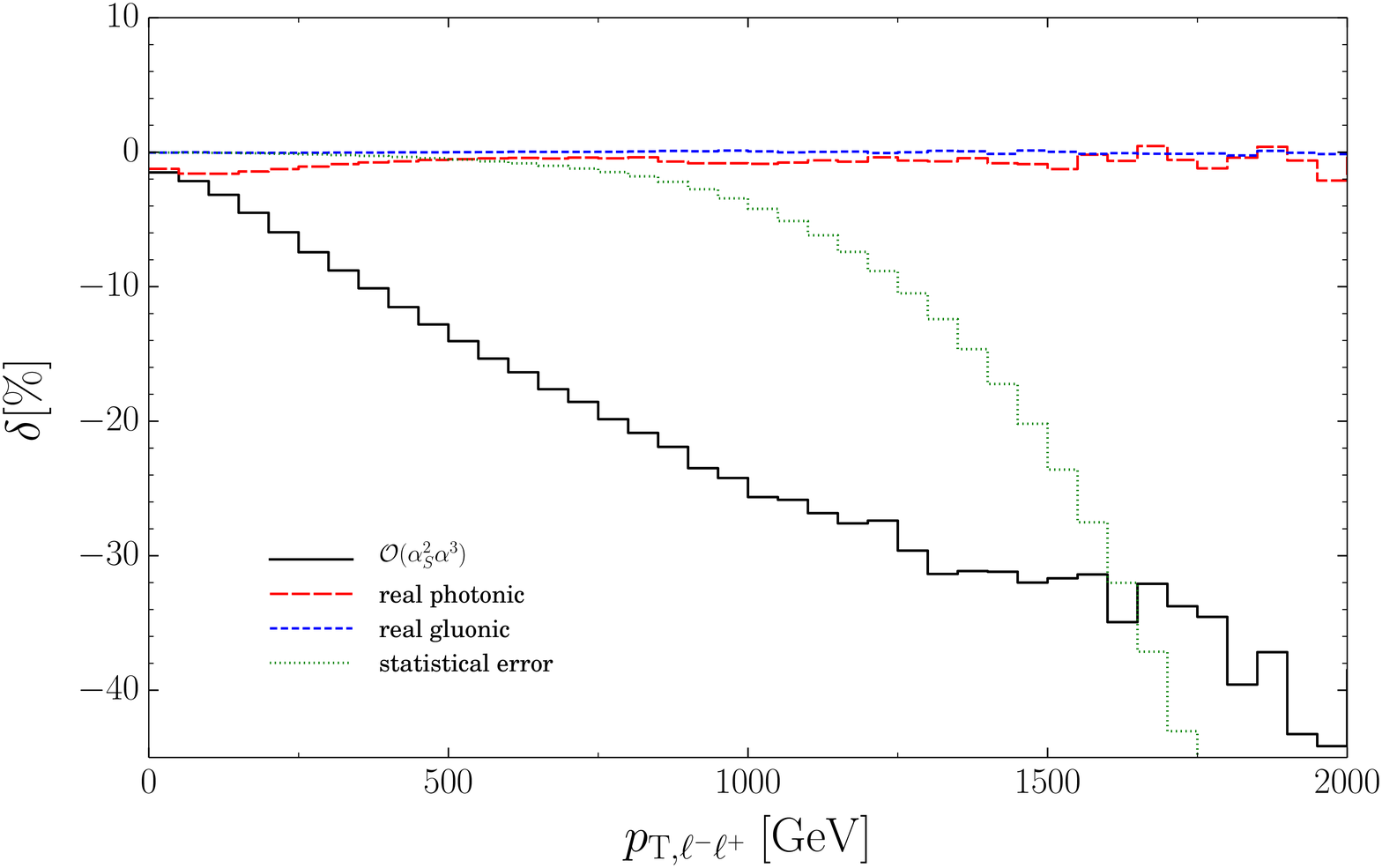}
  \includegraphics[width=8cm]{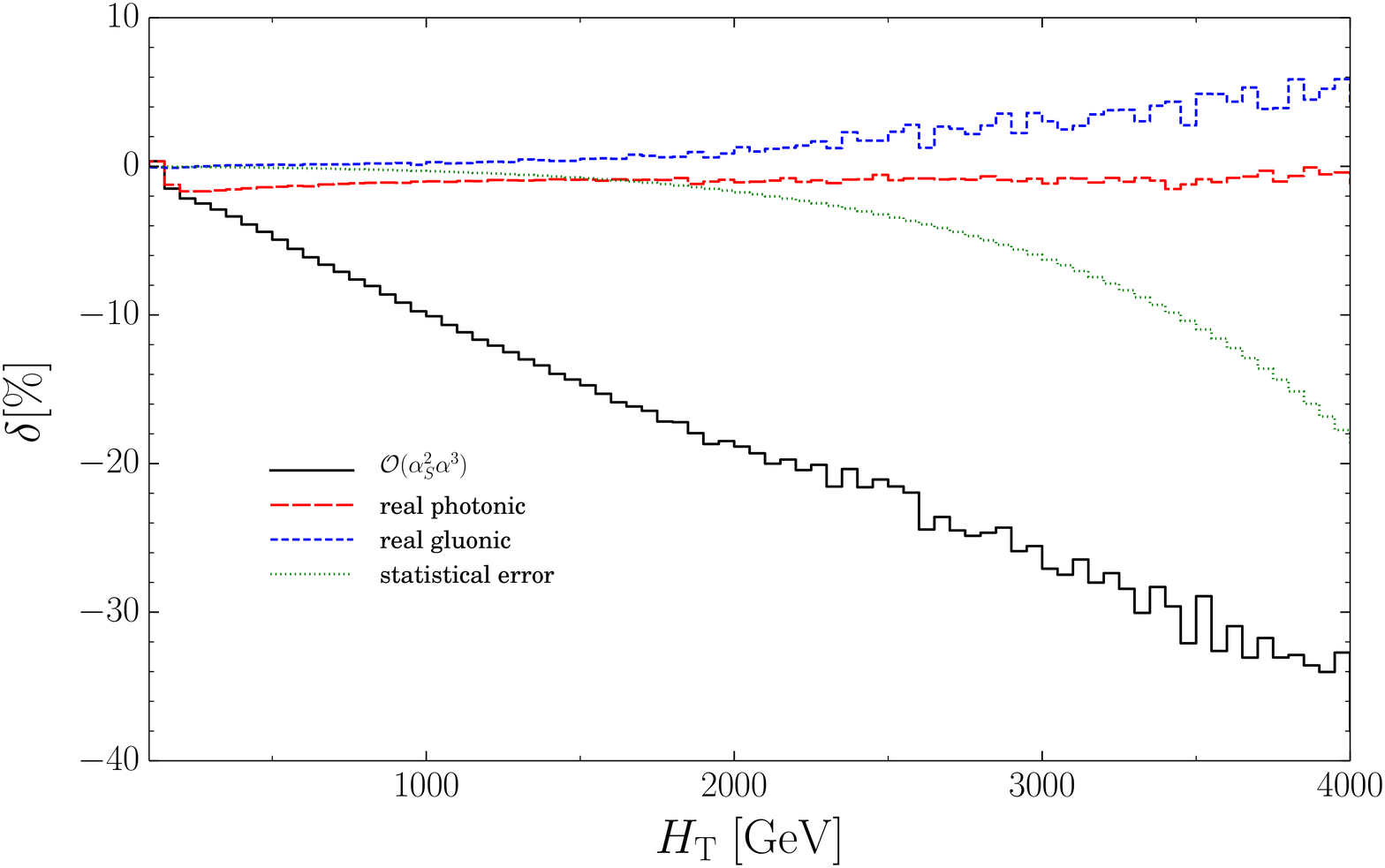}
  \includegraphics[width=8cm]{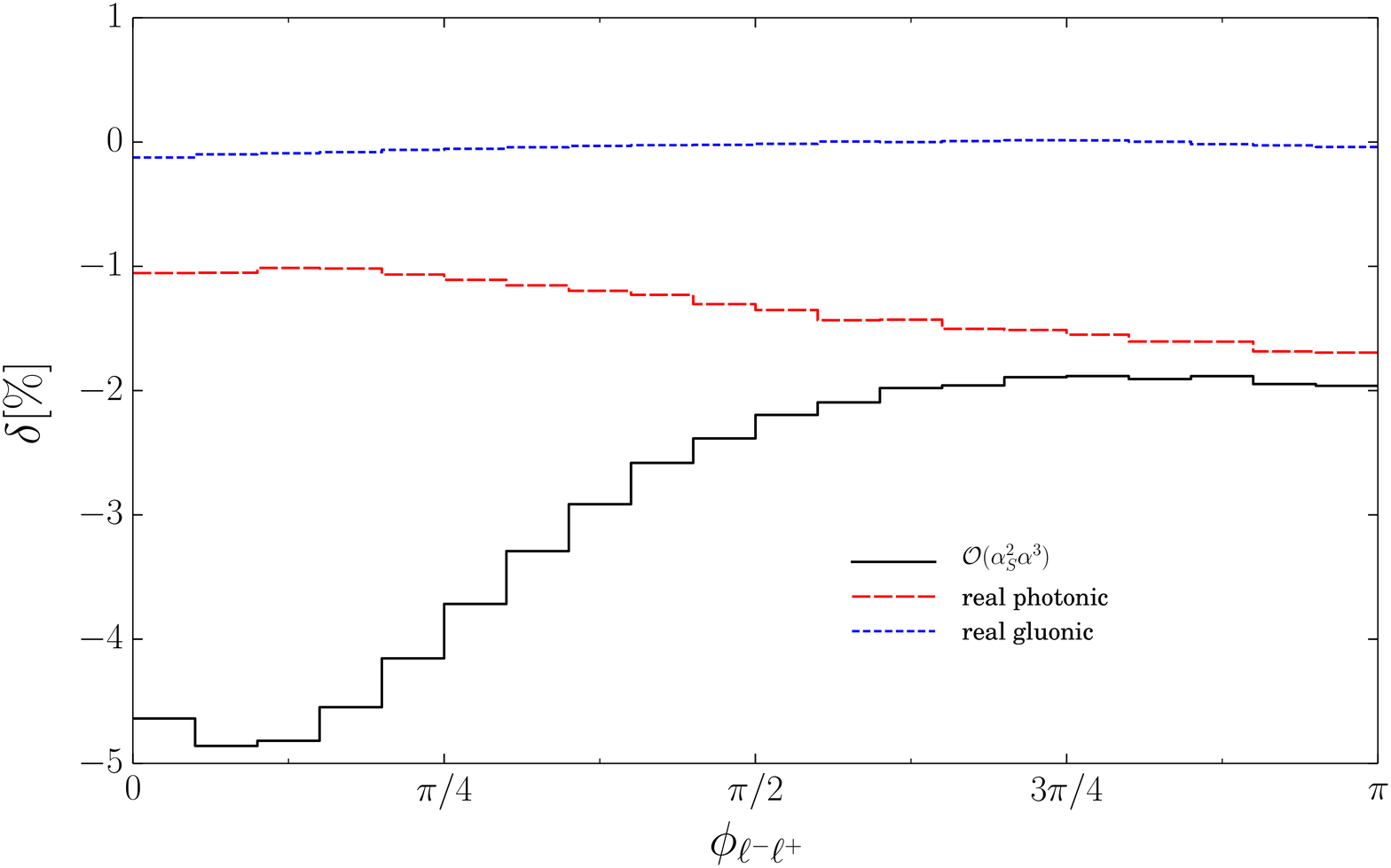}
  \caption{Relative $\ord{(\alphas^2\alpha^3)}$ corrections in per
    cent at the $13 \TeV$ LHC for basic cuts to the distributions in the
    transverse momentum of the harder jet ${\rm j_1}$, the di-jet
    invariant mass $M_{\Pj\Pj}$, the transverse momenta of the
    negatively charged lepton $\Plm$ and the lepton pair $\Plm\Plp$,
    the scalar sum of all transverse momenta, and the relative
    azimuthal angle between the leptons.  Further
    details are described in the text.}
  \label{fig:NLO_LHC13_Std}
\end{figure}

In \reffi{fig:NLO_LHC13_Std} we present the effects of the NLO
electroweak corrections on differential distributions. 
Each plot depicts the correction relative to the LO distribution in
per cent.  The solid (black) curve represents the ratio of the NLO
correction normalised to the LO distribution.  Although unphysical we
show the impact of real contributions (subtracted real contributions
and integrated dipole contributions ignoring the IR poles) from photon
(red, long-dashed) and gluon (blue, short-dashed) emission.  A rough
estimate of the expected experimental statistical error is given by
the green dotted curve, where we anticipate an integrated luminosity
of $300\fba^{-1}$ and count all events in the relevant bin and in all
bins with higher energies.

For the transverse momentum of the hardest jet (upper left plot in
\reffi{fig:NLO_LHC13_Std}) the corrections are small ($\lsim 2\%$) and
negative close to the kinematical threshold. They become more and more
negative with increasing momentum and amount to $-20\%$ for $\pTo{j_1}
= 1\TeV$ and $-30\%$ for $\pTo{j_1} = 2\TeV$. The impact of
real-subtracted photon emission is very small for the whole
$\pTo{j_1}$ range. In contrast the real-subtracted gluon emission is
sensitive to the jet transverse momentum.  With increasing $\pTo{j_1}$
these interference contributions grow in size and reach $+8\%$ of the
LO distribution at
$\pTo{j_1} = 2\TeV$.%
\footnote{We find a noticeable real-subtracted gluon emission in
  distributions where the QCD corrections (see
  \citere{Campbell:2003hd}) are large.}  
The comparison of the real-subtracted contributions with the full NLO
result demonstrates that the virtual corrections dominate where high
energy scales matter.  This behaviour is governed by the Sudakov
logarithms. For the whole $\pTo{j_1}$ range up to $2\TeV$ we find the
NLO corrections to be significant compared to the estimate of the
statistical error.  The relative corrections for the transverse
momentum of the subleading jet (not shown) are similar to those of the
leading jet in shape and size.

The relative EW corrections to the $M_{\Pj\Pj}$ distribution (upper
right plot in \reffi{fig:NLO_LHC13_Std}) stay between $-2\%$ and
$-9\%$ for invariant masses up to $5\TeV$. The corresponding
real-subtracted photonic corrections are also negative and below $2\%$
in absolute value, and the real-subtracted gluonic corrections are
even smaller. Note that for large $M_{\Pj\Pj}$ the EW LO diagrams
contribute $10{-}20\%$ (see \reffi{fig:LO_LHC13_Std}). Since we do
not include EW corrections to these contributions, the full EW
corrections are expected to be even larger than the results presented in
\reffi{fig:NLO_LHC13_Std}. This explains the flattening of the relative
EW corrections in the $M_{\Pj\Pj}$ distribution to some extent.

The relative corrections for the transverse momentum of the negatively
charged lepton and of the lepton pair are presented in the second row
of \reffi{fig:NLO_LHC13_Std}.  The NLO corrections are again negative
but larger in absolute size for $\pTo{\Plm}$ than for the
corresponding $\pTo{\Pj_1}$. For $\pTo{\Plm} = 200\GeV$ we find
corrections of $-10\%$ and for $\pTo{\Plm}$ values around $1\TeV$ the
corrections are of the order of $-30\%$. The contribution from
real-subtracted photon emission behaves similarly as in the
$\pTo{\Pj_1}$ case and remains small over the whole $\pTo{\Plm}$
range.  We observe that the contribution of real-subtracted gluon
emission remains at the per-mille level. For $\pTo{\Plm}\simeq
1400\GeV$ the estimate of the statistical error becomes of the same
size as the NLO corrections.  The relative corrections to the
distribution in the transverse momentum of the positively charged
lepton (not shown) are similar in shape but smaller in magnitude
($-22\%$ at $\pTo{\Plp}\simeq1\TeV$).  The relative corrections to the
distribution in the transverse momentum of the lepton pair are again
dominated by the virtual corrections and reach $-24\%$ at
$\pTo{\Plm\Plp}\simeq1\TeV$.

The relative corrections to the scalar sum of all transverse momenta
$\HT$ as defined in \refeq{eq:def-HT},
shown in the lower left plot of \reffi{fig:NLO_LHC13_Std}, 
exhibit a similar behaviour as for the previously discussed
$p_\mathrm{T}$ distributions.  For $\HT = 2(4) \TeV$ we find the NLO
corrections to be of the order of $-20\%$ ($-35\%$). As for $\pTo{\Pj_1}$
the relative corrections induced by real-subtracted gluon
emission are positive and rise with increasing energy, while the
real-subtracted photon emission remains at the $-1\%$ level.  The
estimate of the statistical error demonstrates that for $300\fba^{-1}$
the corrections are relevant up to $\HT=4{-}5\TeV$.

The relative NLO corrections to the azimuthal angle between the
leptons are shown in the lower right plot of
\reffi{fig:NLO_LHC13_Std}.  Since the angle between the leptons is an
energy-independent observable we do not observe any Sudakov
enhancement. Nevertheless it is interesting that the NLO corrections
induce non-uniform shape changes. For angles $\phill\lsim\pi/2$ the
relative corrections are of the order of $-5\%$. Above $\phill\simeq
\pi/2$ they decrease in absolute value with increasing angle. For
$\phill\simeq\pi$ the NLO corrections reach $-2\%$ and stay constant
for higher $\phill$ values. It is interesting to note that the shape
of the relative corrections is essentially determined by the virtual
corrections.  While the corrections from real-subtracted photon
emission dominate for $\phill \gsim \pi/2$, the contribution from
real-subtracted gluon emission is always below 2\permil.

The relative corrections to the distributions in the rapidities of the
jets and leptons are essentially flat (with a variation $\lsim1\%$).

\subsection{Results for vector-boson-fusion cuts}
\label{sec: NLO VBF cuts}

In this section we present NLO results for the total cross section and
various differential distributions using the input
parameters \refeq{eq:SMinput} and acceptance cuts
\refeq{eq:jetcuts}, \refeq{eq:leptoncuts}, \refeq{eq:DRcuts},
and \refeq{eq:VBFcuts}.  In \refta{tab:NLO_LHC13_VBF} we show the
impact of the NLO corrections on the various subprocesses.
\begin{table}
\begin{center}
\renewcommand{\arraystretch}{1.2}
   \begin{tabular}{|c|c|c|c|}
     \hline
&&&\\[-.4cm]
Process class  & $\sigma^\LO~ [\rm{ fb}] $     & $\sigma_\EW^\NLO~ [\rm{ fb}] $  & $ \frac{\sigma_\EW^\NLO}{\sigma^\LO}-1$ $[\%]$ \\[+.2cm]
\hline 
$u\Pg\to u\Pg \Pl^-\Pl^+$, $d\Pg\to d\Pg \Pl^-\Pl^+$,& \multirow{2}{*}{540.9(3)}  &\multirow{2}{*}{524.0(3) }  & \multirow{2}{*}{-3.12}\\
$\ub \Pg\to \ub \Pg \Pl^-\Pl^+$, $\db \Pg\to \db \Pg \Pl^-\Pl^+$ &&&\\
\hline 
$u\ub \to \Pg\Pg \Pl^-\Pl^+$, $d\db \to \Pg\Pg \Pl^-\Pl^+$ & 22.35(1)    & 21.80(1) & -2.46  \\
\hline 
$\Pg\Pg \to u\ub \Pl^-\Pl^+$, $ \Pg\Pg \to d\db \Pl^-\Pl^+$ & 54.53(2) & 53.34(3)  & -2.18 \\
\hline 
$uu\to uu \Pl^-\Pl^+$, $dd\to dd \Pl^-\Pl^+$, & \multirow{2}{*}{86.22(5)}  &\multirow{2}{*}{ 83.31(7)}  & \multirow{2}{*}{-3.38}\\
$\ub\ub\to \ub\ub \Pl^-\Pl^+$, $\db\db\to \db\db \Pl^-\Pl^+$ &&&\\
\hline 
$u\ub\to u'\ub' \Pl^-\Pl^+$,  $d\db\to d'\db' \Pl^-\Pl^+$, & \multirow{2}{*}{65.98(3)}  &\multirow{2}{*}{63.76(4)}  & \multirow{2}{*}{-3.36}\\
$u\ub'\to u\ub' \Pl^-\Pl^+$,  $d\db'\to d\db' \Pl^-\Pl^+$ &&&\\
\hline 
$u\ub\to d\db \Pl^-\Pl^+$, $d\db\to u\ub \Pl^-\Pl^+$, & \multirow{2}{*}{21.198(7)} &\multirow{2}{*}{20.905(7) }  & \multirow{2}{*}{-1.38 }\\
$u\ub'\to d\db' \Pl^-\Pl^+$,  $d\db'\to u\ub' \Pl^-\Pl^+$ &&&\\
\hline 
$ud\to u'd' \Pl^-\Pl^+$, $\ub\db\to \ub'\db' \Pl^-\Pl^+$,& \multirow{4}{*}{180.61(8)}  &\multirow{4}{*}{176.3(1)}  & \multirow{4}{*}{-2.39}\\
$ud\to ud \Pl^-\Pl^+$, $\ub\db\to \ub\db \Pl^-\Pl^+$&&& \\
$uu'\to uu' \Pl^-\Pl^+$, $\ub\ub'\to \ub\ub' \Pl^-\Pl^+$&&& \\
$dd'\to dd' \Pl^-\Pl^+$, $\db\db'\to \db\db' \Pl^-\Pl^+$&&& \\
     \hline 
     $u\db\to u'\db' \Pl^-\Pl^+$, $\ub d\to \ub'd' \Pl^-\Pl^+$, & \multirow{2}{*}{67.73(2)} &\multirow{2}{*}{65.92(3)}  & \multirow{2}{*}{-2.67}\\
$u\db\to u\db \Pl^-\Pl^+$, $\ub d\to \ub d \Pl^-\Pl^+$&&&\\
     \hline
     \hline gluonic    &  617.8(4) & 599.2(3)  & -3.01  \\
     \hline four-quark &  421.7(1) & 410.2(1)  & -2.73 \\
     \hline sum        & 1039.6(4) & 1009.3(3) & -2.91 \\
     \hline
   \end{tabular}
\end{center}
  \caption{NLO corrections to the total cross section for 
   $\process$ at the $13\TeV$ LHC for VBF cuts. In the first column the
    partonic processes are listed at LO. The second column
   provides the corresponding cross section including the NLO corrections. The third
   column contains the relative contribution to the total cross section
   in per cent.}
  \label{tab:NLO_LHC13_VBF}
\end{table}
The corrections to the partonic total cross sections are again
negative for all process classes. The spread of the corrections for
the different channels is somewhat larger as for basic cuts ranging
from $-3.4\%$ to $-2.2\%$ and the correction for the total hadronic
cross section amounts to $-2.7\%$.  The relative correction for the
channels of type $u\ub\to d\db \Pl^-\Pl^+$ should be regarded with
care. These channels are dominated by LO EW diagrams (see
\refta{tab:LO-LHC13-VBF-xsection}), for which the EW corrections are
not included in our calculation. The full EW corrections might be
different for these channels. This caveat applies to a lesser extent
also to the channels of type $ud\to u'd' \Pl^-\Pl^+$, while for all
other channels and for the sums the EW corrections to the LO EW
diagrams are negligible.

\begin{figure}
  \includegraphics[width=8cm]{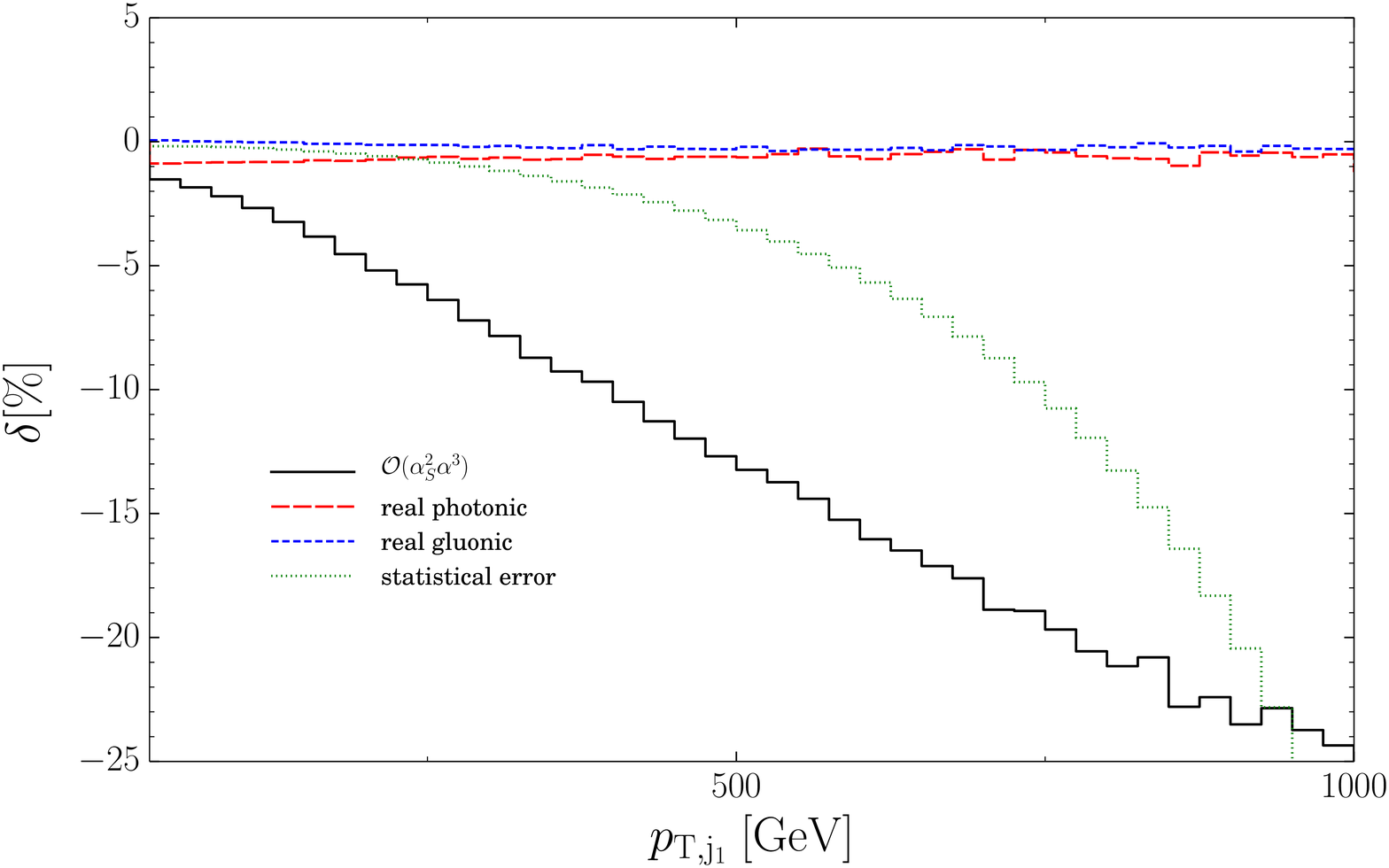}
  \includegraphics[width=8cm]{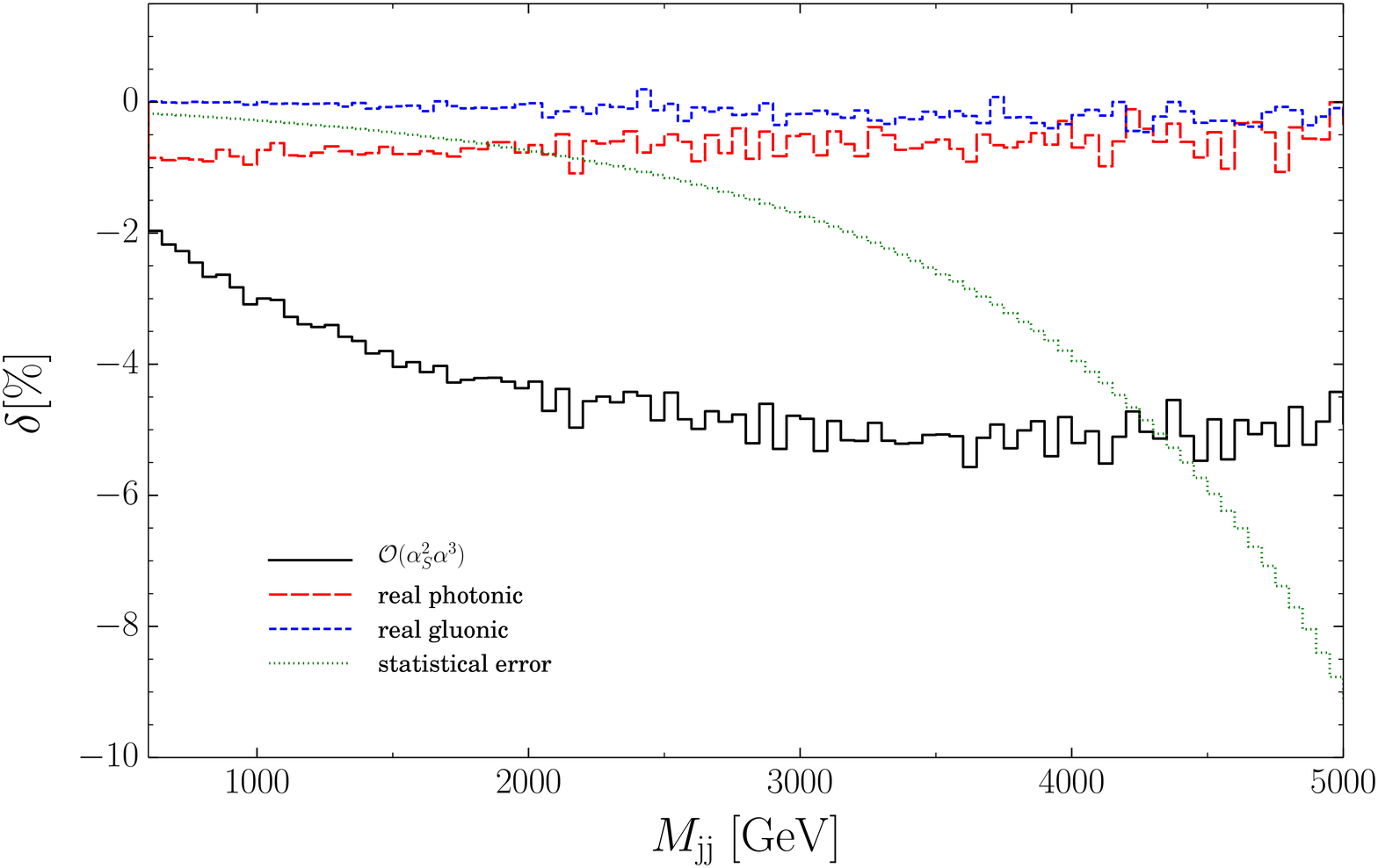}
  \includegraphics[width=8cm]{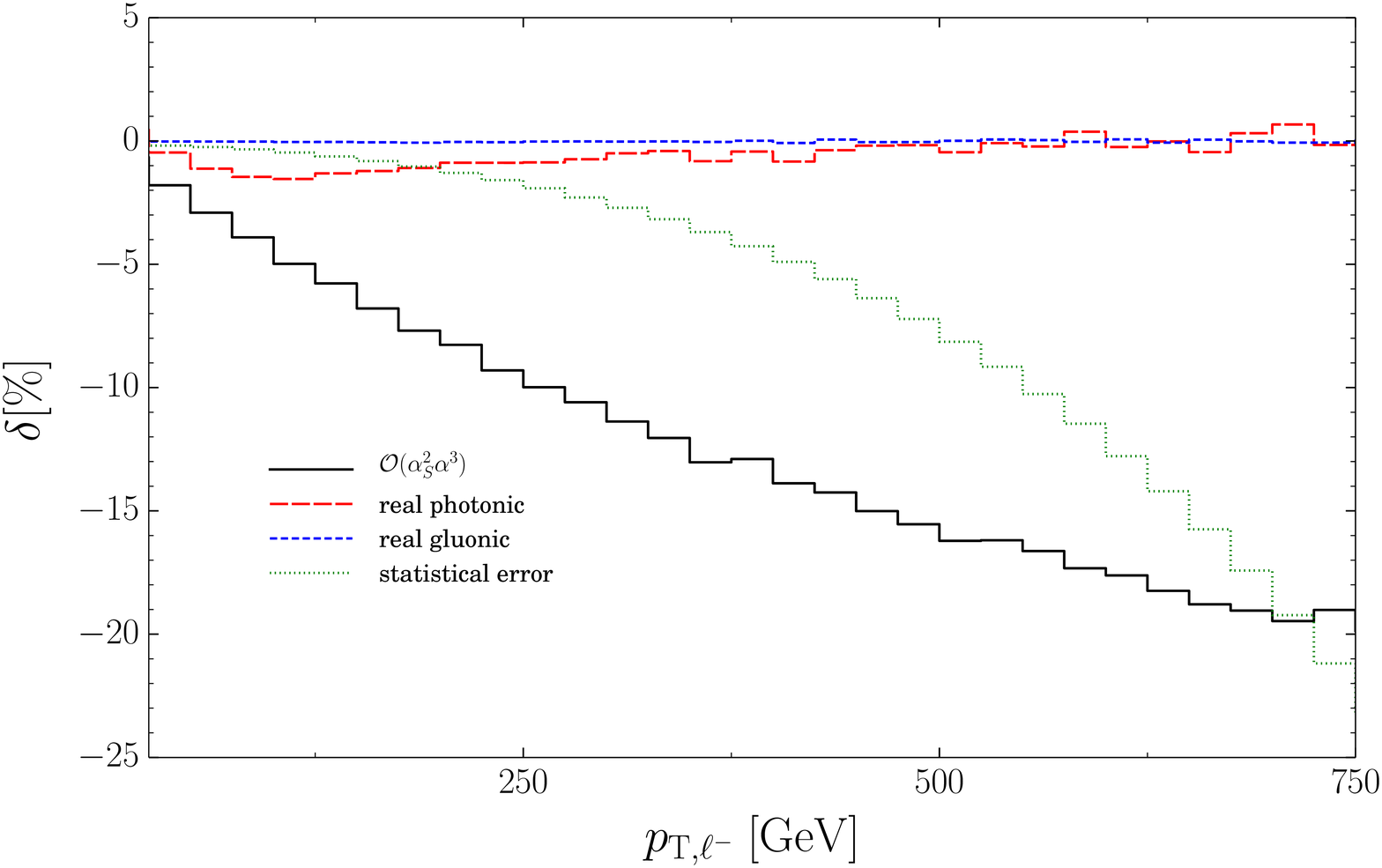}
  \includegraphics[width=8cm]{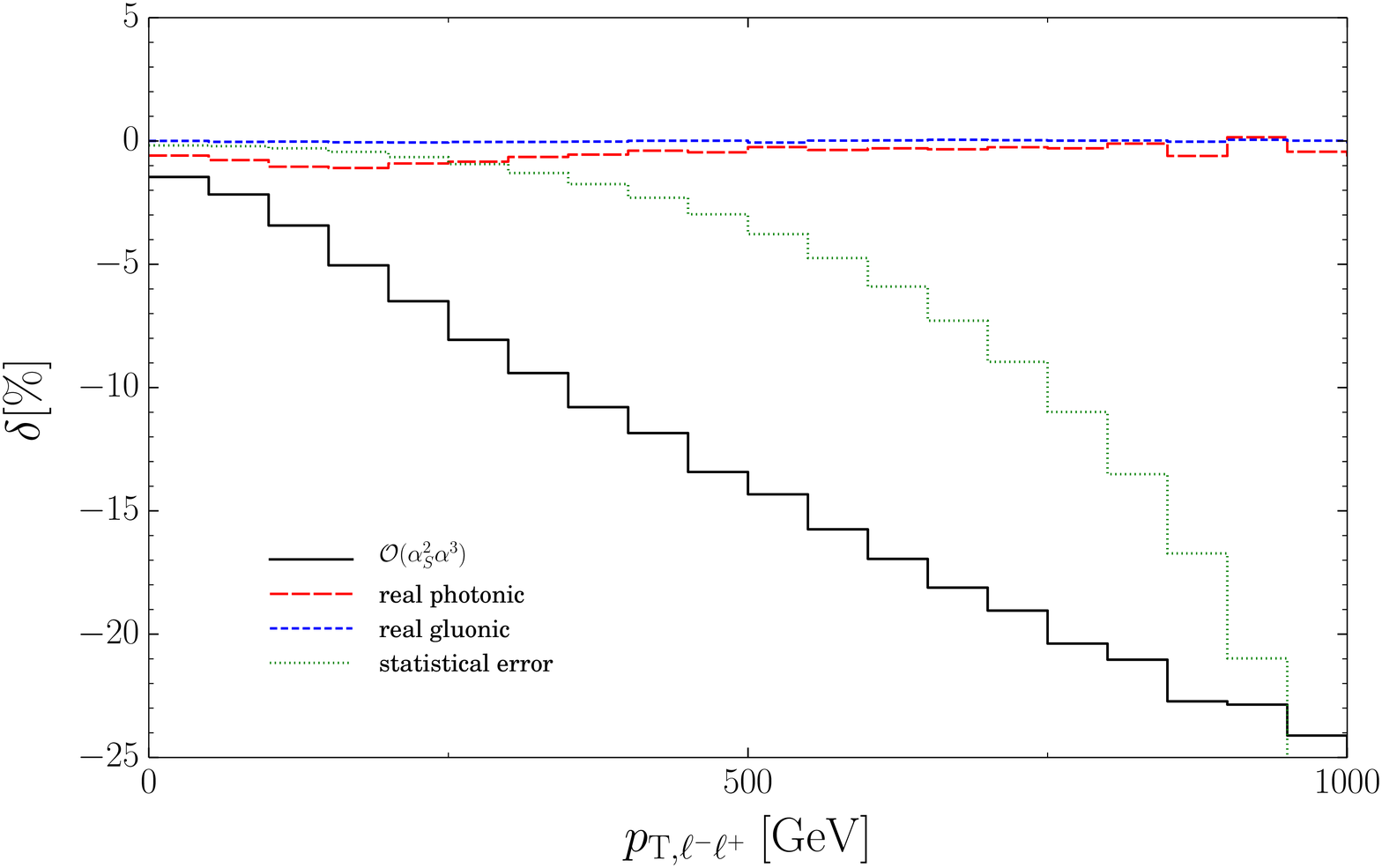}
  \includegraphics[width=8cm]{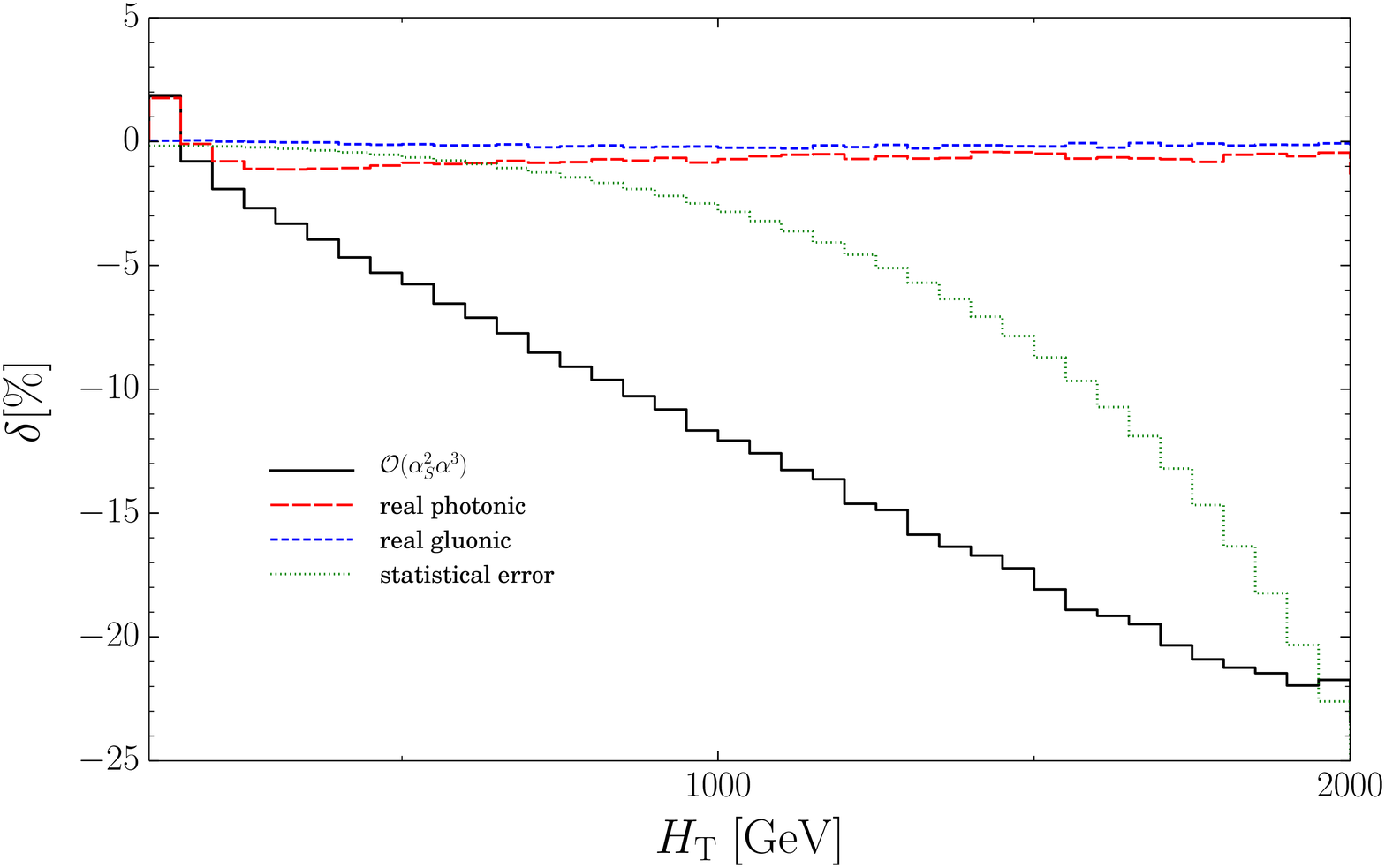}
  \includegraphics[width=8cm]{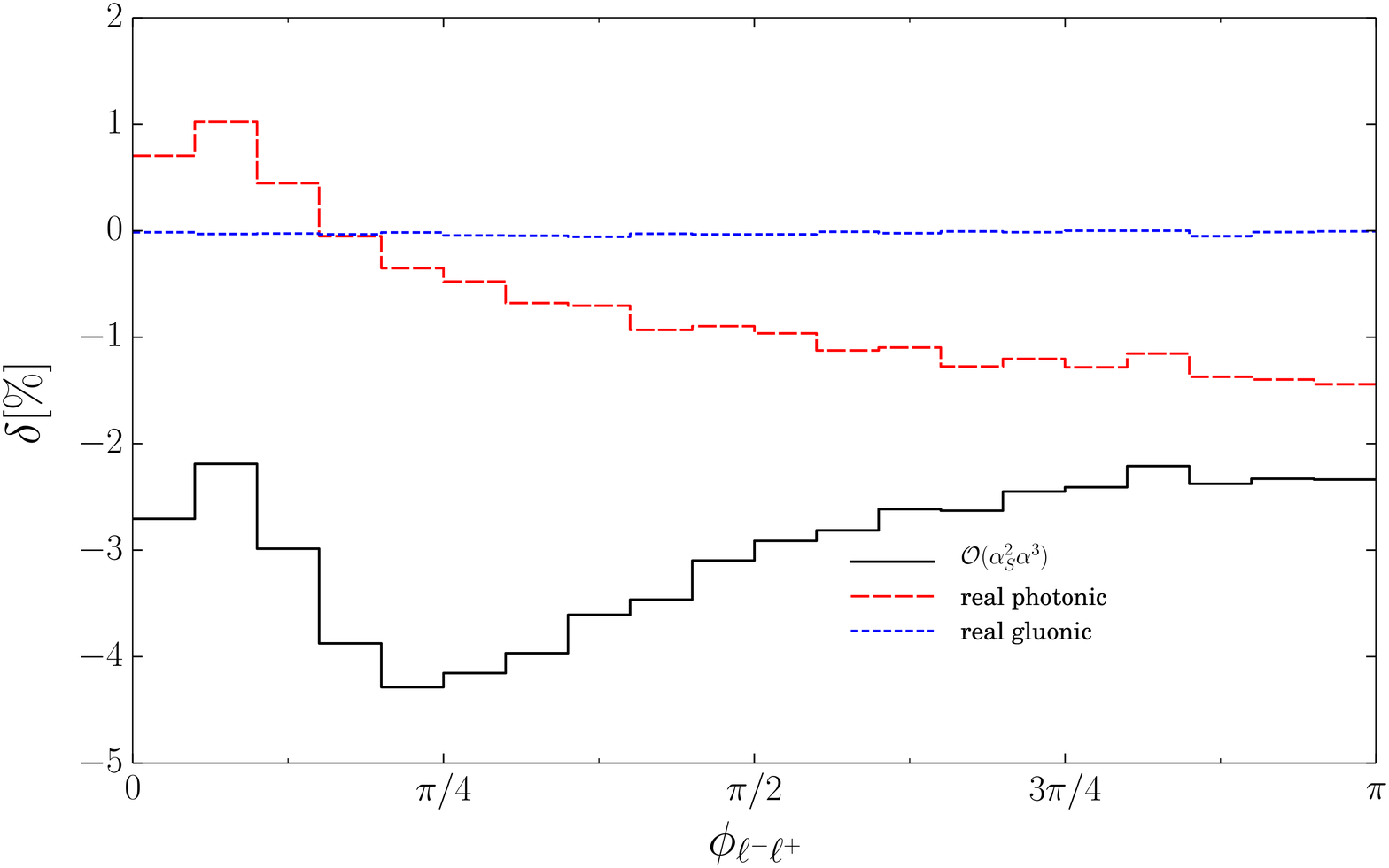}
  \caption{Relative $\ord{(\alphas^2\alpha^3)}$ corrections in per
    cent at the $13 \TeV$ LHC for VBF cuts to the distributions in the
    transverse momentum of the harder jet ${\rm j_1}$, the di-jet
    invariant mass $M_{\Pj\Pj}$, the transverse momenta of the
    negatively charged lepton $\Plm$ and the lepton pair $\Plm\Plp$,
    the scalar sum of all transverse momenta, and the relative
    azimuthal angle between the leptons.  Further
    details are described in the text.}
  \label{fig:NLO_LHC13_VBF}
\end{figure}

In \reffi{fig:NLO_LHC13_VBF} we present the effects of the NLO
electroweak corrections on differential distributions for VBF cuts
using the same style as in \reffi{fig:NLO_LHC13_Std}.
For all considered distributions we find relative corrections from
subtracted real photon and real gluon contributions below $\sim2\%$
and $\sim1\%$, respectively.  Thus, in particular for high energy
scales, the virtual corrections and more precisely the Sudakov
corrections dominate again. The effects of the total
$\ord{(\alphas^3\al^2)}$ corrections are qualitatively similar as for
basic cuts, but quantitatively we find some differences.

For the transverse momentum of the hardest jet (upper left plot in
\reffi{fig:NLO_LHC13_VBF}) the EW corrections are somewhat larger as
for basic cuts and reach $-25\%$ for $\pTo{\Pj_1}=1\TeV$.  For the
transverse momentum of the second hardest jet (not shown), the
corrections are similar as for basic cuts. For the di-jet invariant
mass distribution the corrections are somewhat smaller and do not
exceed $-6\%$ (upper right plot in \reffi{fig:NLO_LHC13_VBF}).
Note, however, that in this case the EW LO diagrams amount up to
$50\%$ and the missing EW corrections to these contributions could
enhance the EW corrections even by a factor $2$. 

The relative corrections for the transverse momentum of the negatively
charged lepton are shown in the middle left plot of
\reffi{fig:NLO_LHC13_VBF}.  The NLO corrections are somewhat smaller
in absolute size than for the same distribution with basic cuts and do
not exceed $-20\%$ even for $\pTo{\Plm}=750\GeV$.  For
$\pTo{\Plm}\simeq 700\GeV$ the estimate of the statistical error
becomes of the same size as the NLO corrections.  The relative
corrections to the distribution in the transverse momentum of the
positively charged lepton (not shown) are again smaller in magnitude
($-17\%$ at $750\GeV$). The distribution in the transverse momentum of
the lepton pair (middle left plot in \reffi{fig:NLO_LHC13_VBF})
receives EW corrections of $-25\%$ at $\pTo{\Plm\Plp}\simeq 1\TeV$.

The relative corrections to the scalar sum of all transverse momenta
$\HT$ shown in the lower left plot of \reffi{fig:NLO_LHC13_VBF} behave
similar as those for the $\pTo{\Pj_1}$ distribution. They are larger
as for basic cuts and are relevant up to $\HT=2\TeV$ for
$300\fba^{-1}$.

The relative NLO corrections to the azimuthal angle between the
leptons are depicted in the lower right plot of
\reffi{fig:NLO_LHC13_VBF}.  For VBF cuts the real photonic corrections
vary between $1\%$ and $-1\%$.  The complete EW corrections range
between $-2\%$ and $-4\%$.

While the relative real corrections to the distributions in the
rapidities of the jets and leptons are still relatively flat, the
virtual corrections induce variations at the level of two per cent.

\section{Conclusions}
\label{sec: conclusions}

We have calculated the electroweak corrections to the production of
lepton pairs in association with two jets at the LHC. More precisely,
we have computed the corrections of absolute order
$\ord{(\alphas^2\alpha^3)}$, consisting of electroweak corrections to
the {squared} leading-order QCD diagrams and QCD corrections to the
leading-order QCD--EW interferences, for the full process $\process$
including all off-shell effects.  The calculation has been performed
with the recursive matrix element generator \recola\ and the one-loop
integral library \collier\ and demonstrates the strength of these
codes for multiparticle NLO calculations.  For the phase-space
integration we use multi-channel
Monte-Carlo techniques, dipole subtraction and the fragmentation
function to separate photons and jets.

We have discussed results for total cross sections and various
distributions within a basic set of cuts and within a set of cuts
typical for vector-boson fusion. 
Electroweak corrections change the total cross sections by a few
per cent and distort angular distributions at a similar level.
However, in the high-energy tails of invariant-mass and
transverse-momentum distributions the electroweak corrections amount
to several tens of per cent and are typically larger than the expected
experimental errors. These large effects can be traced back to the
virtual corrections and thus to the Sudakov logarithms.

\subsection*{Acknowledgements}
This work was supported in part by the Deutsche Forschungsgemeinschaft
(DFG) under reference number DE~623/2-1. The work of L.H. was
supported by the grant FPA2011-25948.

\appendix
\section{Appendix}
\label{app: photon fragmentation}

In the following we describe our implementation of the photon
fragmentation in the context of the dipole method for a consistent
recombination of photons and jets. We follow closely
\citeres{Denner:2009gj,Denner:2010ia} but instead of regularising IR
singularities by infinitesimal photon and fermion masses we apply
dimensional regularisation. Furthermore, we have generalised the
results to include the $\alpha$-dependent subtraction terms introduced
in \citeres{Nagy:1998bb,Nagy:2003tz,Campbell:2004ch} which allow to
vary the size of the subtraction region in phase space.

In the framework of the subtraction procedure \cite{Catani:1996vz} the
NLO cross-section for the process $\process$ can
schematically be written as
\beq
\dd\sigma_{\rm NLO} = \int_{n+1} \Big[\dd \sigma_{\rm real} - \dd \sigma_{\rm dipole}\Big]\;\;  + \;\;\int_n~  \Big[\dd \sigma_{\rm virtual} +  \int_1~  \dd \sigma_{\rm dipole} \Big],
\label{eq:NLO-x-section}
\eeq
where adding and subtracting the dipole contribution $\dd \sigma_{\rm
  dipole}$ renders the $n$- and the $(n+1)$- dimensional phase-space
integrals separately IR-finite (in our case $n=4$).  When calculating
photonic NLO corrections for processes with final-state quarks and
gluons, however, the above procedure suffers from IR-unsafe
configurations in $\dd \sigma_{\rm real}$ arising from the
recombination of a gluon with a hard photon to form a jet. Such jets
that are dominated by hard photons are called {\it hard-photon jets}
in the following.

Since a recombined $\Pg\gamma$ configuration with a soft gluon forms
part of the NLO QCD corrections to the process $\Pp\Pp \to \Pl^-\Pl^+
\Pj \gamma$, the corresponding IR singularity is not cancelled by the
QED dipole term $\dd \sigma_{\rm dipole}$ constructed for the NLO QED
corrections to the process $\process$.  To solve this problem we
slightly modify the jet algorithm such that hard-photon jets are
discarded. For the recombination of a QCD parton with energy $E_a$ and
a photon with energy $E_\gamma$ we consider the photon-jet energy
fraction
\beq
z_\gamma = \frac{E_\gamma}{E_\gamma + E_a}.
\eeq
The type of the recombined particle is then determined as a function
of $z_\gamma$.  For $z_\gamma < \zgacut$ we treat the recombined
particle as a jet, for $z_\gamma > z_{\gamma, {\rm cut}}$ as a photon.
Using this criterion to reject hard-photon jets,
\refeq{eq:NLO-x-section} becomes
\ba
\dd\sigma_{\rm NLO} &=& \int_{n+1} \theta(\zgacut - z_{\gamma})\Big[\dd \sigma_{\rm real} - \dd \sigma_{\rm dipole}\Big]\;\; + \nonumber\\
     && \int_n~  \Big[\dd \sigma_{\rm virtual} + \int_1~ \left( \dd \sigma_{\rm dipole} - 
             \dd \sigma_{\rm dipole}^{\gamma \;\rm coll}(\zgacut)\right) - \dd \sigma_{\rm frag}(\zgacut) \Big].
\label{eq:NLO-x-section2}
\ea
Inside the $n$-dimensional phase-space integral the rejection of
jets dominated by hard photons leads to a modification of the original dipole
contribution by the additional term 
\beq
  \dd \sigma_{\rm dipole}^{\gamma \;\rm coll}(\zgacut) = \theta(z_{\gamma}-\zgacut)\dd \sigma_{\rm dipole}.
\eeq
Further, since the final state is not fully inclusive anymore,
contributions in which the final-state quark fragments into a
hard photon have to be subtracted by means of 
\beq
  \dd \sigma_{\rm frag}(\zgacut) = \sum\limits_i \dd \sigma_{\rm born}\int_{\zgacut}^1\dd z_{\gamma}\;D_{q_i\to\gamma}(z_{\gamma})\,,
\eeq
where $D_{q_i\to\gamma}$ is the quark-to-photon fragmentation function and
the sum covers all quarks $q_i$ in the final state.

Since photon fragmentation is related to final-state singularities,
only dipoles with final-state emitter are affected and are considered
in the following.  The results for gluon emission from final-state
quarks of \citeres{Catani:1996vz,Nagy:2003tz} can easily be
transferred to the photon case by replacing $\alpha_s\to\alpha$ and
$C_F\to Q^2$, and by substituting colour-correlated matrix elements by
their charge-weighted counterparts
\beq
  Q_i\sigma_iQ_j\sigma_j |\mathcal{M}_{\rm born}|^2.
\eeq
Here, $Q_{i,j}$ represent the charges of the emitter and spectator
quarks ($Q_{i,j}=2/3$ for up-type and $Q_{i,j}=-1/3$ for down-type
quarks), while $\sigma_{i,j}$ denotes the corresponding charge flow
($\sigma_{i,j}=+1$ for incoming quarks and outgoing anti-quarks,
$\sigma_{i,j}=-1$ for incoming anti-quarks and outgoing quarks). The
integrated dipole contribution $\int_1\dd\sigma_{\rm dipole}$ for {\it
  final-state emitter with final-state spectator} can be obtained from
eq.~(5.32) of \citere{Catani:1996vz} and adding the $\alpha$-dependent
terms of eq.~(11) of \citere{Nagy:2003tz} reads
\beq
\mathcal{V}_{qg}(\alpha,\eps)= Q_q^2
\left(\frac{1}{\eps^2}+\frac{3}{2\eps}+5-\frac{\pi^2}{2}-\ln^2\alpha+\frac{3}{2}(\alpha-1-\ln\alpha)\right)\,.  
\eeq
For {\it final-state emitter with initial-state spectator} the
$\alpha$-extended version of eq.~(5.57) of \citere{Catani:1996vz} can
be obtained from eqs.~(5.50) and (5.54) of this paper by restricting
the integration range of $x$ to the interval $[1-\alpha,1]$ resulting
in 
\ba 
\mathcal{V}_{qg}(x;\alpha,\eps)&=& Q_q^2\biggl\{
\left(\frac{2}{1-x}\ln\frac{1}{1-x}\right)_{\!1-\alpha}\!-\frac{3}{2}\left(\frac{1}{1-x}\right)_{\!1-\alpha}\!+\Theta(x-1+\alpha)\frac{2}{1-x}\ln(2-x)\nonumber\\
&&{}+\delta(1-x)\left[\mathcal{V}_{qg}(1,\eps)-\frac{3}{2}-\ln^2\alpha-\frac{3}{2}\ln\alpha\right]\biggr\}.
\ea 
The $(1-\alpha)$\,-\,distribution is a modification of the usual
$+$-distribution and defined as
\beq
\int_0^1 \dd x ~f(x)_{1-\alpha} g(x) = \int_{1-\alpha}^1 \dd x ~f(x)\left[g(x) -g(1)\right].
\eeq

The additional dipole subtraction term $\int_1\dd\sigma_{\rm
  dipole}^{\gamma\;\rm coll}$ can be represented in a completely
analogous manner.  For {\it final-state emitter with final-state
  spectator} it results from eqs.~(5.28) and (5.29) of
\citere{Catani:1996vz} by restricting the $y$ integration by $\alpha$
and the $z$ integration by $z_{\gamma,\mathrm{cut}}$,
\ba
  \mathcal{V}_{qg}^{\gamma\;\rm coll}(x;\alpha,\eps,\zgacut)&=& Q_q^2
  \int_0^{1-\zgacut}\dd z\left(z(1-z)\right)^{-\eps}
  \int_0^{\al}\rd y \,y^{-\eps-1}(1-y)^{1-2\eps}\nn\\
  &&\qquad\times\left[\frac{2}{1-z+yz}-(1+z)-\eps(1-z)\right]\nn\\
  &=&Q_q^2\int_0^{1-z_{\gamma,\rm cut}}\dd z  \left\{\frac{1+z^2}{1-z}\left(-\frac{1}{\eps}+\ln[z(1-z)]+\ln\alpha\right)+\alpha(1+z)+1-z\right.\nonumber\\
  &&\hspace{2.5cm}\left.{}-\frac{2}{z(1-z)}\ln\left(\frac{1-z+\alpha
        z}{1-z}\right)\right\}\,.
\label{eq:FEFS}
\ea
Similarly the contribution from {\it final-state emitter with
initial-state spectator} to the additional dipole subtraction term
is obtained from eqs. (5.50) and (5.54) of \citere{Catani:1996vz} upon
using appropriate integration boundaries
\ba
\mathcal{V}_{qg}^{\gamma\;\rm coll}(x;\alpha,\eps,\zgacut)&=& 
Q_q^2
\theta(1-x)\theta(x-1+\al)(1-x)^{-1-\eps}
  \int_0^{1-\zgacut}\dd z\left(z(1-z)\right)^{-\eps}
  \nn\\
  &&\qquad\times\left[\frac{2}{2-z-x}-(1+z)-\eps(1-z)\right]
\nn\\
  &=&
 Q_q^2
\int_0^{1-z_{\gamma,\rm cut}}\dd z \, \biggl\{
\left[\frac{1}{1-x}\left(\frac{2}{2-z-x}-1-z\right)\right]_{1-\alpha}\nonumber\\
&&\qquad{}+\de(1-x)\biggl[\frac{1+z^2}{1-z}\left(-\frac{1}{\eps}+\ln[z(1-z)]+\ln\alpha\right)\nonumber\\
&&\hspace{2.5cm}\left.{}+(1-z)-\frac{2}{1-z}\ln\left(\frac{1-z+\alpha}{1-z}\right)\biggr]\right\}\,.
\label{eq:FEIS}
\ea

Equations \refeq{eq:FEFS} and \refeq{eq:FEIS} contain collinear
singularities which are cancelled by the contribution $\dd \sigma_{\rm
  frag}$ once the photon fragmentation function $D_{q_i\to\gamma}$ is
renormalised, reflecting the IR-safety of our event definition. To
this end we follow closely the procedure in
\citeres{Denner:2009gj,Denner:2010ia} and relate $D_{q_i\to\gamma}$ to
the parametrisation $D_{q\to \gamma}^{\rm ALEPH,~\overline{MS} }$
measured at ALEPH \cite{Buskulic:1995au}. We find
\beq
D_{q\to\gamma} = \frac{\alpha Q_q^2}{2\pi} P_{q\to \gamma} (z_\gamma) 
\left[\frac{(4\pi)^\eps}{\Gamma(1-\eps)} \frac{1}{\eps} +\ln\left(\frac{\mu^2}{\muF^2}\right)\right] + D_{q\to \gamma}^{\rm ALEPH,~\bar{MS} }(z_\gamma,\muF),
\label{eq:frag}
\eeq
where we identify 
\beq
P_{q\to \gamma}(\zg) = \frac{1+(1-\zg)^2}{\zg}  \xrightarrow{\zg = 1-z} \frac{1+z^2}{1-z} 
\eeq
and 
\ba
D_{q\to \gamma}^{\rm ALEPH,~\overline{MS} }(1-z,\muF) 
&=& \frac{\alpha Q_q^2}{2\pi}\left[ \frac{1+z^2}{1-z} \ln\left(\frac{\muF^2}{z^2\mu_0^2} \right) + C\right].
\ea
The $\muF$-dependence of the photon fragmentation function cancels by construction. The constants $\mu_0$ and $C$ are fit parameters extracted from the experimental
measurement \cite{Buskulic:1995au}. They are given as
\beq
\mu_0 = 0.14\GeV, \quad C=-13.26.
\eeq

\end{document}